\documentclass[a4paper,fleqn]{cas-dc}
\usepackage{enumitem}
\usepackage[numbers]{natbib}
\usepackage[figuresright]{rotating}
\usepackage{moreverb}
\usepackage{amsmath,amssymb,amsfonts}
\usepackage{algorithmic}
\usepackage{url}
\usepackage{hyperref}

\usepackage{tabularx}
\usepackage{pifont}
\usepackage{subfig}
\usepackage{amssymb}
\usepackage{fontawesome}
\usepackage{longtable}
\usepackage{graphicx}
\usepackage{textcomp}
\usepackage{xcolor}
\usepackage{tcolorbox}
\usepackage{ragged2e}
\usepackage{balance}
\usepackage{colortbl}
\usepackage{multirow}
\usepackage{color}
\usepackage{subfig}
\usepackage{subcaption} 
\usepackage{booktabs} 

\usepackage{CJKutf8}
\usepackage{siunitx}
\usepackage{rotfloat}
\usepackage{float}
\usepackage{lscape}
\usepackage{amssymb,mathrsfs,amsmath}
\usepackage{adjustbox}
\usepackage{pifont} 
\newcommand{\cmark}{\ding{51}} 
\newcommand{\xmark}{\ding{55}} 
\newcommand{\pmark}{\textbf{$\triangle$}} 

\usepackage{afterpage}
\usepackage{graphicx} 
\usepackage{caption}   
\usepackage{subcaption} 
\usepackage{colortbl}  
\def\tsc#1{\csdef{#1}{\textsc{\lowercase{#1}}\xspace}}
\tsc{WGM}
\tsc{QE}
\tsc{EP}
\tsc{PMS}
\tsc{BEC}
\tsc{DE}

\begin{document}
\begin{sloppypar}

\let\WriteBookmarks\relax
\def\floatpagepagefraction{1}
\def\textpagefraction{.001}
\shorttitle{Containerization in Multi-Cloud Environment}
\shortauthors{Waseem et al.}
\title [mode = title]{Containerization in Multi-Cloud Environment: Roles, Strategies, Challenges, and Solutions for Effective Implementation}

\author[tuni]{Muhammad Waseem\corref{cor1}}
\ead{muhammad.waseem@tuni.fi}

\author[lanc]{Aakash Ahmad}
\ead{a.ahmad13@lancaster.ac.uk}

\author[whu]{Peng Liang}
\ead{liangp@whu.edu.cn}

\author[lut]{Muhammad Azeem Akbar}
\ead{azeem.akbar@lut.fi}

\author[oulu]{Arif Ali Khan}
\ead{Arif.Khan@oulu.fi}

\author[tieto]{Iftikhar Ahmad}
\ead{iftikhar.ahmad@tietoevry.com}

\author[solita]{Manu Setälä}
\ead{manu.setala@solita.fi}

\author[jyu]{Tommi Mikkonen}
\ead{tommi.j.mikkonen@jyu.fi}

\address[tuni]{Faculty of Information Technology and Communication Sciences, Tampere University, Tampere, Finland}
\address[lanc]{School of Computing and Communications, Lancaster University Leipzig, Leipzig, Germany}
\address[whu]{School of Computer Science, Wuhan University, Wuhan, China}
\address[lut]{Software Engineering Department, Lappeenranta-Lahti University of Technology, Lappeenranta, Finland}
\address[oulu]{M3S Empirical Software Engineering Research Unit, University of Oulu, Oulu, Finland}
\address[tieto]{TietoEVRY Oy, Tampere, Finland}
\address[solita]{Solita Oy, Tampere, Finland}
\address[jyu]{Faculty of Information Technology, University of Jyväskylä, Jyväskylä, Finland}

\begin{abstract}
Containerization in multi-cloud environments has received significant attention in recent years both from academic research and industrial development perspectives. However, there exists no effort to systematically investigate the state of research on this topic. The aim of this research is to systematically identify and categorize the multiple aspects of containerization in multi-cloud environment. 
We conducted the Systematic Mapping Study (SMS) on the literature published between January 2013 and July 2024. One hundred twenty one studies were selected and the key results are: (1) Four leading  themes on containerization in multi-cloud environment are identified: ‘Scalability and High Availability’, ‘Performance and Optimization’, ‘Security and Privacy’, and ‘Multi-Cloud Container Monitoring and Adaptation’. (2) Ninety-eight patterns and strategies for containerization in multi-cloud environment were classified across 10 subcategories and 4 categories. (3) Ten quality attributes considered were identified with 47 associated tactics. (4) Four catalogs consisting of challenges and solutions related to security, automation, deployment, and monitoring were introduced. The results of this SMS will assist researchers and practitioners in pursuing further studies on containerization in multi-cloud environment and developing specialized solutions for containerization applications in multi-cloud environment. 
\end{abstract}


\maketitle

\section{Introduction}
\label{Sec_Introduction}
The use of containers in multi-cloud environment has been widespread in the industry for many years \cite{naik2016building}. Containers are standalone and executable packages of software that include everything needed to run an application, such as code, system tools, libraries, and settings \cite{merkel2014docker}. Containers allow the packaging of an application and its required dependencies, which makes it easy to move the application across different environments with minimal modification \cite{docker2021}. On the other hand, multi-cloud environment is the distribution of cloud assets \cite{petcu2013multi}. Using containers in multi-cloud environment allows organizations to achieve  flexibility, agility, and cost-efficiency. Developers can build applications in a consistent environment and easily move them between multiple cloud platforms, thereby leveraging the unique strengths (e.g., extensive infrastructure, seamless integration, advanced data analytic) of each platform.  

\begin{figure}[h]
    \centering
    \includegraphics[width=0.9\linewidth]{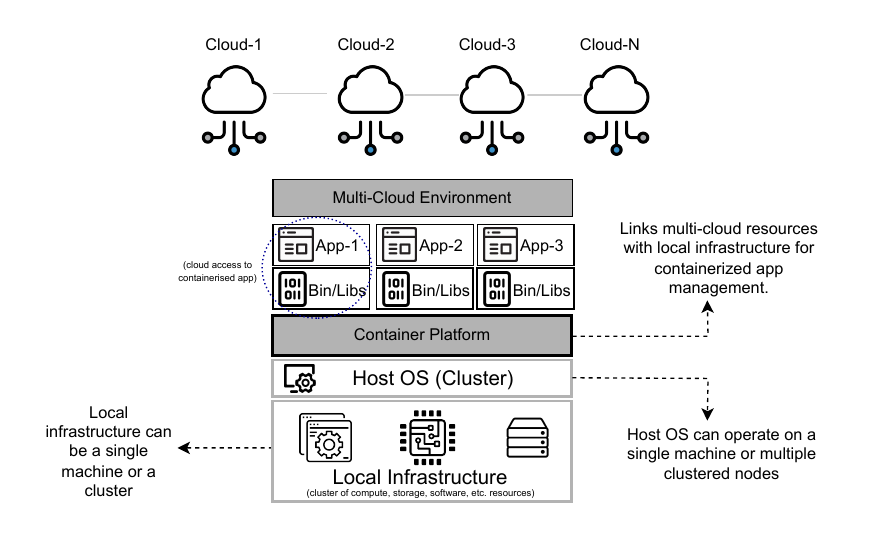}
    \caption{Context: Containerization in multi-clouds}
    \label{fig:context}
\end{figure}

Containerization in multi-cloud is illustrated in Figure \ref{fig:context}. This figure provides an overview of applications deployed within various cloud configurations, including public, private, and hybrid models. Each cloud hosts multiple applications encapsulated in containers, which can support different types of applications, such as those in healthcare or finance, that require flexibility in deployment and scalability across cloud providers. These containers complete various tasks with their required binaries and libraries, emphasizing the portability and isolation that containerization provides. A container platform layer, which could be exemplified by systems like Kubernetes or Docker, manages and orchestrates these containers across the different cloud environments. For instance, a healthcare application can use a private cloud for processing sensitive data for compliance and a public cloud for data analysis to achieve cost-efficiency and scalability. This architecture enables a flexible and scalable approach, allowing efficient application deployment and operations in a multi-cloud setting, while maintaining consistency and resilience across platforms.

However, utilizing containerization in multi-cloud environment is not without challenges that organizations may face \cite{saif2022multi}. These challenges may arise during different phases and activities of container-based application development including architectural design phase, system implementation, or when establishing an automated development infrastructure. Furthermore, the challenges can also present themselves during system testing, the coding process, and the deployment phase. For instance, during the architectural design phase, challenges may include ensuring seamless integration of containerized components with existing systems and selecting appropriate container orchestration patterns, strategies, and tools that can effectively manage containers across multiple clouds \cite{raj2018automated} \cite{baby2015multicloud}, and while establishing an automated development infrastructure, organizations may encounter difficulties in creating efficient Continuous Integration/Continuous Deployment (CI/CD) pipelines that handle multi-cloud deployment scenarios while maintaining consistent performance and security standards \cite{debroy2018building}. Similarly, in the system implementation phase, challenges may include resolving potential compatibility issues between containers and various cloud platforms, as well as optimizing resource utilization to control costs effectively \cite{helali2021survey}. 

\textbf{Motivation}: Our study is part of the QLEAP project (2022–2024) funded by Business Finland, which investigates the use of containers in multi-cloud environment specifically for architecture design \cite{QLEAP}. The project brings together a consortium of four companies (Bittium\footnote{\url{https://www.bittium.com/}}, M-Files\footnote{\url{https://www.m-files.com/}}, Solita\footnote{\url{https://www.solita.fi/en/}}, and Vaadin\footnote{\url{https://vaadin.com/}}), each with distinct requirements for containerization, and is further supported by industry leaders Nokia\footnote{\url{https://www.nokia.com/}} and TietoEVRY\footnote{\url{https://www.tietoevry.com/}}. This study fulfills an industry demand for optimized container strategies, addressing critical gaps in deployment consistency, security, and scalability that are necessary for practical applications in multi-cloud contexts. Despite the growing interest in containerization for multi-cloud environment, there remains a significant gap in systematically organized knowledge to address these challenges comprehensively. Our study not only fulfills an immediate industry demand but also aims to provide a structured view of containerization practices, addressing the needs of both practitioners and researchers in navigating this fragmented knowledge space.
\textcolor{black}{Recent research (e.g., see Selected Studies sheet in \cite{replpack}) has highlighted the growing significance of container utilization in multi-cloud environment}. These studies have explored various aspects, including container roles and strategies, architectural patterns, Quality Attributes (QAs), and the tools and frameworks employed. Additionally, these studies have explored the challenges associated with automation, deployment, monitoring, and security of containerization applications in multi-cloud environment. Despite the breadth of knowledge available, such valuable information is dispersed across different publications, spanning scientific research papers and gray literature. This fragmentation of knowledge about containerization applications in multi-cloud environment presents a significant navigational challenge for practitioners tasked with real-world implementations. They need to thoroughly scrutinize numerous aspects, such as patterns and strategies for containerization applications in multi-cloud environment, to locate specific information relevant to their use cases, whether it be a pattern, a challenge, or a solution to an existing problem. Consequently, this spreading of knowledge restricts the development of a holistic understanding of the containerization applications in multi-cloud environment, leaving practitioners underprepared to devise comprehensive, secure solutions.

Furthermore, this state of affairs of containerization in multi-cloud also poses challenges for the academic community. Researchers bear the obligation of navigating through this vast array of information to uncover the precise aspects they seek - be it a unique pattern, an unexplored challenge, or an innovative solution. Moreover, the distributed nature of this knowledge delays the formation of a cohesive and unified understanding of the subject matter. Recent studies (e.g., \cite{bordeleau2020towards} \cite{mihai2022digital} \cite{sarker2022ai}) have notably highlighted that challenges in design, development, monitoring, and testing of containerization applications in multi-cloud environment are deeply interconnected with the software development life cycle.

To bridge this knowledge gap and support the practical demands of the QLEAP project consortium, our study systematically identifies and categorizes the various facets of container utilization in multi-cloud environment. These facets include container roles, implementation strategies, architectural patterns, and quality attributes. In addition, we explore the challenges and solutions associated with automation, deployment, monitoring, and security, providing practitioners and researchers with actionable insights to tackle these critical issues effectively. To achieve this objective, we employed a systematic approach to conduct a mapping study. We finally selected 86 studies that focus on containerized applications in multi-cloud environment for further analysis. We adopted the following Goal-Question-Metric (GQM) framework \cite{caldiera1994goal}:

\begin{itemize}
    \item \textbf{Goal}:To enhance the understanding and efficiency of deploying and managing container-based applications in multi-cloud environment.
    \item \textbf{Questions}:To explore specific aspects such as container roles, implementation strategies, architectural patterns, quality attributes, challenges, and solutions.
    \item \textbf{Metrics}:To evaluate the effectiveness of our findings based on the identified roles, strategies, patterns, attributes, challenges, solutions, and tools.
\end{itemize}

The empirical findings and challenge-solution catalogs from this Systematic Mapping Study (SMS) provide significant benefits for developers of container-based applications. By examining implementation strategies, architectural patterns, and quality attributes, practitioners gain insights to improve practices. Detailed analyses of challenges and solutions in automation, deployment, monitoring, and security equip practitioners to address issues and enhance quality attributes. These challenge-solution catalogs facilitate knowledge sharing, training, and a shared understanding of containerized applications in multi-cloud environments, empowering developers to optimize their approaches and improve efficiency and quality. In response to the real-world needs identified in the QLEAP project and the broader research gap, our SMS presents the following \textbf{key contributions}:

\begin{itemize}

\item We systematically identified and categorized challenges and their solutions related to security, automation, deployment, and monitoring for containerization in a multi-cloud environment.

\item We systematically identified and categorized an extensive list of patterns and strategies for container-based applications in multi-cloud environment. 

\item We systematically identified and categorized quality attributes and tactics for containerized applications in multi-cloud environment.

\item We systematically identified and classified a wide range of tools and frameworks for containerized applications in multi-cloud environment.

\item We have publicly released a dataset, which is available online \cite{replpack}, to enable researchers and practitioners to access, replicate, and validate all collected data from our SMS. This dataset includes detailed hierarchies of the developed catalogs, identified and classified container implementation strategies, and the roles of containers in multi-cloud environment. It also encompasses patterns, strategies, quality attributes, tactics, tools, and frameworks for containerized-based applications in multi-cloud environment.
\end{itemize}

The paper is structured as follows: Section \ref{Sec_Methodology} describes the research methodology employed. Section \ref{Sec_Results} presents the results and findings of this SMS. Section \ref{Sec_Discussion} discusses the implications of the SMS results. Section \ref{sec:SMSthreats} clarifies the potential threats to the validity of this SMS.  Section \ref{Sec_RelatedWork} reviews relevant prior work and highlights the differences between our work and previous research. Section \ref{Sec_Conclusion} draws conclusions and outlines future research directions.

\section{Methodology}
\label{Sec_Methodology}
\begin{figure*}[h] 
    \centering 
    \includegraphics[width=1.0\textwidth]{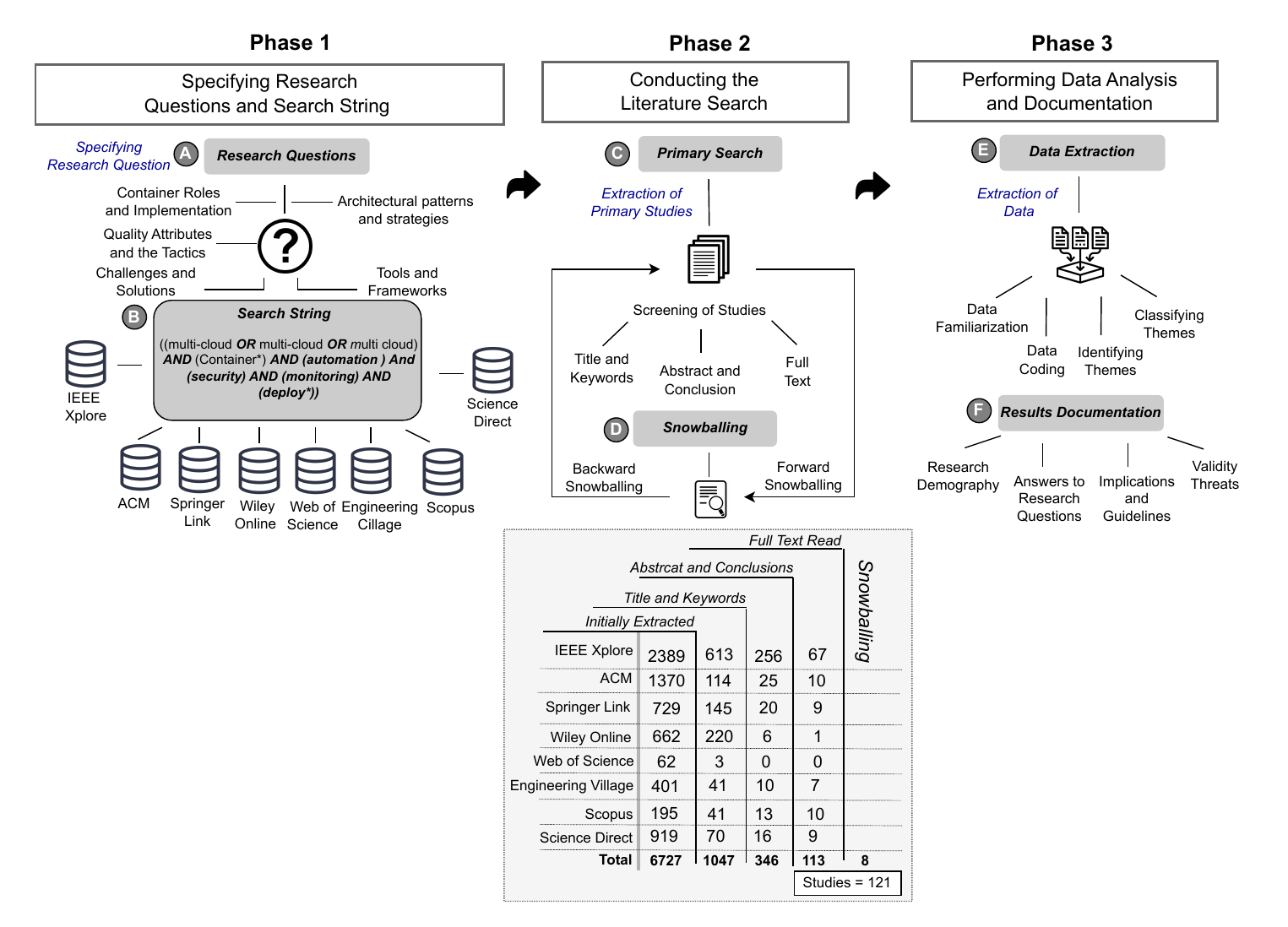} 
    \caption{Schematic representation of the research methodology implemented in this study} 
    \label{fig:ResearchMethod} 
\end{figure*}

We conducted a Systematic Mapping Study (SMS) by following the guidelines in \cite{AR27} and augmenting them with SLR strategies \cite{AR26}. Our SMS consists of three phases: specifying research questions and search string, conducting the literature search, and performing data analysis and documentation. Figure \ref{fig:ResearchMethod} illustrates the SMS execution process. Reliability and consistency are critical aspects of any empirical study. In this study, we followed established statistical approaches, including inter-rater agreement (IRA) computation, to ensure the reliability of our data coding and extraction processes. Additionally, we applied systematic quality assessment techniques to evaluate the studies included in our SMS.

\subsection{Research Questions}

To conduct this SMS, we formulated six Research Questions (RQs) based on the GQM approach presented in the Introduction section (see Section~\ref{Sec_Introduction}) aligns with the objective of our SMS, as outlined in Table \ref{tab:research_questions}. The table provides a structured view of the RQs with their corresponding rationale and categorization.


\begin{table*}[t]
\footnotesize
\caption{Research questions and their rationale}
\label{tab:research_questions}
\begin{tabularx}{\textwidth}{|c|>{\hsize=1\hsize}X|>{\hsize=1\hsize}X|}
\hline
\rowcolor{gray!50} \textbf{\#} & \textbf{Research Questions} & \textbf{Rationale} \\
\hline
\rowcolor{gray!25} \multicolumn{3}{|c|}{\textbf{Container Roles and Implementation}} \\
\hline
RQ1 & What roles do containers play, and what strategies can be employed for implementing container-based applications in multi-cloud environment? & To investigate the role of containers and explore implementation strategies in multi-cloud environment, with the objective of enhancing understanding and facilitating efficient deployment and management of applications across multiple cloud environments. \\
\hline
RQ2 & What architectural patterns and strategies are utilized in the implementation of container-based applications in multi-cloud environment? & To explore and document the various architectural patterns and strategies that are utilized in implementing container-based applications in multi-cloud environment, thus providing a roadmap for efficient and scalable application development and deployment. \\
\hline
RQ3 & What are the quality attributes and the tactics associated with their implementation in container-based applications in multi-cloud environment? & To examine and analyze the quality attributes required for container-based applications in multi-cloud environment and the corresponding tactics employed in their implementation, enabling the identification of best practices and guidelines for developing robust and reliable applications in such environments. \\
\hline

\rowcolor{gray!25} \multicolumn{3}{|c|}{\textbf{Challenges and Solutions}} \\
\hline
RQ4& Challenges in Container-Based Applications in Multi-Cloud Environment
\begin{itemize}
    \item \textbf{RQ4.1}:What are the challenges related to security in container-based applications in multi-cloud environment?
    \item \textbf{RQ4.2}: What are the challenges related to automation in container-based applications in a multi-cloud environment?
    \item \textbf{RQ4.3}: What are the challenges related to deployment in container-based applications in a multi-cloud environment?
    \item \textbf{RQ4.4}: What are the challenges related to monitoring in container-based applications in a multi-cloud environment?
\end{itemize}
 & To comprehensively evaluate and gain insights into the challenges surrounding security (RQ4.1), automation (RQ4.2), deployment (RQ4.3), and monitoring (RQ4.4) concerning container-based applications in multi-cloud environment. These sub-RQs aim to provide valuable insights into the obstacles and intricacies that organizations encounter while implementing and managing container-based applications in a multi-cloud environment.\\
\hline

RQ5 &Solutions to Address Challenges in Container-Based Applications in Multi-Cloud Environment
\begin{itemize}
    \item \textbf{RQ5.1}:How to address the challenges related to security in container-based applications within a multi-cloud environment?
    \item \textbf{RQ5.2}: How to address the challenges related to automation in container-based applications within a multi-cloud environment?
    \item \textbf{RQ5.3}: How to address the challenges related to deployment in container-based applications within a multi-cloud environment?
    \item \textbf{RQ5.4}:  How to address the challenges related to monitoring in container-based applications within a multi-cloud environment?
\end{itemize}
&  These sub-RQs seek to investigate solutions aimed at tackling the challenges of security (RQ5.1), automation (RQ5.2), deployment (RQ5.3), and monitoring (RQ5.4) within container-based applications operating in multi-cloud environment. By providing actionable guidance and recommendations, the goal is to elevate the overall management and performance of these applications in the context of multi-cloud environment. \\
\hline
\rowcolor{gray!25} \multicolumn{3}{|c|}{\textbf{Tools and Frameworks}} \\
\hline
RQ6 & What tools and frameworks are used to implement container-based multi-cloud applications? & To investigate and identify the specific tools and frameworks utilized in the implementation of container-based multi-cloud applications, providing an overview of the technological landscape and aiding developers and organizations in making informed decisions when selecting appropriate tools for their application deployment and management. \\
\hline
\end{tabularx}
\end{table*}

These RQs are designed to address important gaps in the current understanding of container-based applications in multi-cloud environment. RQ1–RQ3 focus on understanding the roles of containers, implementation strategies, architectural patterns, and quality attributes. These questions aim to provide a comprehensive perspective, building on and organizing fragmented insights from prior studies. RQ4 and RQ5 focus on specific challenges and corresponding solutions related to security, automation, deployment, and monitoring in container-based multi-cloud environment. These questions aim to systematically address issues that are often only partially explored in existing works. Finally, RQ6 investigates tools and frameworks used for implementation, helping bridge the gap between academic research and practical applications. Together, these RQs provide a structured framework that enhances understanding and supports the development and management of container-based applications in multi-cloud environment, advancing the state-of-the-art in this domain

\subsection{Search String Composition}
Initially, we considered the PICO (Population, Intervention, Comparison, Outcome) framework  \cite{schardt2007utilization} \cite{ralph2020empirical} for constructing our search string. However, given the complexity and breadth of our topic, applying PICO strictly required an extensive array of terms and logical operators, exceeding the query limits of databases like ScienceDirect. Therefore, we developed the search string for this study based on authors' knowledge and iterative trial searches. This approach allowed us to tailor the search strategy to the specific goal of our study, ensuring comprehensive coverage within practical constraints.

\subsection{Study Search and Selection Process}
The study search and selection process for this study is divided into two phases. In the first phase, we conducted a primary search using a specific search string on various databases (see Table \ref{tab:stringDatabase}  and the Summary sheet in \cite{replpack}). In the second phase, we applied the snowballing technique on primary studies that were selected from the first phase.

\begin{table}[t]
\centering
{\small{}%
\scriptsize
\caption{Search string and targeted search area for databases}
\label{tab:stringDatabase}
\begin{tabular}{|c|c|}
\hline
\rowcolor{gray!50} \multicolumn{2}{|c|}{\textbf{Search String}}\tabularnewline
\hline
\rowcolor{gray!25} \multicolumn{2}{|c|}{\emph{((multicloud OR multi-cloud OR multi cloud) AND }}\tabularnewline
\rowcolor{gray!25} \multicolumn{2}{|c|}{\emph{(container*) AND (automation) And (security) AND (monitoring) AND (deploy*))}}\tabularnewline
\hline
\multicolumn{2}{|c|}{\textbf{Databases}}\tabularnewline
\hline
\textbf{Database} & \textbf{Targeted search area}\tabularnewline
\hline
ACM Digital Library & Paper title, abstract\tabularnewline
\hline
IEEE Xplore & Paper title, keywords, abstract\tabularnewline
\hline
Scopus & Paper title, keywords, abstract\tabularnewline
\hline
SpringerLink & Paper title, abstract\tabularnewline
\hline
ScienceDirect & Paper title, keywords, abstract\tabularnewline
\hline
Wiley Online Library & Paper title, abstract\tabularnewline
\hline
Engineering Village & Paper title, abstract\tabularnewline
\hline
Web of Science & Paper title, keywords, abstract\tabularnewline
\hline
\end{tabular}}
\end{table}

\subsubsection{Primary Search}
\label{sec:PrimarySearch}
The primary search involved querying digital databases (see Table \ref{tab:stringDatabase}) using customized search string. The search strings were executed concurrently on eight databases, as illustrated in Figure \ref{fig:ResearchMethod}, spanning between January 2013 and July 2024 when we started this SMS. We chose 2013 as the starting point because it is the year when container technologies like Docker emerged~\cite{DockerContainers2023}, marking a key phase in containerization and cloud computing. This period also saw a rise in relevant research to containers in multi-cloud environment, making it ideal to understand the development of containerized applications in multi-cloud environment over the past decade. We followed the steps mentioned below to select the relevant studies.

\begin{table*}[t]
\footnotesize
\centering
\caption{Inclusion and exclusion criteria for selecting primary studies in this SMS}
\centering
\begin{tabular}{|p{2cm}|p{5.4cm}|p{5.4cm}|}
\hline 
\rowcolor{gray!50} \textbf{Selection Criteria} & \textbf{Inclusion Criteria } & \textbf{Exclusion Criteria} \\
\hline 
\emph{Language } & English & Non-English \\
\hline 
\emph{Study Type\scriptsize } & Primary studies including peer-reviewed journal articles, book chapters, conference papers, workshops, and symposium papers & Secondary studies or non-peer-reviewed content (e.g., blogs, webpages, videos, white papers, technical reports, grey literature)\\
\hline 
\emph{Study Focus} &  \textcolor{black}{Studies that explicitly address the intersection of containerization and multi-cloud environments. This includes studies that investigate the use of container technologies (e.g., Docker, Kubernetes) in the context of multi-cloud deployment, management, or architecture. Studies that focus solely on containerization without a multi-cloud context, or vice versa, are excluded.
}& \textcolor{black}{Studies that address only traditional cloud computing, single-cloud environments, or containerization without reference to multi-cloud contexts, and vice versa. Specifically, studies that examine container technologies without situating them in multi-cloud scenarios, or studies focused on multi-cloud architectures without discussing container-based solutions, are excluded.}
\\
\hline 
\emph{Study Duration} & Studies published between January 2013 and July 2024 & Studies published before January 2013 or after July 2024\\
\hline 
\end{tabular}
\label{tab:InclusionExclusion}
\end{table*}

\begin{itemize}
    \item \textit{Step 1: Extraction of Studies}: We ran custom search strings in the selected databases (see Table \ref{tab:stringDatabase}) to retrieve study titles, author names, publication years, venues, publication types, and abstracts. This initial search yielded 6,727 studies from eight databases. Following initial retrieval, we did not apply Cohen's Kappa \cite{cohen1960coefficient} \cite{cohen1968weighted} as this step is automated and does not involve subjective decision-making.

    \item \textit{Step 2: Title and Keyword Screening}: This step involves reviewing the titles and keywords of the collected studies to assess their relevance to the research topic. \textcolor{black}{To ensure alignment with our inclusion criteria, we manually reviewed study titles and abstracts to include only those that explicitly discuss containerization within multi-cloud contexts. Studies that focused solely on single-cloud containerization, general cloud orchestration without containers, or container use cases unrelated to multi-cloud environments were excluded—even if they matched keywords in the search string.} Initially, we removed any duplicate studies obtained from different databases (e.g., IEEE Xplore, Scopus) by sorting them in ascending order. This resulted in the elimination of several thousand duplicate studies. We divided the remaining studies between two authors (R1, R2), who independently applied the inclusion and exclusion criteria to identify studies relevant to our research objective, as detailed in Table \ref{tab:InclusionExclusion}. To assess the consistency of study selection and mitigate bias, a Cohen’s Kappa analysis \cite{cohen1960coefficient} \cite{cohen1968weighted} was performed on a random sample of 400 studies (10\% of the total dataset of 4,000 studies), as illustrated in Formula 1:
    \begin{equation}
    \kappa = \frac{P_o - P_e}{1 - P_e}
    \end{equation}
    where \( P_o \) represents the relative observed agreement among authors, and \( P_e \) the hypothetical probability of chance agreement. The results, recorded in Table \ref{tab:kappa_agreement}.a, indicated a Cohen’s Kappa of \( \kappa = 0.531 \), which reflects moderate agreement between R1 and R2 \cite{sim2005kappa}. Specifically, R1 and R2 agreed on 360 studies being relevant (``YES'') and 20 studies being irrelevant (``NO''). Disagreements occurred in 20 studies, which were systematically resolved through discussions among all authors, leading to a final consensus. Using a random subset is a common practice in systematic reviews to measure inter-rater agreement without overburdening resources, ensuring reliability while maintaining efficiency \cite{AR26, chandler2019cochrane}. Following this process, the number of studies was reduced to 1,047 for further analysis.

    \item \textit{Step 3: Abstract-based Screening}: During this step, we conducted a thorough review of the abstracts of the collected studies to determine their relevance to our research topic. Two researchers (R1 and R2) independently examined each abstract, assigning a status of ``relevant'', ``irrelevant'', or ``doubtful'' with respect to our research goals. To reduce subjective bias and to quantify the consensus between the researchers, we performed a Cohen’s Kappa analysis on a random subset of 105 studies (approximately 10\% of the 1,047 studies remaining after the Title and Keyword Screening phase).

    The outcomes of this assessment are depicted in Table \ref{tab:kappa_agreement}.b. The analysis showed substantial agreement between R1 and R2, with 88 studies being unanimously considered relevant and 5 studies unanimously considered irrelevant. Disagreements occurred in 12 cases (7 studies where one researcher marked ``relevant” while the other marked ``irrelevant”, and 5 studies initially classified as ``doubtful”). The kappa value, calculated based on observed agreements and disagreements, was \( \kappa = 0.521 \), calculated based on observed agreements and disagreements, indicates moderate agreement \cite{sim2005kappa, AR26}. 

To compute Cohen’s Kappa during the abstract-based screening phase, the original classifications of ``relevant", ``irrelevant”, and ``doubtful” were simplified into a binary scale (``YES” or ``NO”) due to the binary nature of Cohen’s Kappa computation. Specifically:
    \begin{itemize}
        \item Studies classified as ``relevant” were mapped to ``YES”.
        \item Studies classified as ``irrelevant” were mapped to ``NO”.
        \item Studies classified as ``doubtful” were resolved through discussions between R1 and R2. If consensus was reached on ``relevant”, the study was marked as ``YES”; otherwise, it was marked as ``NO”.
    \end{itemize}

 Using subsets for kappa computation ensures that inter-rater reliability can be rigorously evaluated while optimizing the use of resources, as recommended by systematic review guidelines \cite{AR26, chandler2019cochrane}. After resolving all disagreements through systematic discussions among the authors, a final consensus was reached for the subset. Consequently, upon completing this step, the number of studies was reduced to 346 for more detailed scrutiny.

    \item \textit{Step 4: Full-Text Screening}: In this step, we reviewed the full texts of 346 studies that had been shortlisted through abstract-based screening. To ensure consistency in our review process, we randomly selected a subset of 35 studies (10\% of the total 346 studies) for a Cohen’s Kappa analysis to measure the agreement between two researchers (R1 and R2). The results, recorded in Table \ref{tab:kappa_agreement}.c, indicated a Cohen’s Kappa value of \( \kappa = 0.578 \), which represents moderate agreement between the two researchers \cite{sim2005kappa}. Specifically, R1 and R2 agreed on 30 studies as relevant (``YES”) and 2 studies as irrelevant (``NO”). Disagreements occurred in 3 studies (2 marked as ``YES” by one researcher and ``NO” by the other, and 1 marked as ``NO” by both but flagged as uncertain). All disagreements were systematically resolved through discussions among the authors, leading to a final consensus. This subset-based approach, coupled with systematic resolution of disagreements, provides confidence in the reliability of the full-text screening process \cite{chandler2019cochrane}. At the end of this step, we obtained a total of 110 studies, which were deemed relevant for further analysis.
\end{itemize}


\begin{table*}[ht]
\centering
\caption{Kappa agreement across different screening steps}
\label{tab:kappa_agreement}

\renewcommand{\arraystretch}{1.2} 
\setlength{\tabcolsep}{4pt} 

\begin{minipage}{0.32\textwidth}
    \centering
   \textbf{(a) Title \& Keyword \\ Screening (400 Studies)}
    \begin{tabular}{|l|c|c|}
    \hline
    \rowcolor{gray!50} 
      & \multicolumn{1}{c|}{R1 Yes} & \multicolumn{1}{c|}{R1 No} \\ \hline
    R2 Yes & 360     & 20     \\ \hline
    R2 No  & 10      & 10     \\ \hline
    \multicolumn{1}{|r|}{Total} & 370     & 30     \\ \hline
    \end{tabular}
\end{minipage}%
\hfill 
\begin{minipage}{0.32\textwidth}
    \centering
    \textbf{(b) Abstract-Based \\Screening (105 Studies)}
    \begin{tabular}{|l|c|c|}
    \hline
    \rowcolor{gray!50}
      & \multicolumn{1}{c|}{R1 Yes} & \multicolumn{1}{c|}{R1 No} \\ \hline
    R2 Yes & 88     & 7     \\ \hline
    R2 No  & 5      & 5     \\ \hline
    \multicolumn{1}{|r|}{Total} & 93     & 12     \\ \hline
    \end{tabular} 
\end{minipage}%
\hfill 
\begin{minipage}{0.32\textwidth}
    \centering
    \textbf{(c) Full-Text\\ Screening (35 Studies)}
    \begin{tabular}{|l|c|c|}
    \hline
    \rowcolor{gray!50} 
      & \multicolumn{1}{c|}{R1 Yes} & \multicolumn{1}{c|}{R1 No} \\ \hline
    R2 Yes & 30     & 3     \\ \hline
    R2 No  & 2      & 0     \\ \hline
    \multicolumn{1}{|r|}{Total} & 32     & 3     \\ \hline
    \end{tabular}
\end{minipage}

\end{table*}

\subsubsection{Snowballing} 
In Phase 2, we utilized the snowballing technique, as described in \cite{AR30}, to examine references within 110 primary studies to identify additional relevant studies. This strategy was augmented by forward snowballing, where we gathered studies that cited the selected studies, and backward snowballing, which involved using references within the selected studies. Notably, we encountered several dozen studies during the forward and backward snowballing that had been excluded in primary search. This phase resulted in the addition of 11 more studies, bringing the final count to 121.

\subsubsection{Quality Assessment of Selected Studies} \label{sec:QualityAssessment}
We assessed the quality of the selected studies based on a set of predefined criteria, including relevance to multi-cloud containerization, clarity of research design, empirical validation (e.g., experiments, case studies), reporting quality (e.g., discussion of limitations and challenges), publication in peer-reviewed venues, and the generalizability of findings to multi-cloud environment. \textcolor{black}{Each study was assessed using six predefined quality criteria: relevance to multi-cloud containerization, clarity of research design, data validity (e.g., empirical evidence), reporting quality (e.g., discussion of limitations), peer-review status, and generalizability of findings. Each criterion was scored using a 3-point scale: 1 (fully satisfied), 0.5 (partially satisfied), or 0 (not satisfied), giving a total score out of 6. The scoring was applied consistently across all studies based on observable features. For example, studies addressing both containerization and multi-cloud contexts were scored 1 for relevance; studies with explicit research methods scored higher on research design; and those lacking empirical evidence were scored lower for data validity. Scores for all individual criteria are documented in the “QA of Selected Studies” sheet in our replication package, providing transparency and allowing independent verification of the quality assessment process.} Studies scoring below 3 were flagged for further review and discussion but retained to ensure alignment with the study's objectives. This quality assessment was conducted after the study selection process to ensure consistency and thoroughness across all 121 studies. Detailed results are available in the replication package \cite{replpack}, specifically in the sheet titled ``QA of Selected Studies''.

\subsection{Data Extraction and Analysis}
\label{DataExtractionandAnalysis}
\textbf{Data Extraction}: The data extraction form was designed based on predefined data items (see Table~\ref{tab:DataExtractionItems}) that were formulated to address the RQs specified in Table \ref{tab:research_questions}. To ensure the reliability of the extracted data items, the pilot data extraction was conducted on ten studies by the first author, and all the other authors assessed the extracted data. Subsequently, the first author employed a revised set of data items (e.g., D11, D12, D16), determined after evaluating the extracted data items, for the formal data extraction from the selected studies. To mitigate the personal bias and ambiguity, all authors engaged in discussions regarding the extracted data. The data items labeled as D1 to D8 present a summary of the demographics of the primary studies selected, while D9 to D22 are specifically employed to address RQ1 to RQ6. A concise description of each data item is presented in Table~\ref{tab:DataExtractionItems}. Finally, Google Sheets were used to record and further analyze the extracted data.

\textbf{Data Analysis}: We employed descriptive statistics to analyze the quantitative data from data items D1, D5, D7, D8, and D9. For the remainder of the data, which primarily consist of qualitative free-text descriptions (e.g., study aim, roles of containers, challenges, and solutions), we conducted a thematic analysis in accordance with the guidelines outlined in \cite{braun2006}. Our thematic analysis process consists of the following steps:

\begin{enumerate}
    \item \textbf{Data Familiarization}: We conducted a thorough review of the selected studies by repeatedly reading through them and meticulously noting key points related to study aims (D7), contributions (D8), roles of containers (D9), implementation strategies (D10), architectural patterns (D11), quality attributes (D12), tactics (D13), motivations (D14), and challenges and solutions (D15-D21) in automation, deployment, monitoring, and security challenges and solutions, as well as the tools, languages, and frameworks (D22) employed.
    \item \textbf{Generation of Initial Codes}: After developing a thorough understanding of the data, we created an initial set of codes derived from the information extracted concerning the data items identified in the previous step.
    \item \textbf{Identification of Types and Emerging Themes}: In this step, we conducted a two-tiered analysis. Initially, we examined the codes to ascertain their types. Subsequently, we developed subcategories based on these types, and then formulated overarching categories that encompass the related subcategories.
    \item \textbf{Critical Evaluation of Types, Subcategories, and Categories}: All authors actively participated in rigorously reviewing and refining the coded data, including types, subcategories, and categories. During this collaborative process, we redefined, merged, or dropped certain themes based on collective input and discussion.
    \item \textbf{Defining and Naming Categories}: At this point, we provided explicit definitions for each of the identified themes and further refined them, ensuring that the terminology used for the categories was precise and unambiguous.
\end{enumerate}

\begin{table*}
\centering
\footnotesize
\caption{Data items to be extracted in this SMS}
\label{tab:DataExtractionItems}
\begin{tabular}{|c|>{\hsize=1.2\hsize}l|>{\hsize=1.3\hsize}l|c|}
\hline
\rowcolor{gray!50} \textbf{Code} & \textbf{Data Item} & \textbf{Description} & \textbf{RQ} \\ \hline
D1 & Index & The ID of the study & \multirow{11}{*}{Demographics} \\ \cline{1-3}
D2 & Publication Year & Publication year of the study &  \\ \cline{1-3}
D3 & Publisher & The publisher of the study &  \\ \cline{1-3}
D4 & Venue & The name of the publishing venue &  \\ \cline{1-3}
D5 & Publication Type & Journal, conference, workshop, and book chapter &  \\ \cline{1-3}
D6 & Authors' Affiliation & The affiliation of the authors &  \\ \cline{1-3}
D7 & Study Aim & The aim of the paper &  \\ \cline{1-3}
D8 & Study Contribution & The contributions made by the paper &  \\ \hline
D9 & Role of Containers & Role of containers in the study & \multirow{2}{*}{RQ1} \\ \cline{1-3}
D10 & Container Implementation strategy & Strategy used for container implementation &  \\ \hline
D11 & Architectural Pattern and Strategies & Architectural patterns and strategies reported in the study & RQ2 \\ \hline
D12 & Quality Attributes & Quality attributes discussed in the study & \multirow{2}{*}{RQ3} \\ \cline{1-3}
D13 & Tactics & Tactics used in the study &  \\ \hline
D14 & Automation Challenges & Challenges in automation discussed in the study & RQ4.2 \\ \hline
D15 & Automation Solutions & Solutions for automation challenges discussed in the study & RQ5.2 \\ \hline
D16 & Deployment Challenges & Challenges in deployment discussed in the study & RQ4 \\ \hline
D17 & Deployment Solutions & Solutions for deployment challenges discussed in the study & RQ5 \\ \hline
D18 & Monitoring Challenges & Challenges in monitoring discussed in the study & RQ4 \\ \hline
D19 & Monitoring Solutions & Solutions for monitoring challenges discussed in the study & RQ6 \\ \hline
D20 & Security Challenges & Challenges in security discussed in the study & RQ4,1 \\ \hline
D21 & Security Solutions & Solutions for security challenges discussed in the study & RQ5.1 \\ \hline
D22 & Tools, Languages, and Frameworks & Tools, programming languages, and frameworks & RQ6 \\ \hline
\end{tabular}
\end{table*}

 Two researchers participated in the process to reduce personal bias. The most important activity of this process is brainstorming sessions that were mainly conducted while reviewing, defining, and naming the research themes. In these sessions, both researchers discussed and validated the research themes found.

Concerning data item D9 (Roles of Containers) and D10 (Container Implementation strategy), we used the open coding and constant comparison techniques from Grounded Theory \cite{glaser1967discovery} to analyze the qualitative data extracted from the selected studies. 

Finally, we provided a replication package \cite{replpack} with the results of each phase of the study selection process (e.g., Phase 1, Phase 2) and detailed results (e.g., Contributions, Patterns, QAs, Challenges, Solutions) for verification and validation purposes of this SMS.

\section{Results}
\label{Sec_Results}
This section presents the results of the SMS, derived from analyzing data from chosen studies. Section \ref{DemographicsMapping} provides a summary of the demographic data, classifications, and the mapping of research themes identified in the studies. Section \ref{RQ1} presents the results concerning container roles and implementation strategies. Section \ref{sec:RQ2} reports on the patterns and strategies employed in container-based applications in multi-cloud environment. Section \ref{Sec:RQ3} presents the QAs and related tactics for containerized applications in multi-cloud environment. Section \ref{SecurityChallengesandSolutionsFramework} presents the security challenges and solutions framework, while Section \ref{sec:Automationframework} presents the automation challenges and solutions framework. Section \ref{DeploymentFramework} presents the deployment challenges and solutions framework, and Section \ref{sec:monitoring} presents the monitoring challenges and solutions framework. Finally, the tools and frameworks are reported in Section \ref{sec:toolsFramework}.

\subsection{Demographics, Research Aims, and Contributions}
\label{DemographicsMapping}

\begin{figure*}[] 
    \centering 
    \includegraphics[width=0.7\textwidth]{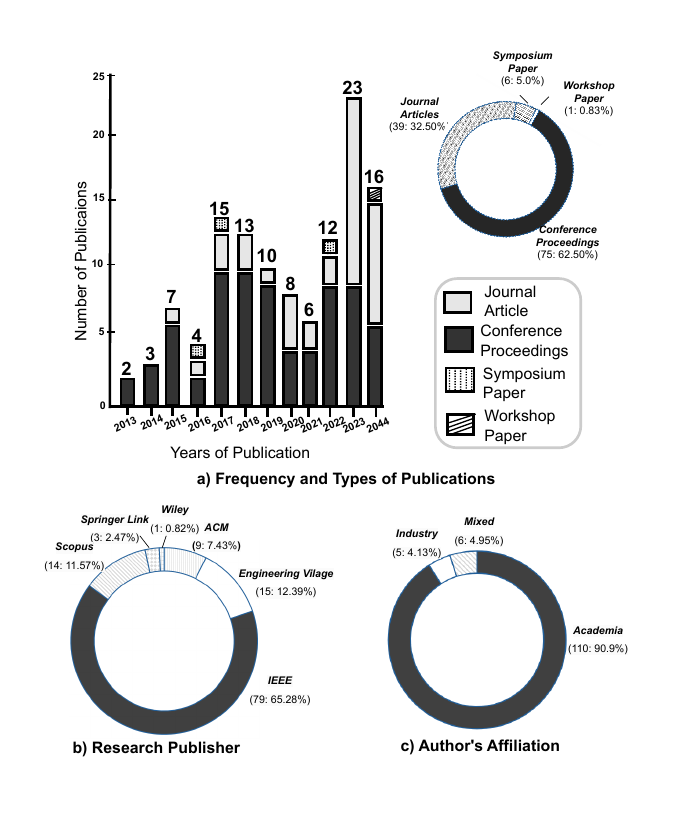} 
    \caption{Demographics of the selected studies, including: a) Frequency and Types of Publications; b) Research Publisher; c) Author's Affiliation} 
    \label{fig:Demography} 
\end{figure*}

                \textbf{Yearly Distribution of Studies}: Figure \ref{fig:Demography}-a illustrates the annual distribution of studies retrieved from eight databases between January 2013 and July 2024. We found only two primary studies on containerization in multi-cloud environment that meet our criteria in 2013. The bars display yearly study counts and trends. A peak in 2017, followed by a decrease and occasional fluctuations, suggests an early spike in interest followed by more targeted research. Notably, the reduction in publication numbers in 2020 and 2021 may be partially attributed to the global disruptions caused by the COVID-19 pandemic, which affected research outputs across various fields. However, there is a notable increase in studies in 2023 and continuing into 2024. This could be due to delayed research projects getting back on track and more funding in tech sectors after the pandemic, showing a renewed focus on improving containerization in multi-cloud environment.

\textbf{Publishers Distribution}: Figure \ref{fig:Demography}-b shows the distribution of studies by publisher. IEEE accounts for 65.29\%, highlighting its major contribution. Scopus follows with 11.57\%, while Engineering Village, ACM, and Springer Link represent smaller portions at 12.40\%,  7.44\%, and 2.48\%, respectively. The chart emphasizes the dominance of IEEE in the identified studies, suggesting its importance as a primary source of relevant literature. This distribution also suggests to researchers where to primarily search and publish on containerization in multi-cloud environment.

\textbf{Study Types}: Figure \ref{fig:Demography}-a provides a breakdown of the types of selected studies. It is apparent that conferences dominate as the most common venue, comprising 62.5\% of the total publications. Journals, often considered venues for mature and rigorously peer-reviewed research, make up 32.5\%. This indicates a preference for conferences, likely due to faster publication and networking opportunities Symposiums and workshops represent for 5.0\% and 0.83\% respectively, suggesting that these are less common avenues, possibly due to their more specialized or focused nature.

\textbf{Authors' Affiliations}: Figure \ref{fig:Demography}-c illustrates the distribution of authors' affiliations in collected publications. 90.90\% of the affiliations are associated with academia. This highlights the central role of academic institutions in producing and disseminating research. implying that industry professionals focus more on practical applications. The low industry affiliation may reflect the SMS's focus on academic databases, where industry participation is limited, indicating typical academic database trends and suggesting broader industry inclusion in future studies. Mixed affiliations, at 4.95\%, indicate a small proportion of collaborations between academia and industry. This low rate of cross-affiliation collaboration may suggest that there is potential for increased synergy between academic and industrial entities in future research endeavors.
\begin{tcolorbox} 
\textbf{Takeaway 1}: \textcolor{black}{Research on containerization in multi-cloud environments has gained more attention in recent years, especially after 2022. Most of the studies are published by IEEE and come from academic authors. This shows that the topic is mainly explored in academic settings. However, there is little contribution from industry, and most studies are presented at conferences rather than in journals. This suggests a need for more collaboration between academia and industry and more detailed research through journal publications.}
\end{tcolorbox}

\begin{figure*}[h] 
    \centering 
    \includegraphics[width=1.0\textwidth]{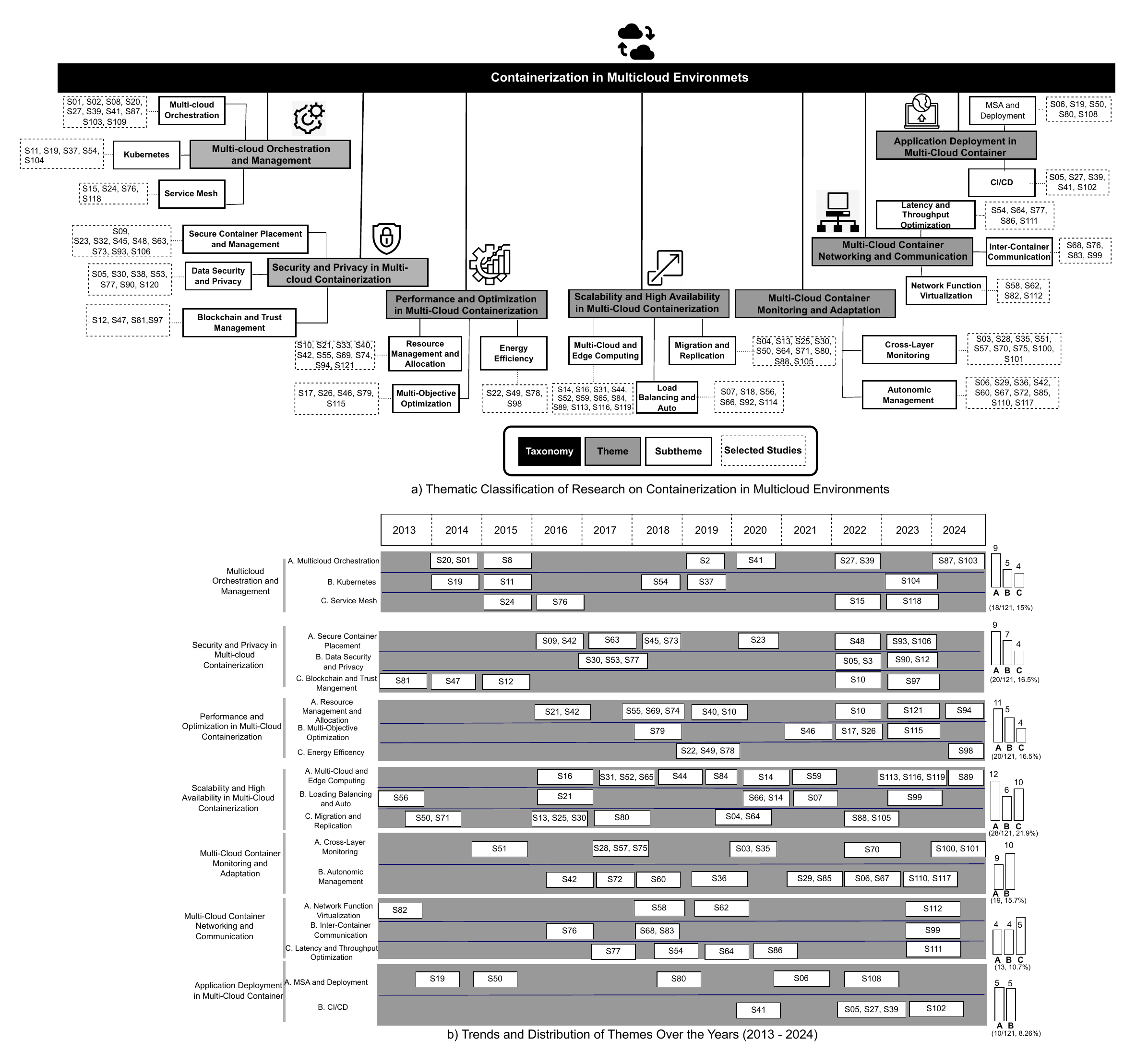} 
    \caption{Selected studies: a) Thematic Classification and b) Trends and Distribution of Themes Over the Years} 
    \label{fig:StudiesAim} 
\end{figure*}

\textbf{Classification of Research Theme}: The identified research themes and their notable trends are illustrated in Figure \ref{fig:StudiesAim} and detailed below. Figure \ref{fig:StudiesAim}-(a) presents the taxonomical classification of existing research. This taxonomy systematically identifies, names, and represents various topics based on their similarities or distinctions. Figure \ref{fig:StudiesAim}-(b) depicts the trends and temporal distribution of the themes over the years (2013–2024) based on the studies identified in our review.
\begin{itemize}
    \item \textbf{Taxonomy of Research Themes}:  The prominent research themes identified include “Scalability and High Availability” with 20 studies, and three other themes: “Security and Privacy” in  “Multi-Cloud Container Monitoring and Adaptation”, and “Performance and Optimization”, each attracting attention in 15 studies. This highlights that the core areas of interest in multi-cloud environment are proficient container management, scalability, and performance optimization. “Security and Privacy” also stands as a considerable theme, as demonstrated by the 15 studies focusing on areas such as secure container handling, data security, and blockchain applications. This emphasizes the necessity of protecting data and applications in the context of containers spreading across multi-cloud environment. Moreover, it is noted that several studies encapsulate more than one theme. For instance, study S19 is catalogued under both `Multi-cloud Orchestration and Management' and  `Application Deployment in Multi-cloud Container', signifying that it encompasses aspects of both Kubernetes and microservices architecture.
    \item \textbf{Trends and Temporal Distribution of Themes}: To highlight prominent trends and their progression over the years (2013–2024), we adopted the thematic and sub-thematic representation from Figure 4(a) and illustrated them via a timeline and relative distribution of each theme in Figure 4-(b). Specifically, Figure 4-(b) demonstrates how certain research themes have progressed and matured (i.e., evolved as topics) over the years. For instance, the theme multi-cloud orchestration and management is addressed in a total of 18 studies, accounting for approximately 15\% of the total reviewed studies. Our analysis indicates that one of the earliest research efforts on multi-cloud orchestration, published during 2014–2015, focused on the component-based and model-driven adaptation of applications for multi-cloud containers [S1, S20, S8]. Following this initial phase of application adaptation for multi-cloud orchestration, subsequent research explored techniques for benchmarking and enabling resource-efficient application orchestration [S2, S41, S39]. More recent studies, published between 2022 and 2024, have investigated a variety of topics, including but not limited to DevOps-based development [S27] and monitoring orchestrated applications in multi-cloud environments [S39, S87]. Our review revealed no published studies on the theme of multi-cloud orchestration and management during 2013, 2017, or 2021. Two possible reasons for this are: (i) our literature search did not identify any relevant studies (see Section \ref{sec:PrimarySearch} - primary search), or (ii) the studies from those years did not meet the qualitative evaluation criteria (see Section \ref{sec:QualityAssessment} - quality assessment) for inclusion in the review.
\end{itemize}
\begin{tcolorbox} 
\textbf{Takeaway 2}: \textcolor{black}{The research themes show that when containerization is applied in multi-cloud environments, unique challenges emerge—such as the need for orchestration tools that can manage service consistency across diverse providers, and security mechanisms that adapt to varying cloud configurations. Studies have evolved from basic scalability and performance concerns to addressing how to efficiently monitor, deploy, and manage containerized applications across different cloud platforms using DevOps and federation-aware solutions. This highlights how multi-cloud settings demand more adaptable and interoperable container strategies than single-cloud systems.}
\end{tcolorbox}

\textbf{Major Contributions of the Selected Studies}: Table \ref{tab:StudiesContribution} provides an overview of the contributions of the selected studies. The contributions are primarily categorized into five main categories and 22 subcategories (see the Contributions sheet in \cite{replpack}), through thematic analysis. The selected studies mainly focus on proposing and validating security solutions, reference architectures, cloud management frameworks, interoperability, orchestration, and tools for optimizing and deploying containerized applications in multi-cloud environment. Notably, the largest contributions come from the \textit{Security Approaches} category, representing 27.25\% of the total studies, which highlights the significant attention to cybersecurity \& access control and cloud analysis approaches. Within this category, the \textit{Cybersecurity and Access Control} subcategory alone accounts for 9.09\%, emphasizing the importance of securing multi-cloud environment. The studies related to the \textit{Reference Architectures and Models for Cloud} category, comprising 21.42\% of the studies, focus on understanding and designing architectures specifically tailored for multi-cloud containerization. Within this category, \textbf{General Cloud Architectures} (4.92\%) and \textbf{Fault Tolerance and Self-Healing} (4.13\%) are key subcategories. These subcategories address critical challenges such as ensuring resilience and performance in cloud-native architectures. The studies on \textbf{Optimization and Deployment Approaches and Tools} category accounts for 18.16\% of the studies, indicating a strong research focus on optimizing container deployment and resource scheduling. Within this category, \textbf{Optimization \& Scheduling Approaches} (8.26\%) represent the second-largest subcategory across all studies, highlighting the practical need for effective resource allocation in multi-cloud systems. Finally, \textbf{Cloud Management and Interoperability Frameworks}, accounting for 12.38\%, primarily aim to improve resource utilization, service management, monitoring, and deployment in multi-cloud environments. This includes contributions from \textbf{Resource and Service Management Frameworks}, which represent 4.95\% of the studies.

\begin{table*}[t]
\centering
\footnotesize
\caption{Selected studies' contributions on containerization in multi-cloud environment}
\label{tab:StudiesContribution}
\begin{tabular}{|c|l|l|}
\hline
\rowcolor{gray!50} \multicolumn{1}{|c|}{\textbf{Category}} & \textbf{Subcategory} & \textbf{Study}\textbf{ \#} \\ \hline
\multirow{7}{*}{\begin{tabular}[c]{@{}c@{}}Reference Architectures \\ and \\ Models for Cloud\end{tabular}} & General Cloud Architectures (6, 4.92\%) & \cite{S26, S33, S35, S42, S76, S117} \\ \cline{2-3} 

& IoT \& Service Mesh Architecture (5, 4.13\%) & \cite{S05, S59, S96, S99, S108}\\ \cline{2-3}

 & Fault Tolerance \& Self-Healing (5, 4.13\%) & \cite{S12, S55, S67, S74, S121} \\ \cline{2-3}   
 
 & Migration \& Testing  Architecture (3, 2.47\%)& \cite{S17, S79, S114} \\ \cline{2-3} 
 & Optimization \& Reliability Model (3, 2.47\%) & \cite{S50, S66,S119} \\ \cline{2-3}
 & Domain-Specific Language (2, 1.65\%) & \cite{S80, S82} \\ \cline{2-3}
 
 & Performance Modeling (2, 1.65\%) & \cite{S28, S62} \\ 
 \hline
 
\multirow{5}{*}{\begin{tabular}[c]{@{}c@{}}Security Approaches\end{tabular}} 

 & Cybersecurity \& Access Control (11, 9.09\%) & \cite{S11, S30, S34, S71, S73} \\ \cline{2-3} 
 
& Cloud Analysis Approaches (7, 5.78\%) & \cite{S18, S24, S37, S65, S69, S86} \\ \cline{2-3} 

 & Security and Performance Solution (6, 4.95\%) & \cite{S10, S23, S38, S77} \\ \cline{2-3} 

  & Security Approaches (5, 4.13\%) & \cite{S06, S25, S40, S83}\\ \cline{2-3} 

 & Provisioning \& Restoration Approach (4, 3.30\%) & \cite{S63, S70, S78, S84}  \\ \hline

\multirow{4}{*}{\begin{tabular}[c]{@{}c@{}}Cloud Management \\ and \\ Interoperability Framework\end{tabular}} & Resource \& Service Management Framework (6, 4.95\%) & \cite{S48, S49, S47, S15, S58, S87} \\ \cline{2-3} 
 & Monitoring \& Deployment Framework (5, 4.13\%) & \cite{S19, S45, S46, S08, S91, S94} \\ \cline{2-3} 
 & Evaluation \& Modeling Framework (4, 3.30\%) & \cite{S20, S85, S101, S105}\\ \cline{2-3} 
 & Interoperability Framework (2, 1.65\%) & \cite{S03, S81} \\ \hline

\multirow{4}{*}{\begin{tabular}[c]{@{}c@{}}Orchestration \\ and\\  Deployment Approach\end{tabular}} & Benchmarking \& Optimization Approach (6, 4.13\%) & \cite{S01, S02, S41, S51, S68, S107} \\ \cline{2-3} 
 & Service Management Approach (4, 3.30\%)& \cite{S07, S14, S29,S116} \\ \cline{2-3} 
 & Simulation \& Integration Approach (3, 2.47\%) & \cite{S27, S31, S56} \\ \cline{2-3} 
 & Storage \& Services Solution (3, 2.47\%) & \cite{S64, S75, S95} \\ \hline

\multirow{3}{*}{\begin{tabular}[c]{@{}c@{}}Optimization and \\ Deployment\\ Approaches and Tool\end{tabular}} & Optimization \& Scheduling Approach (10, 8.26\%)& \cite{S22, S32, S36, S39, S88, S92, S102, S109, S111} \\ \cline{2-3} 
 & Multi-Cloud Platform (6, 4.95\%)& \cite{S09, S16, S43, S60, S72, S22, S32, S36, S39, S113} \\ \cline{2-3} 
 & Deployment Approach and Tool (6, 4.95\%) & \cite{S21,S52,S57,S61,S100,S112} \\ \hline
\end{tabular}
\end{table*}

\subsection{Container Roles and Implementation Strategies (RQ1)}
\label{RQ1}
\textcolor{black}{In this section, we present the roles that containers play in multi-cloud environments and the strategies used to implement them. Based on thematic analysis of the selected studies, we classified the findings into six major roles and five strategic categories. Each category is supported with examples to help readers relate to real-world use cases. The goal is to provide a holistic understanding of how containers are applied and managed across clouds. Readers can interpret the tables as a catalog of practical use scenarios and corresponding strategic techniques.}
\subsubsection{Container Roles}
\label{ContainerRoles}
We classify the roles of containers in a multi-cloud environment into six categories, as shown in Table \ref{tab:RoleofContainers}. Each category is briefly reported below.

\begin{tcolorbox} 
\textbf{Takeaway 3}: \textcolor{black}{The contributions of the reviewed studies highlight how multi-cloud environments introduce specific demands on containerization, particularly in areas such as cross-platform security, cloud-aware architecture design, and adaptive resource scheduling. Unlike generic solutions, these studies propose container orchestration frameworks, interoperability tools, and security models tailored to the challenges of managing distributed containerized applications across heterogeneous cloud providers.}
\end{tcolorbox}

\textbf{1. Container-Based Resource and Service Management}: This category includes studies discussing the use of containers for resource management and hosting. The studies underline how the use of containers enables an efficient multi-tenancy solution, sharing of resources, and reduction in complexity due to efficient management of computational resources. It also supports OS-level virtualization and multi-tenant service management, enhancing general resource utilization. In the context of hosting services, containers provide scalable and isolated environments for application components and services. They support microcloud scaling, ensure efficient service delivery, and are vastly used in edge computing for hosting applications closer to end users. Each of these capabilities together shows why containers are very important for optimizing resource management and offering flexible and reliable hosting of services in cloud environments.

\textbf{2. Container-based Application Deployment}: Containers provide a way to package and distribute applications, along with their dependencies, across different environments. This category reports on the use of containers for deploying various types of applications, such as web apps, Platform as a Service (PaaS) solutions, data-intensive applications, and IoT apps.

\textbf{3. Containerization in App Development and IoT}:  
This category focuses on application development and IoT integration using containers in multi-cloud environment. Key areas include improving deployment speed, enhancing DevOps integration for better portability and efficiency, and enabling standardized software units. Containers are also use to address issues like privacy, compliance, and vendor lock-in. Notable highlights include their role in orchestrating container placement in fog computing and integration with tools like OpenVAS and Chef for enhanced security and automation.

\textbf{4. Performance and Efficiency through Containers}:  In this category, we classified studies that discuss how containers improve performance and efficiency. The selected studies highlights their role in enabling scalable application development and enhancing performance isolation for SaaS. Additionally, studies reported that containers contribute to cloud platform optimization, boosting overall cloud performance, and improving cost efficiency specifically while using microservices. Further discussions include evaluating the performance impact of containers across various systems.

\textbf{5. Container Technology Features}: This category also gathers the studies that report the special characteristics and advantages of container technology, to be specific: lightweight virtualization, where the containers improve resource efficiency in edge computing, and applications are brought closer to end users. Additionally, their compatibility with orchestration tools like Kubernetes simplifies container management and deployment across diverse environments.

\textbf{6. Other Container Uses}: This category report several roles of container technology beyond traditional use cases. Key roles include container-based data storage, support for system architecture, and enhancing blockchain reliability in multi-cloud environment. Additionally, containers are used for evaluating scheduling algorithms and enabling virtual network composition, demonstrating their versatility in addressing a wide range of technical challenges.

\begin{table*}[t]
\centering
\footnotesize
\caption{Roles of containers in multi-cloud environment}
\label{tab:RoleofContainers}
\begin{tabular}{|c|l|l|}
\hline
\rowcolor{gray!50} \textbf{Category} & \textbf{Role of Containers}                                  & \textbf{ID} \\ \hline

\multirow{10}{*}{Container-Based Resource and Service Management}        & Container-Based Multi-Tenancy Solution            & \cite{S04} \\ \cline{2-3} 
              & Container-Driven Resource Sharing \& Multi-Tenancy      & \cite{S06}       \\ \cline{2-3} 
              & Multi-Tenant Service Management via Containers          & \cite{S17}       \\ \cline{2-3} 
              & Container-Enabled OS-Level Virtualization               & \cite{S40}       \\ \cline{2-3} 
              & Container-Based Complexity Reduction                    & \cite{S41}        \\ \cline{2-3}
                      & Containerized Service Hosting                           & \cite{S15}       \\ \cline{2-3} 
                      & Multi-Cloud Workload Migration via Containers           & \cite{S60, S65}  \\ \cline{2-3} 
              & Seamless Workload Migration via Containers              & \cite{S65}       \\ \cline{2-3} 
              & Container-Based Multi-Cloud Platform Execution          & \cite{S79}       \\ \cline{2-3}
              & Container-Driven MicroCloud Scaling \& Isolation        & \cite{S42}       \\ \hline
              
\multirow{9}{*}{Container-based Application Deployment}     & Containerized Web-App Deployment                  & \cite{S01} \\ \cline{2-3} 
              & Container-Driven App Deployment                         & \cite{S05, S08}  \\ \cline{2-3} 
              & Containerized PaaS Prototyping                          & \cite{S09}       \\ \cline{2-3} 
              & Data-Intensive App Deployment via Containers            & \cite{S28}       \\ \cline{2-3} 
              & Containerized App Management                            & \cite{S29}       \\ \cline{2-3} 
              & Container-Based Aneka Service Deployment                & \cite{S51}       \\ \cline{2-3} 
              & Container-Driven Microservices Deployment \& Scaling    & \cite{S62}       \\ \cline{2-3} 
              & Container-Based Multi-Cloud App Deployment \& Scaling   & \cite{S74}       \\ \cline{2-3} 
              & Containerized IoT App Deployment                        & \cite{S85}       \\ \hline

\multirow{9}{*}{Containerization in App Development \& IoT} & App Development \& IoT Integration via Containers & \cite{S16} \\ \cline{2-3} 
              & Container-Based OpenVAS \& Integration of Chef            & \cite{S30}       \\ \cline{2-3} 
              & Containers as Standardized Software Units               & \cite{S32}       \\ \cline{2-3} 
              & Container Placement \& Orchestration in Fog Computing   & \cite{S33}       \\ \cline{2-3} 
              & Container Deployment Speed \& DevOps Integration        & \cite{S36}       \\ \cline{2-3} 
              & Containerizing Apps for Privacy \& Compliance           & \cite{S37}       \\ \cline{2-3} 
              & Container-Driven Portability \& Efficiency Improvement  & \cite{S39}       \\ \cline{2-3} 
              & Preventing Vendor Lock-In via Containers                & \cite{S63}       \\ \cline{2-3} 
              & Container-Based Extensible App Development              & \cite{S76}       \\ \hline

\multirow{6}{*}{Performance \& Efficiency via Containers}   & Scalable App Building via Containers              &\cite{S18} \\ \cline{2-3} 
              & Container-Driven SaaS Performance Isolation             & \cite{S23}       \\ \cline{2-3} 
              & Cloud Performance Enhancement via Containers            & \cite{S48}       \\ \cline{2-3} 
              & Evaluating Performance Impact of Containers             & \cite{S59}       \\ \cline{2-3} 
              & Cost Efficiency of Microservices via Containers         & \cite{S68}       \\ \cline{2-3} 
              & Container-Based Cloud Platform Optimization             & \cite{S69}       \\ \hline

\multirow{5}{*}{Container Technology Features}              & Container Usage in Edge Computing                 & \cite{S22} \\ \cline{2-3} 
              & Container-Enabled Lightweight Virtualization            & \cite{S24}       \\ \cline{2-3} 
              & Key Features of Container Technology                    & \cite{S25}       \\ \cline{2-3} 
              & Container-Based Edge Computing Virtualization           & \cite{S35}       \\ \cline{2-3} 
              & Container Leverage in Kubernetes                        & \cite{S70}       \\ \hline

\multirow{5}{*}{Other Container Uses}                       & Container-Based Azure Data Storage                & \cite{S54} \\ \cline{2-3} 
              & Container \& VM Support in Architecture                 & \cite{S55}       \\ \cline{2-3} 
              & Blockchain-Based Multi-Cloud Reliability via Containers & \cite{S66}       \\ \cline{2-3} 
              & Evaluating Scheduling Algorithm via Containers          & \cite{S67}       \\ \cline{2-3} 
              & Container-Based Virtual Network Composition             &\cite{S72}       \\ \hline
\end{tabular}
\end{table*}
\begin{tcolorbox} 
\textbf{Takeaway 4}: \textcolor{black}{The reviewed studies show that containers take on distinct roles in multi-cloud environments, particularly in addressing challenges such as provider heterogeneity, cross-cloud orchestration, and decentralized application deployment. Containers are used for multi-tenant resource management, cross-cloud service hosting, and workload migration. They also support edge and fog computing, compliance, and vendor lock-in prevention—roles less prominent in single-cloud contexts. This shows that multi-cloud scenarios introduce unique architectural and operational needs that shape how containers are implemented.}
\end{tcolorbox}


\subsubsection{Implementation Strategies}
\label{ContainerImplementationStrategies}
Container implementation strategies in multi-cloud environment refer to the various approaches and techniques used to deploy, manage, and orchestrate containerized applications across multiple cloud platforms. These strategies ensure efficient, secure, and scalable management of containerized workloads. Table \ref{tab:ImplementationStrategies} provides an overview of these strategies, categorized into subcategories. Overall, we identified 76 strategies classified into 11 subcategories and 5 categories. Notably, these strategies were identified from only 76 out of 121 selected studies (62.81\%). The remaining studies do not explicitly discuss the strategies, or it is unclear what strategies they employed. One possible reason for this could be that many studies focus on high-level discussions of container technology without detailing specific implementation strategies. Additionally, in some cases, the strategies might have been implicitly integrated into broader architectural frameworks, making them difficult to extract or clearly identify.

\textbf{1. Container Deployment and Management}: This category focuses on how containers are deployed and managed in multi-cloud environment.  We identified 23 strategies within this category, grouped into three subcategories: \textit{Deployment Strategies},\textit{ Container Lifecycle Management}, and \textit{Data Management}. Deployment strategies, with 10 strategies identified, focus on optimizing the deployment of containers in multi-cloud environment, including approaches such as Kubernetes for containerized applications and Docker image management. Nine strategies related to container lifecycle management address resource monitoring, image optimization, and privacy, with examples like multi-cloud container resource usage monitoring and container image reorganization. The data management subcategory, comprising 4 strategies, focuses on the storage and management of container data in multi-cloud environment, featuring solutions such as Firebase Cloud for real-time databases and blockchain-based container log management. While deployment strategies are well-represented, fewer strategies focus on data management, indicating potential areas for further research in managing data across multi-cloud platforms.

\textbf{2. Container Orchestration}: We identified a total of 20 strategies in this category, classified into two subcategories: \textit{Cloud Orchestration} and \textit{Microservices Orchestration}. The \textit{Cloud Orchestration} subcategory, comprising 13 strategies, focuses on managing containers and resources across multi-cloud environment. These strategies include AI/ML-driven container orchestration, cross-cloud orchestration tools, and Kubernetes CI/CD for multi-cloud setups. The \textit{Microservices Orchestration} subcategory, consisting of 7 strategies, emphasizes the orchestration of microservices using technologies like Kubernetes and Docker. Key strategies include Kubernetes orchestration for multi-cloud containers and edge computing with Docker and Kubernetes. These approaches are essential for enabling scalable and efficient management of containerized applications, with \textit{Cloud Orchestration} strategies focusing on adaptive resource management, while \textit{Microservices Orchestration} supports effective multi-cloud microservices deployment.

\textbf{3. Container Security}: We identified a total of 11 strategies within the \textit{Container Security} category, classified into two subcategories: \textit{Security Policies} and\textit{ Container Placement}. The Security Policies subcategory consists of eight strategies and mainly focuses on enhancing container security through mechanisms such as access control, secure connectivity, and deep learning-based policy generation. Key strategies include implementing defense mechanisms, integrating HIP for Docker security, and managing access in multi-cloud environments using signed URLs. These approaches help address security challenges, safeguard containerized applications, and ensure compliance in multi-cloud infrastructures. The Container Placement subcategory consists of three strategies and mainly focuses on optimizing container placement to balance security and efficiency. These strategies include secure placement using deep learning and the dynamic relocation of microservices across clouds. By enabling secure and adaptive placement, these strategies enhance resource utilization, reduce latency, and maintain consistent application performance across cloud environments.

\textbf{4. Performance Optimization and Scaling Strategies}: We identified a total of 12 strategies within the \textit{Performance Optimization and Scaling Strategies} category, divided into two subcategories: Container \textit{Performance Optimization and Scaling Strategies}. The \textit{Container Performance Optimization} subcategory, consisting of seven strategies, focuses on improving container performance in multi-cloud environments. Key approaches include reactive auto-scaling and resource provisioning, AI-driven resource allocation, and multi-cloud resource monitoring. These strategies aim to enhance performance by optimizing resource usage and reducing latency across cloud platforms. The Scaling Strategies subcategory, with five strategies, centers on improving the scalability of containerized applications. Notable approaches include managing distributed workloads with Kubernetes, utilizing hybrid abstractions for container scaling, and implementing MicroCloud architectures for efficient scaling. Together, these strategies support the dynamic adjustment of containerized applications to fluctuating resource demands in multi-cloud setups while maintaining performance and efficiency.

\textbf{5. Cloud Service and Networking}: We identified a total of 10 strategies within the \textit{Cloud Service and Networking} category, classified into two subcategories: \textit{The Cloud Service Integration} subcategory, with five strategies, focuses on integrating services across multi-cloud platforms. Key approaches include multi-cloud runtime environments, abstraction layers for cloud service standardization, and the integration of OpenStack and IoT in multi-cloud setups. These strategies aim to simplify and streamline the seamless integration of services across different cloud platforms. The \textit{Multi-Cloud Networking Strategies} subcategory, also comprising five strategies, addresses networking challenges in multi-cloud environments. Notable strategies include edge scheduling with location awareness, federated frameworks for cloud storage, and TOSCA-based deployment models for multi-cloud setups. These approaches ensure efficient and secure communication between clouds, supporting smooth service operation and data transfer across diverse cloud infrastructures.


\begin{table*}[]
\centering
\footnotesize
\caption{Container implementation strategies in multi-cloud environment}
\label{tab:ImplementationStrategies}
\begin{tabular}{|c|c|l|l|}
\hline
\rowcolor{gray!50}   \textbf{Category} &
  \multicolumn{1}{c|}{\textbf{Subcategory}} &
  \multicolumn{1}{c|}{\textbf{Implementation Strategies}} &
  \multicolumn{1}{c|}{\textbf{ID}}   
  \\ \hline

\multirow{10}{*}{Container Deployment and Management} &
  
  \multirow{6}{*}{Deployment Strategies} &
  Automated Orchestration for Web Applications &
  \cite{S02} \\ \cline{3-4} 
 &                                       & Kubernetes for Containerized APPs            & \cite{S05} \\ \cline{3-4} 
 &                                       & Multi-cloud App Development Framework                & \cite{S07} \\ \cline{3-4} 
 &                                       & Modular Approach to Container Implementation         & \cite{S08} \\ \cline{3-4} 
 &                                       & Deployment with Docker Images                        & \cite{S09} \\ \cline{3-4} 
&                                       & Docker Image Creation and Management    & \cite{S31} \\ \cline{3-4}
&                                       & Real-time Kubernetes workloads in multi-cloud & \cite{S87} \\ \cline{3-4} 
&                                       & AWS multi-cloud container app deployment & \cite{S115} \\ \cline{3-4} 
&                                       & Multi-cloud container policy enforcement & \cite{S100} \\ \cline{3-4} 
&                                       & Probes and orchestration for cloud containers &  \cite{S101} \\ \cline{2-4}

 & \multirow{4}{*}{Container Lifecycle Management} & Container Image Optimization                         & \cite{S32} \\ \cline{3-4} 
 &                                       & Efficient Image Reorganization                       & \cite{S36} \\ \cline{3-4} 
 &                                       & Orchestrating with Docker and Kubernetes             & \cite{S37} \\ \cline{3-4} 
 &                                       & Executable Images on Multi-cloud Platforms            & \cite{S79} \\ \cline{3-4} 
&                                       & Privacy monitoring for containerized clouds            & \cite{S90} \\ \cline{3-4} 
 &                                       & Multi-cloud container resource usage monitoring           & \cite{S91} \\ \cline{3-4} 
&                                       & Container ecosystem for multi-cloud deployment           & \cite{S93} \\ \cline{3-4} 
&                                       & Containers in AWS ECS for cloud services          & \cite{S107} \\ \cline{3-4} 
&                                       & Container management for multi-cloud MEC           & \cite{S113} \\ \cline{3-4} 
 &                                       &IoT container deployment in multi-cloud          & \cite{S108}  \\ \cline{2-4} 

&
  \multirow{3}{*}{Data Management} &
  Firebase Cloud for Real-Time Database &
  \cite{S35} \\ \cline{3-4} 
 &                                       & High Availability with SpyStorage MCSS               & \cite{S75} \\ \cline{3-4} 
 &                                       & Data Storage with Object Store Service               & \cite{S76} \\ \hline

\multirow{10}{*}{Container Orchestration} &
  \multirow{7}{*}{Cloud Orchestration} &
  SLA-Based Container Strategies &
  \cite{S04} \\ \cline{3-4} 
&                                       & Task Division and Scaling in Container Strategy      & \cite{S12} \\ \cline{3-4} 
&                                       & Location-Aware Service Brokering                     & \cite{S14} \\ \cline{3-4} 
&                                       & Docker Containers for Multi-tenant Services          & \cite{S17} \\ \cline{3-4} 
&                                       & Resource Acquisition for Aneka Containers            & \cite{S51} \\ \cline{3-4} 
&                                       & Cross-Cloud Container and VM Management              & \cite{S60} \\ \cline{3-4} 
&                                       &  Container Orchestration Technologies Overview              & \cite{S33} \\ \cline{3-4} 
&                                       & Multi-cloud dynamic resource adaptation            & \cite{S89} \\ \cline{3-4} 
&                                       & AI/ML orchestration in multi-cloud containers             & \cite{S103} \\ \cline{3-4} 
&                                       & AI-driven container orchestration across clouds             & \cite{S116} \\ \cline{3-4} 
&                                       & Cross-cloud container orchestration tools             & \cite{S117} \\ \cline{3-4} 

&                                       & Kubernetes CI/CD for multi-cloud setups	        & \cite{S102} \\ \cline{2-4} 
 
 &

  \multirow{3}{*}{Microservices Orchestration} & Microservices Architecture in IoT Cloud Apps & \cite{S16} \\ \cline{3-4} 
 &                                       & ADAPT Deployment of Microservices                    & \cite{S27} \\ \cline{3-4} 
&                                       &  MSA Benchmark with Docker and Kubernetes                    & \cite{S62} \\ \cline{3-4} 
&                                       &  Kubernetes for power-efficient multi-cloud                   & \cite{S98} \\ \cline{3-4} 
&                                       &  Edge computing with Docker and Kubernetes                  & \cite{S99} \\ \cline{3-4} 
&                                       &  Kubernetes for edge-cloud orchestration	                 & \cite{S114} \\ \cline{3-4} 
&                                       & Kubernetes orchestration for multi-cloud containers  & \cite{S109} \\ \hline

\multirow{7}{*}{Container Security} &
  \multirow{2}{*}{Container Placement} &  Secure Placement with Deep Learning & \cite{S06} \\ \cline{3-4} 
   &                                    &  Container Placement Strategies  & \cite{S39} \\ \cline{3-4}

 &                                       & Dynamic microservice relocation across clouds                     & \cite{S88} \\ \cline{2-4}

 & \multirow{5}{*}{Security Policies}    & SLA-Oriented Performance Isolation                   & \cite{S23} \\ \cline{3-4} 
 &                                       & Access Control and Defense for Container Security    & \cite{S25} \\ \cline{3-4} 
 &                                       & HIP Integration for Docker Security                  & \cite{S77} \\ \cline{3-4} 
 &                                       & Secure Connectivity in Multi-tenant Environment      & \cite{S72} \\ \cline{3-4} 
 &                                       & Signed URLs for Access Control in Multi-cloud        & \cite{S71} \\ \hline

\multirow{8}{*}{Performance and Scaling Strategies} &
  \multirow{4}{*}{Performance Optimization} &
  LXC Virtual Cluster for Performance &
  \cite{S21} \\ \cline{3-4} 
 &                                       & Scalable Multi-layer Architecture with Containers    & \cite{S28} \\ \cline{3-4} 
 &                                       & Reactive Auto-scaling and Resource Provisioning      & \cite{S48} \\ \cline{3-4} 
 &                                          & Istio for Control Plane Impact and Latency Reduction & \cite{S59} \\ \cline{3-4} 
 &                                          & Auto-scaling containers across clouds     & \cite{S92} \\ \cline{3-4} 
&                                          & Multi-cloud resource monitoring and provisioning    & \cite{S94} \\ \cline{3-4}

 &                                       & AI resource allocation in multi-cloud containers & \cite{S121} \\ \cline{2-4} 
 
 &

  \multirow{4}{*}{Scaling Strategies} &
  Edge Data Streaming and Real-Time Database &
  \cite{S35} \\ \cline{3-4} 
 &                                       & MicroCloud Architecture for Efficient Scaling        & \cite{S42} \\ \cline{3-4} 
 &                                       & Distributed Workloads with Kubernetes                & \cite{S70} \\ \cline{3-4} 
 &                                       & Hybrid Abstractions for Container and Host           & \cite{S40} \\ \hline

\multirow{7}{*}{Cloud Service and Networking} &
  \multirow{3}{*}{Cloud Service Management} & Multi-Tenant Service with Docker and Cloudant & \cite{S17} \\ \cline{3-4} 
 &                                       & Abstraction Layer for Cloud Service Standardization   & \cite{S63} \\ \cline{3-4} 
&                                       & OpenStack and IoT in multi-cloud                      & \cite{S96} \\ \cline{3-4} 

 &                                       & Data indexing for multi-cloud exposure  & \cite{S95} \\ \cline{2-4}

 & \multirow{4}{*}{Networking}           & Edge Scheduling with Location-Awareness              & \cite{S22} \\ \cline{3-4} 
 &                                       & Openstack Networking with Kuryr                      & \cite{S55} \\ \cline{3-4} 
 &                                       & Federated Framework for Cloud Storage                & \cite{S64} \\ \cline{3-4} 
 &                                       & Host Identity Protocol for Docker Networking         & \cite{S77} \\ \hline


\end{tabular}
\end{table*}

\begin{tcolorbox}    
\textbf{Takeaway 5}: \textcolor{black}{The reviewed studies identify a wide range of implementation strategies that reflect the specific challenges of containerizing applications in multi-cloud environments. These include orchestrating containers across heterogeneous platforms, ensuring secure placement and access control under varying provider policies, and optimizing performance and scaling through AI-driven resource provisioning and hybrid abstractions. Unique to multi-cloud settings are strategies for managing distributed services, networking across clouds, and integrating edge and IoT services.}
\end{tcolorbox}

\subsection{Pattern and Strategies (RQ2)}
\label{sec:RQ2}

\textcolor{black}{
Table \ref{tab:PatternsStrategies} presents a thematic classification of patterns and strategies employed in container-based applications within multi-cloud environments. These patterns were identified through qualitative coding and grouped into multiple categories and subcategories based on their architectural, management, security, resilience, and migration focus. Each row corresponds to a pattern extracted from the literature, and studies that reported multiple distinct patterns appear in more than one subcategory (e.g., \cite{S24,S71,S75,S15}). Table \ref{tab:PatternsStrategies} is structured to allow readers to quickly identify and compare solution strategies across different technical domains. To improve interpretability, we provide a brief overview of each category below.
}

\textbf{1. Cloud Architectural Patterns and Models}:
We identified 48 patterns within this category, grouped into four subcategories: \textit{Architecture Patterns}, \textit{Communication and Networking Patterns}, \textit{Deployment Patterns}, and \textit{Service Models}. The Architecture Patterns subcategory, with 22 patterns, is the largest, focusing on designing scalable and flexible cloud systems. Leading patterns include \textit{Microservice Architecture} and \textit{Service-Oriented Architecture (SOA)}, which emphasize modularity and independence of services in multi-cloud environment. Emerging patterns such as \textit{AI-Driven Edge-Cloud Architecture} and \textit{Blockchain-Based Architecture} reflect the incorporation of advanced technologies in cloud design. The Communication and Networking Patterns subcategory consists of 11 patterns, with notable strategies such as \textit{Service Mesh} and \textit{Service Chaining (SC)} providing essential frameworks for ensuring secure and efficient communication between distributed services across cloud environments. These patterns are critical for enabling seamless integration and operation of cloud services in multi-cloud setups. The Deployment Patterns subcategory, containing 8 patterns, focuses on strategies for deploying applications across multiple cloud platforms. Key patterns include \textit{black-Green Deployment}, which supports smooth application updates by alternating between two environments, and \textit{Object Store Service}, which ensures efficient storage of unstructured data across clouds. Advanced orchestration patterns like \textit{Decentralized Orchestration Architecture} and \textit{Policy-Driven Orchestration Architecture} further enhance deployment efficiency by automating resource management across clouds. Lastly, the Service Models subcategory comprises 7 models that define various cloud service delivery mechanisms. The leading models, \textit{Platform-as-a-Service (PaaS)} and \textit{Software-as-a-Service (SaaS)}, provide essential platforms for building, deploying, and managing applications in the cloud without the need for managing underlying infrastructure. Other models, like \textit{Function-as-a-Service (FaaS)} and \textit{Serverless Container-Oriented Architecture}, emphasize serverless computing, allowing developers to focus on code rather than infrastructure management.

\textbf{2. Cloud Management and Resource Allocation Strategies}: We identified 30 strategies within this category, grouped into four subcategories: \textit{Multi-Cloud Management Strategies}, \textit{Container Management Strategies}, \textit{Edge Computing and IoT Strategies}, and \textit{Resource Management Strategies}. The \textit{Multi-Cloud Management Strategies} subcategory, with 8 strategies, focuses on managing and optimizing resources across multiple cloud environments. Leading strategies include the \textit{Multi-cloud Computing Strategy}, \textit{Hybrid/Multi-cloud Approach}, and \textit{MAPE-K control loop}, all of which emphasize seamless integration and resource optimization across cloud platforms. The \textit{Container Management Strategies} subcategory includes 7 approaches that focus on effectively managing containers in cloud environments. Key examples are \textit{Docker Container Images} and \textit{Lightweight Virtualization}, both of which help simplify container operations and improve resource efficiency in multi-cloud settings. The \textit{Edge Computing and IoT Strategies} subcategory consists of 8 strategies that explore how edge computing and IoT devices can be integrated with cloud systems. Noteworthy strategies such as \textit{Mobile Edge Computing (MEC)} and \textit{Fog Computing} aim to bring computation and storage closer to IoT devices, enhancing system responsiveness and performance. Lastly, the \textit{Resource Management Strategies} subcategory comprises 7 strategies dedicated to optimizing resource usage in cloud environments. Prominent approaches like \textit{Resource Allocation based on SLA levels}, \textit{Container Orchestration, Kubernetes, Minikube}, and \textit{AI-Driven Edge-Cloud Architecture} focus on smart resource distribution and coordination across cloud and edge platforms.

\textbf{3. Cloud Security and Resilience Strategies}: We identified 20 strategies within this category, divided into two subcategories: \textit{Security and Resiliency Patterns} and \textit{Fault-tolerance Strategies}. The \textit{Security and Resiliency Patterns} subcategory, comprising 14 strategies, focuses on enhancing security and data resilience in cloud environments. Prominent strategies include \textit{Security-by-Design approach}, \textit{Encryption of files}, and \textit{Centralized/External ACM Service}, which aim to secure data sharing and access control. Emerging patterns like \textit{Blockchain-Driven SLA Management Architecture} and \textit{Hybrid Malware Detection Architecture} highlight the use of advanced technologies to ensure resilient cloud infrastructures. The \textit{Fault-tolerance Strategies} subcategory contains 6 strategies that focus on ensuring system reliability in case of failures. Key strategies include \textit{Fault-tolerance with Redundant Engines} and \textit{Multi-cloud Systems Fault-tolerant Workflow}, both of which emphasize maintaining uninterrupted service in multi-cloud environment by building in redundancy and fault-tolerant processes. The \textit{RAFT Consensus Algorithm} ensures consistency across distributed systems, further improving the resilience of containerized applications in a multi-cloud environment.
.

\textbf{4. Cloud Migration Strategies}: We identified 4 strategies within this category, which focus on various approaches to migrating applications and services to the cloud. These strategies include \textit{Service-oriented Migration}, \textit{Application-centric Migration}, \textit{Image-based Migration}, and \textit{Migration to a Virtualized Container}. Each of these strategies provides a unique method for transferring workloads and data from on-premises or legacy systems to cloud environments. The emphasis is on ensuring a smooth transition with minimal disruption to services, whether through the migration of individual services, entire applications, or the use of virtualized containers. These strategies are essential for organizations looking to leverage cloud technologies while preserving the integrity of their existing systems during the migration process.

\begin{tcolorbox}

\textbf{Takeaway 6}: \textcolor{black}{The studies reveal that when containerized applications are deployed in multi-cloud environments, architecture and management patterns must be adapted to support distributed, heterogeneous infrastructure. Patterns like Microservices and SOA are extended with multi-cloud orchestration and federation capabilities, while service mesh and chaining are tailored for secure, cross-cloud communication. Deployment and fault-tolerant patterns—such as black-Green deployment, decentralized orchestration, and blockchain-driven resilience.}
\end{tcolorbox}

\begin{table*}[]
\footnotesize
\caption{Patterns and strategies for container-based applications in multi-cloud environment}
\label{tab:PatternsStrategies}
\resizebox{\textwidth}{!}{%
\begin{tabular}{|c|c|p{8cm}|p{8cm}|}
\hline
\rowcolor{gray!50} \textbf{Categories} & \textbf{Subcategories} & Patterns & \textbf{ID} \\ \hline
\multirow{30}{*}{\begin{tabular}[c]{@{}c@{}}Cloud Architectural Patterns and Models\end{tabular}} & \multirow{11}{*}
{Architecture Patterns} & Microservice Architecture & \begin{tabular}[c]{@{}l@{}}\cite{S02, S09, S16, S77, S80,  S24, S31, S33, S68, S98, S99, S112}\end{tabular} \\ \cline{3-4} 
 &  & Service-Oriented Architecture (SOA) & \cite{S51, S52, S86, S96} \\ \cline{3-4} 
 &  & Multitier Architecture, Layered Architecture & \cite{S02, S04, S54, S90, S91, S100, S101} \\ \cline{3-4} 
 &  & Client-Server Architecture & \cite{S03, S54} \\ \cline{3-4} 
 &  & Component-based Architecture & \cite{S22} \\ \cline{3-4} 
 &  & Event-driven Architecture & \cite{S44} \\ \cline{3-4} 
 &  & Multi-agent Architecture & \cite{S74, S48} \\ \cline{3-4} 
 &  & Master-worker Nodes Architecture & \cite{S33} \\ \cline{3-4} 
 &  & Netflix Zuul-based Seeker Component & \cite{S17} \\ \cline{3-4} 
 &  & Multi-tenant Cloud Service Architecture & \cite{S17} \\ \cline{3-4} 
    &  & Stateful Engine Architecture & \cite{S17} \\ \cline{3-4} 
    &  & Distributed Multi-Cloud Native Architecture & \cite{S117} \\ \cline{3-4}
 &  & Plugin-Based Deployment Automation Architecture & \cite{S118} \\ \cline{3-4}
  &  & Cloud-Edge Integrated Robotics Architecture & \cite{S119} \\ \cline{3-4}

   &  & Hybrid Malware Detection Architecture & \cite{S120} \\ \cline{3-4}
  &  & Cloudfront-Enhanced Container Architecture & \cite{S115} \\ \cline{3-4}
  &  & AI-Driven Edge-Cloud Architecture & \cite{S116} \\ \cline{3-4}
  &  & Metadata-Driven Architecture & \cite{S95} \\ \cline{3-4}
&  & Blockchain-Based Architecture & \cite{S97} \\ \cline{3-4}
 &  & Cross-Layered Edge-Cloud Architecture & \cite{S109, S113} \\ \cline{2-4}

 & \multirow{10}{*}{\begin{tabular}[c]{@{}c@{}}Communication and Networking Patterns\end{tabular}} & Service Mesh & \cite{S05} \\ \cline{3-4} 
 &  & Service Chaining (SC) & \cite{S50} \\ \cline{3-4} 
 &  & Multi-Cloud APIs & \cite{S71} \\ \cline{3-4} 
 &  & Third-Party  APIs & \cite{S71} \\ \cline{3-4} 
 &  & Service Request Broker & \cite{S07} \\ \cline{3-4} 
 &  & Publish and Subscribe Communication Protocol & \cite{S45} \\ \cline{3-4} 
 &  & Synchronous Network Communication Protocols & \cite{S01} \\ \cline{3-4} 
 &  & Asynchronous Network Communication Protocols & \cite{S01} \\ \cline{3-4} 
 &  & Network Function Virtualization (NFV) & \cite{S50} \\ \cline{3-4} 
&  &Service Discovery & \cite{S76} \\ \cline{3-4} 
 &  &  Blockchain-Based Identity and Access Management (IAM) Architecture &\cite{S110} \\ \cline{3-4} 
 &  & Federated Cloud-Edge Architecture & \cite{S114} \\ \cline{2-4}

 & \multirow{5}{*}{Deployment Pattern} & black-Green deployment & \cite{S27, S30} \\ \cline{3-4} 
 &  & Object store service & \cite{S76} \\ \cline{3-4} 
 &  & Multi-cloud Deployment Model & \cite{S10} \\ \cline{3-4} 
 &  & Distributed Deployment Model & \cite{S10} \\ \cline{3-4} 
 &  & Multi-Cloud Deployment Architecture & \cite{S107, S102} \\ \cline{3-4} 
 &  & Decentralized Orchestration Architecture & \cite{S103} \\ \cline{3-4} 
 &  & Policy-Driven Orchestration Architecture & \cite{S104} \\ \cline{3-4} 
 &  & DevOps-Oriented Architecture & \cite{S108} \\ \cline{2-4} 

 & \multirow{4}{*}{Service Models} & Platform-as-a-Service (PaaS) & \cite{S24, S41} \\ \cline{3-4} 
 &  & Software-as-a-Service & \cite{S79} \\ \cline{3-4} 
 &  & Infrastructure-as-a-Service (IaaS) & \cite{S24} \\ \cline{3-4} 
 &  & Function-as-a-Service (FaaS) & \cite{S24, S105} \\ \cline{3-4} 
 &  & Multi-Cloud Orchestration Architecture & \cite{S102} \\ \cline{3-4} 
 &  & Serverless Container-Oriented Architecture & \cite{S41} \\ \hline

\multirow{24}{*}{\begin{tabular}[c]{@{}c@{}}Cloud Management and  Resource Allocation Strategies\end{tabular}} & \multirow{10}{*}{\begin{tabular}[c]{@{}c@{}}Multi-Cloud  \\ Management Strategies\end{tabular}} & Multi-cloud computing strategy & S78, S81 \\ \cline{3-4} 
 &  & Hybrid cloud architecture & \cite{S78} \\ \cline{3-4} 
 &  & Hybrid/Multi-cloud Approach & \cite{S05, S83} \\ \cline{3-4} 
 &  & Introducing    multi-cloud Middleware & \cite{S52, S53} \\ \cline{3-4} 
 &  & MAPE-K control loop & \cite{S46, S48} \\ \cline{3-4} 
 &  & Connecting to multiple cloud service providers & \cite{S76} \\ \cline{3-4} 
 &  & Multi-cloud   load balancing & \cite{S23, S76} \\ \cline{3-4} 
 &  & Multi-Cloud BP provisioning & \cite{S84} \\ \cline{3-4} 
 &  & Cross-level orchestration of cloud services & \cite{S84} \\ \cline{3-4} 
 &  & Cross-level monitoring and adaptation of BPs & \cite{S84} \\ \cline{2-4} 
 & \multirow{7}{*}{\begin{tabular}[c]{@{}c@{}}Container Management \\ Strategies\end{tabular}} & Linux Container   (LXC) project & \cite{S24} \\ \cline{3-4} 
 &  & Container engine & \cite{S24} \\ \cline{3-4} 
 &  & Docker Container Images & \cite{S24} \\ \cline{3-4} 
 &  & Lightweight virtualization & \cite{S24} \\ \cline{3-4} 
 &  & Portable application packaging & \cite{S24} \\ \cline{3-4} 
 &  & One-container-per-app approach & \cite{S24} \\ \cline{3-4} 
 &  & Data volumes and data volume containers & \cite{S24} \\ \cline{2-4} 
 & \multirow{5}{*}{\begin{tabular}[c]{@{}c@{}}Edge Computing \\ and \\ IoT Strategies\end{tabular}} & Mobile Edge Computing (MEC) & \cite{S22, S70, S43, S54} \\ \cline{3-4} 
 &  & Fog computing & \cite{S32} \\ \cline{3-4} 
 &  & Edge Services & \cite{S35} \\ \cline{3-4}

 &  & Connecting IoT Edge Devices to the Cluster & \cite{S05} \\ \cline{3-4} 
 &  & Cloud of Things (CoT) & \cite{S05} \\ \cline{3-4} 
 &  & Multi-Access Edge Computing (MEC) Architecture & \cite{S113} \\ \cline{3-4} 
 
   &  & Cross-Layered Edge-Cloud Architecture & \cite{S109} \\ \cline{3-4} 
   
   &  & Cloud-Edge Integrated Robotics Architecture & \cite{S119} \\ \cline{3-4} 
   
 &  & AI-Driven Edge-Cloud Architecture & \cite{S116} \\ \cline{2-4} 

 & \multirow{2}{*}{\begin{tabular}[c]{@{}c@{}}Resource Management \\ Strategies\end{tabular}} & Resource allocation based on different SLA levels & \cite{S23} \\ \cline{3-4} 
   &  & Container Orchestration, Kubernetes, Minikube &\cite{S24, S29, S33} \\ \cline{3-4} 
  
   &  & Rate-Based Stream Processing Architecture & \cite{S111} \\ \cline{3-4} 
 &  & Multi-Agent Resource Optimization Architecture 
 & \cite{S121}\\ \hline

\multirow{16}{*}{\begin{tabular}[c]{@{}c@{}}Cloud Security\\ and \\ Resilience Strategies\end{tabular}} & \multirow{13}{*}{\begin{tabular}[c]{@{}c@{}}Security\\ and \\ Resiliency Patterns\end{tabular}} & Security-by-Design   approach & \cite{S51} \\ \cline{3-4} 
 &  & Encryption of files & \cite{S71} \\ \cline{3-4} 
 &  & Standardized APIs for file transfer & \cite{S71} \\ \cline{3-4} 
 &  & Centralized/External ACM Service & \cite{S71} \\ \cline{3-4} 
 &  & Distributed ACM Service & \cite{S71} \\ \cline{3-4} 
 &  & Dynamic Switching between Authentication Methods & \cite{S71} \\ \cline{3-4} 
 &  & ACM based on Signed URLs & \cite{S71} \\ \cline{3-4} 
 &  & Attribute-Based   Encryption and Signature & \cite{S75} \\ \cline{3-4} 
 &  & Secure data sharing & \cite{S75} \\ \cline{3-4} 
 &  & Data access control & \cite{S75} \\ \cline{3-4} 
 &  & Local encryption and signing & \cite{S75} \\ \cline{3-4} 
 &  & Authorized Tokens & \cite{S75} \\ \cline{3-4} 
&  & Byzantine quorum protocol & \cite{S75}\\ \cline{3-4} 
&  & Blockchain-Driven SLA Management Architecture & \cite{S106} \\ \cline{3-4} 
&  & Hybrid Malware Detection Architecture & \cite{S120} \\ \cline{2-4}

 & \multirow{3}{*}{\begin{tabular}[c]{@{}c@{}}Fault-tolerance Strategies\end{tabular}} & Fault-tolerance with Redundant Engines & \cite{S17} \\ \cline{3-4} 
 &  & Multi-cloud systems fault-tolerant workflow & \cite{S67} \\ \cline{3-4} 
 &  & RAFT Consensus Algorithm & \cite{S13} \\ \hline

\multirow{4}{*}{\begin{tabular}[c]{@{}c@{}}Cloud Migration\\ Strategies\end{tabular}} & \multirow{4}{*}{} & Service-oriented Migration & \cite{S15} \\ \cline{3-4} 
 &  & Application-centric Migration & \cite{S15} \\ \cline{3-4} 
 &  & Image-based Migration & \cite{S15} \\ \cline{3-4} 
 &  & Migration to a Virtualized Container & \cite{S15} \\ \hline
\end{tabular}}
\end{table*}


\subsection{Quality Attributes and Tactics (RQ3)}
\label{Sec:RQ3}
\textcolor{black}{
Table \ref{tab:QATactics} presents a thematic classification of QAs and their associated implementation tactics for containerized applications in multi-cloud environments. These QAs and tactics were derived through thematic analysis of data extracted from 121 primary studies and are organized according to the ISO 25010 standard. This standard defines high-level software quality characteristics and their sub-attributes, which served as a guiding framework for our classification. The table includes nine core QAs, namely Performance (Efficiency), Security, Compatibility, Scalability, Reliability, Portability, Flexibility, Maintainability, and Usability. For each QA, the table lists representative related terms, example study references, and commonly applied tactics. Due to space constraints, only five example tactics per QA are shown in Table \ref{tab:QATactics}, while the complete list of 70 tactics is available in the replication package~\cite{replpack}. These tactics are distributed across multiple studies and highlight recurring practices for achieving specific quality goals.  To help readers navigate this classification, we briefly explain below how each QA is addressed in the literature and what patterns emerge in terms of implementation strategies.
}

\textbf{Performance (Efficiency)}: Performance or efficiency is the most dominant QA, frequently discussed in 60 (49.58\%) of the selected studies, along with related terms or characteristics such as time behavior, resource utilization, and capacity. We also listed several tactics that can be used to achieve performance (or efficiency) in containerized applications, such as efficient resource optimization and allocation, use of machine learning techniques for performance optimization, utilization of container technology, location-aware service brokering, and performance-oriented Service Level Agreements (SLA). For instance, optimizing resources ensures that the system uses the minimum possible resources while delivering the required output. Using ML techniques for performance optimization can enhance system performance by learning and adjusting the operational parameters. Similarly, utilizing container technology can help maintain optimal performance by encapsulating the application and its environment. These results also indicate that performance is a critical QA, with a strong focus in containerized applications, and the majority of studies advocate for advanced tactics to achieve optimal performance.

\textbf{Security}: This QA has been discussed in 38 studies along with various tactics. We also identified several characteristics of security that have been highlighted in the selected studies, such as confidentiality, authenticity, access control, authorization, privacy, trustworthiness, and integrity.  Some of the identified tactics to improve the security of containerized applications from the selected studies include encryption and strong authentication mechanisms, implementation of firewalls, VPNs, or SDNs for network security, utilization of ML techniques to detect and prevent security threats and attacks, deployment of applications on diverse cloud providers, and consideration of users' security specifications. For example, encryption provides a secure way of transmitting data, and ML can detect unusual behavior that could signal a threat or attack.

\textbf{Compatibility}: This QA has been reported in 15 studies, along with characteristics such as interoperability and co-existence. Several tactics that can be employed to achieve compatibility include the implementation of lightweight communication protocols and modes, standardization of interfaces for seamless integration, componentization and modular design for interoperability, prototyping and exploring interoperability approaches for multi-cloud deployment, and interoperability standardization and federation between clouds. Standardizing interfaces can ensure that different software components can interact with each other seamlessly, enhancing compatibility. The results suggest that compatibility is essential for ensuring smooth integration and operation in multi-cloud environment.

\textbf{Scalability}: Scalability refers to the ability to expand capacity or performance without losing the efficiency or functionality of a software system. We identified 40 studies that report scalability along with other characteristics like capacity, extensibility, elasticity, and throughput. We also identified ten tactics, five of which are listed in Table \ref{tab:QATactics}, to achieve scalability. For example, scalability can be achieved through tactics like combining containerization and microservices for enhanced scalability, implementing elastic resource allocation, enabling migration between multi-cloud services, combining container-based deployment with runtime monitoring and optimization, and leveraging orchestration, federated networks, and geographic placement. Elastic resource allocation ensures that the system can seamlessly scale up or down in response to changing demands.

\textbf{Reliability}: This QA is about the system's correct operation and consistent performance without failure over a specified period, while ensuring the availability of the system. We identified 33 studies that report on reliability, along with characteristics such as maturity, availability, and fault tolerance. We identified nine tactics overall (see Table \ref{tab:QATactics} and the QA and Tactics Sheet in \cite{replpack}) to achieve reliability from the selected studies. These include deploying redundant engines for fault tolerance, distributing resources and replicating applications for improved response time, enforcing redundancy and distributed services for availability, deploying parallel search and multi-cloud distribution, and building fault-tolerant systems to enhance system reliability.

\textbf{Portability}: Portability refers to a system’s ability to function correctly across different platforms (e.g., operating systems, hardware configurations). We identified 6 studies that report on portability, along with characteristics such as installability, replaceability, and adaptability, as well as 5 tactics that can help achieve portability. For example, portability can be achieved using tactics such as virtual machine-based packaging, Docker image-based packaging, cloud-portable containerization, dynamic cross-level adaptation and provisioning, and adaptive rule-based system modification. These tactics ensure that an application can be easily transferred from one computing environment to another.

\textbf{Flexibility}: This QA represents the system's capacity to accommodate new features based on user requirements with minimal effort. We identified 10 studies (see Table \ref{tab:QATactics} and the QA and Tactics Sheet in \cite{replpack}) that report this QA, along with five tactics that can help achieve it. For example, flexibility can be improved through tactics such as service-oriented architecture, on-demand dynamic allocation of resources across different cloud platforms, containers, container orchestration, model-driven development, risk analysis, and cross-cloud service orchestration. Using multi-cloud environment, for instance, offers the flexibility to choose services from different providers as per specific needs.

\textbf{Maintainability}: This QA has been reported in 7 studies, along with several characteristics such as modularity, reusability, and analyzability. We also identified 6 tactics from these studies that can help achieve maintainability. For example, tactics like microservices for easier maintenance, Infrastructure Provisioning as Code (IaC), containerization and image management, configuration management and templating, version control and change management, and continuous integration and deployment automation can enhance maintainability. For example, IaC allows developers to manage infrastructure more efficiently and minimize human errors, thus increasing maintainability.

\textbf{Usability}: This QA refers to the ease with which users can interact with a system to perform various operations. We identified 6 studies that report on usability, along with several characteristics such as learnability, and operability, as well as 6 other tactics. According to the selected studies, usability can be enhanced by employing tactics such as a feedback-loop controller for multi-cloud infrastructure, container orchestration (e.g., Kubernetes), user-centric interface design, responsive and adaptive user experience, accessible design and compliance, and error handling and feedback mechanisms. For instance, a feedback-loop controller is effective in a multi-cloud context as it dynamically adjusts resources across different cloud platforms based on user interactions, ensuring a seamless and responsive user experience tailored to the multi-cloud setup.
\begin{tcolorbox} 
\textbf{Takeaway 7}: \textcolor{black}{QAs and their associated tactics from the reviewed studies reveal that ensuring performance, security, reliability, and other QAs in containerized applications requires context-specific adaptations when operating in multi-cloud environments. Unlike single-cloud setups, these environments introduce heterogeneity in infrastructure, distributed control, and varying security models across providers. As a result, tactics such as SLA-aware resource optimization, federated access control, decentralized fault-tolerance mechanisms, and cross-platform provisioning are essential to achieving QA goals.}
\end{tcolorbox}


\begin{table*}
\footnotesize
\caption{Quality attributes and tactics for containerized applications in multi-cloud environment}
\label{tab:QATactics}
\resizebox{\textwidth}{!}{%
\begin{tabular}{|c|c|l|l|}
\hline
\rowcolor{gray!50} \textbf{Quality Attribute} & \textbf{Related Terms} & \textbf{ID} & \multicolumn{1}{c|}{\textbf{Tactics}} \\ \hline
\multirow{5}{*}{\begin{tabular}[c]{@{}c@{}}Performance\\ (Efficiency)\end{tabular}} & \multirow{5}{*}{\begin{tabular}[c]{@{}c@{}}Time behavior, \\ Resource utilization, \\ Capacity, \\ Throughput, \\ Response time\end{tabular}} & \multirow{5}{*}{\begin{tabular}[c]{@{}l@{}} \cite{S01, S02, S03, S06, S09, S14, S15},\\ \cite{S16, S20, S21, S23, S25, S27, S29}, \\ \cite{S34, S35, S36, S37, S38, S39, S41}, \\ \cite{S43, S44, S47, S49, S52, S54, S57}, \\ \cite{S58, S59, S60, S62, S64, S65, S66}\end{tabular}} & Efficient resource optimization and allocation \\ \cline{4-4} 
 &  &  & Utilizing machine learning techniques for performance optimization \\ \cline{4-4} 
 &  &  & Container technology utilization \\ \cline{4-4} 
 &  &  & Location-aware service brokering \\ \cline{4-4} 
 &  &  & Performance-oriented Service Level Agreements (SLA) \\ \hline
\multirow{5}{*}{Security} & \multirow{5}{*}{\begin{tabular}[c]{@{}c@{}}Confidentiality, \\ Authenticity, \\ Access control, \\ Authorization, \\ Privacy, \\ Trustworthiness, \\ Integrity\end{tabular}} & \multirow{5}{*}{\begin{tabular}[c]{@{}l@{}} \cite{S01,S05, S06, S09, S15, S23, S24}, \\ \cite{S25, S29, S34, S37, S38, S40, S44},\\ \cite{S47, S52, S54, S55, S57, S59, S61}, \\ \cite{S64, S65, S66, S69, S71, S72, S73, S77, S86}\end{tabular}} & Encryption and strong authentication mechanisms \\ \cline{4-4} 
 &  &  & Implementation of firewalls, VPNs, or SDNs for network security \\ \cline{4-4} 
 &  &  & \begin{tabular}[c]{@{}l@{}}Utilizing machine learning techniques to detect and\\ prevent security threats and attacks\end{tabular} \\ \cline{4-4} 
 &  &  & Deployment of cloud applications on diverse cloud providers \\ \cline{4-4} 
 &  &  & \begin{tabular}[c]{@{}l@{}}Consideration of users' security specifications and \\ addressing CSP incompatibilities\end{tabular} \\ \hline
\multirow{5}{*}{Compatibility} & \multirow{5}{*}{\begin{tabular}[c]{@{}c@{}}Interoperability, \\ Co-existence\end{tabular}} & \multirow{5}{*}{\begin{tabular}[c]{@{}l@{}}\cite{S15, S16, S21, S24, S27, S42,S44},\\ \cite{S45, S49, S63, S64, S71, S77, S80, S82} \end{tabular}} & Implementing lightweight communication protocols  and modes \\ \cline{4-4} 
 &  &  & Standardization of interfaces for seamless integration \\ \cline{4-4} 
 &  &  & Componentization and modular design for interoperability \\ \cline{4-4} 
 &  &  & \begin{tabular}[c]{@{}l@{}}Prototyping and exploring interoperability approaches for \\ multi-cloud deployment\end{tabular} \\ \cline{4-4} 
 &  &  & Interoperability standardization and federation between clouds \\ \hline
\multirow{5}{*}{Scalability} & \multirow{5}{*}{\begin{tabular}[c]{@{}c@{}}Capacity, \\ Extensibility, \\ Elasticity, \\ Throughput, \\ Responsiveness\end{tabular}} & \multirow{5}{*}{\begin{tabular}[c]{@{}l@{}}\cite{S01, S08, S09, S14, S15, S16, S17},\\ \cite{S18,S19, S21, S22, S24, S25, S27},\\ \cite{S29, S31, S35, S41, S42, S44, S45},\\ \cite{S54, S57, S62, S66, S69, S72, S74},\\ \cite{S76, S77, S82, S84}\end{tabular}} & \begin{tabular}[c]{@{}l@{}}Combining Containerization and Microservices for enhanced scalability\end{tabular} \\ \cline{4-4} 
 &  &  & Implementing Elastic Resource Allocation \\ \cline{4-4} 
 &  &  & Enabling Migration between Multi-Cloud Services \\ \cline{4-4} 
 &  &  & \begin{tabular}[c]{@{}l@{}}Combining Container-based Deployment with Run-time Monitoring\\ and Optimization\end{tabular} \\ \cline{4-4} 
 &  &  & \begin{tabular}[c]{@{}l@{}}Leveraging Orchestration, Federated Network,and Geographic Placement\end{tabular} \\ \hline
\multirow{5}{*}{Reliability} & \multirow{5}{*}{\begin{tabular}[c]{@{}c@{}}Maturity,\\ Availability, \\ Fault tolerance, \\ Recoverability\end{tabular}} & \multirow{5}{*}{\begin{tabular}[c]{@{}l@{}}\cite{S01, S02, S04, S10, S13, S17,S19},\\ \cite{S20, S22, S25, S27, S32, S33, S35},\\ \cite{S37, S44, S47, S51,S53, S54, S55},\\ \cite{S56, S57, S60, S64, S65, S66, S67},\\ \cite{S71, S73, S74, S75, S82}\end{tabular}} & Deploying Redundant Engines for Fault-Tolerance \\ \cline{4-4} 
 &  &  & \begin{tabular}[c]{@{}l@{}}Distributing Resources and Replicating Applications for Improved \\ Response Time\end{tabular} \\ \cline{4-4} 
 &  &  & Enforcing Redundancy and Distributed Service for Availability \\ \cline{4-4} 
 &  &  & Deploying Parallel Search and Multi-Cloud Distribution \\ \cline{4-4} 
 &  &  & Building Fault-Tolerant Systems \\ \hline
\multirow{5}{*}{Portability} & \multirow{5}{*}{\begin{tabular}[c]{@{}c@{}}Installability,\\ Replaceability, \\ Adaptability\end{tabular}} & \multirow{5}{*}{\cite{S18, S24, S25, S31, S62, S77}} & Virtual Machine-Based Packaging \\ \cline{4-4} 
 &  &  & Docker Image-Based Packaging \\ \cline{4-4} 
 &  &  & Cloud-Portable Containerization \\ \cline{4-4} 
 &  &  & \begin{tabular}[c]{@{}l@{}}Dynamic Cross-Level Adaptation and Provisioning\end{tabular} \\ \cline{4-4} 
 &  &  & Adaptive Rule-Based System Modification \\ \hline
\multirow{5}{*}{Flexibility} & \multirow{5}{*}{\begin{tabular}[c]{@{}c@{}}Functional completeness, \\ Functional correctness,\\ Functional appropriateness\end{tabular}} & \multirow{5}{*}{\begin{tabular}[c]{@{}l@{}} \cite{S16, S21, S27, S31, S41, S61, S60},\\ \cite{S70, S76, S84}\end{tabular}} & Service-oriented Architecture \\ \cline{4-4} 
 &  &  & Dynamic allocation of resources across different cloud platforms based on demand \\ \cline{4-4} 
 &  &  & \begin{tabular}[c]{@{}l@{}}Container orchestration\end{tabular} \\ \cline{4-4} 
 &  &  & Model-Driven Development, Risk Analysis\\ \cline{4-4} 
 &  &  & Cross-cloud service orchestration, \\ \hline
\multirow{6}{*}{Maintainability} & \multirow{6}{*}{\begin{tabular}[c]{@{}c@{}}Modularity,\\ Reusability, \\ Analyzability,\\ Modfiabality, \\ Testability\end{tabular}} & \multirow{6}{*}{\cite{S08, S17, S25, S27, S62, S74, S82}} & Microservices, easier maintenance \\ \cline{4-4} 
 &  &  & Infrastructure Provisioning as Code (IaC) \\ \cline{4-4} 
 &  &  & Containerization and Image Management \\ \cline{4-4} 
 &  &  & Configuration Management and Templating \\ \cline{4-4} 
 &  &  & Version Control and Change Management \\ \cline{4-4} 
 &  &  & \begin{tabular}[c]{@{}l@{}}Continuous Integration and Deployment Automation\end{tabular} \\ \hline
 
\multirow{6}{*}{Usability} & \multirow{6}{*}{\begin{tabular}[c]{@{}c@{}}Appropriateness \\ recognizability, \\ Learnability, \\ Operability , \\ User interface aesthetics, \\ Accessibility\end{tabular}} & \multirow{6}{*}{\cite{S09, S17, S27, S27, S37, S74}} & Feedback-loop controller for multi-cloud Infrastructure \\ \cline{4-4} 
 &  &  & Container Orchestration (e.g., Kubernetes) \\ \cline{4-4} 
 &  &  & User-Centric Interface Design \\ \cline{4-4} 
 &  &  & Responsive and Adaptive User Experience \\ \cline{4-4} 
 &  &  & Accessible Design and Compliance \\ \cline{4-4} 
 &  &  & Error Handling and Feedback Mechanisms \\ \hline
\end{tabular}}
\end{table*}

\subsection{Security Challenges and Solution Framework (RQ4 and RQ5)}
\label{SecurityChallengesandSolutionsFramework}

\textcolor{black}{This section presents a framework for addressing security challenges in containerized applications in multi-cloud environment. Based on a detailed review, it identifies seven key categories of challenges and solutions, outlined in Table \ref{tab:Security-challenges-solutions}. Each category is briefly reported below:}

\textbf{Data Security in Container-based Applications} category covers multiple challenges related to data security in container-based applications, including data protection, database security, data compliance, and secure data transfer in a multi-cloud environment. Out of the 121 reviewed studies, 11 specifically highlighted these issues and provided corresponding solutions (see Table \ref{tab:Security-challenges-solutions} and the Security Challenges-Solution sheet in \cite{replpack}). The identified solutions focus on implementing robust encryption techniques, secure container orchestration, and compliance with regulatory standards such as GDPR and HIPAA. 

\clearpage
\onecolumn
\footnotesize
\renewcommand{\arraystretch}{1.15}

\begin{longtable}{|c|>{\columncolor{gray!50}}p{8cm}|>{\columncolor{gray!20}}p{7cm}|}
\caption{Identified security challenges and solutions for containerized applications in multi-cloud environment}
\label{tab:Security-challenges-solutions} \\
\hline
\rowcolor{gray!50}
\textbf{ID} & \textbf{Challenge} & \textbf{Solution} \\
\hline
\endfirsthead

\multicolumn{3}{c}%
{{\bfseries \tablename\ \thetable{} -- continued from previous page}} \\
\hline
\rowcolor{gray!50}
\textbf{ID} & \textbf{Challenge} & \textbf{Solution} \\
\hline
\endhead

\hline \multicolumn{3}{r}{{Continued on next page}} \\
\endfoot

\hline
\endlastfoot

\multicolumn{3}{|c|}{\textbf{Category 1: Data Security}} \\ \hline

\cite{S17} & \textsf{Data Protection} & Multi-Cloud Data Protection through Secure Containers \\ \hline

\cite{S29} & \textsf{Database Security} & Enhanced multi-cloud Database Security via Containerization \\ \hline

\cite{S37} & \textsf{Data Compliance} & Multi-cloud Legal Compliance and Data Protection \\ \hline

\cite{S38} & \textsf{Data Security} & Secure CSP Interoperability and Selection in multi-cloud \\ \hline

\cite{S53} & \textsf{Data Ownership and Privacy} & Multi-cloud Security via NoMISHAP Service Abstraction \\ \hline

\cite{S58} & \textsf{Data Transfer Security} & Secure multi-cloud Integration via OpenStack Standardization \\ \hline

\cite{S85} & \textsf{Cloud Data Management} & Multi-cloud Data Security via DRA Framework \\ \hline

\cite{S97} & \textsf{Data leaks and access risks} & Cryptography, smart contracts, encryption \\ \hline

\cite{S103} & \textsf{Decentralized data security challenge} & Secure data handling with decentralized AI \\ \hline

\cite{S106} & \textsf{SLA data integrity challenge} & Blockchain for SLA security and integrity \\ \hline

\cite{S113} & \textsf{Multi-cloud data privacy issue} & Stringent security controls for data protection \\ \hline

\multicolumn{3}{|c|}{\textbf{Category 2: Access and Communication Control}} \\ \hline

\cite{S05} & \textsf{Connectivity and Access Control} & Multi-cloud Containerized Application Security \\ \hline

\cite{S43} & \textsf{Secure Communication} & Middleware Secure Deployment for multi-cloud \\ \hline

\cite{S51} & \textsf{Provider-Agnostic Operations} & Aneka-Based Security and Performance for Containerized multi-cloud Applications \\ \hline

\cite{S99} & \textsf{Microservices privacy and access control} & Encryption, authentication, service mesh for security \\ \hline

\cite{S110} & \textsf{Decentralized IAM security issue} & Blockchain for IAM security and data integrity \\ \hline

\cite{S114} & \textsf{Secure communication across edge-cloud environments} & WireGuard for secure edge-cloud communication \\ \hline

\cite{S116} & \textsf{Multi-domain edge communication security} & Secure communication and AI anomaly detection \\ \hline

\cite{S119} & \textsf{Secure robot-cloud communication} & Secure communication and encryption for data integrity \\ \hline

\multicolumn{3}{|c|}{\textbf{Category 3: Container Security}} \\ \hline

\cite{S06} & \textsf{Co-resident Security} & Secure Placement Strategy via Deep RL \\ \hline

\cite{S25} & \textsf{Container Isolation} & Power and Leakage Management for multi-cloud Security \\ \hline

\cite{S40} & \textsf{Container and Host Security} & Host-Container Cooperation for multi-cloud Defense \\ \hline

\cite{S120} & \textsf{Container security and malware detection} & DockerWatch for malicious activity detection \\ \hline

\cite{S112} & \textsf{Multi-tenant serverless security challenge} & Security best practices for serverless deployments \\ \hline

\multicolumn{3}{|c|}{\textbf{Category 4: Infrastructure and Deployment Management Security}} \\ \hline

\cite{S20} & \textsf{Data Deployment} & Secure Private Data Deployment Strategy \\ \hline

\cite{S44} & \textsf{Infrastructure Orchestration} & Standard Interface Implementation for Serverless Security \\ \hline

\cite{S47} & \textsf{Storage Security} & Multi-cloud Storage Security with MSSF \\ \hline

\cite{S80} & \textsf{Migration Security} & Infrastructure-Aware and Agnostic Secure Migration \\ \hline

\cite{S108} & \textsf{Distributed infrastructure security challenge} & Orchestration security for IoT data protection \\ \hline

\cite{S115} & \textsf{Securing AWS app deployments} & AWS best practices for CI/CD security \\ \hline

\multicolumn{3}{|c|}{\textbf{Category 5: Security Monitoring and Breach Prevention}} \\ \hline

\cite{S30} & \textsf{System Vulnerabilities} & Automated Security Evaluation and Bootstrapping \\ \hline

\cite{S45} & \textsf{Breach Detection and Prevention} & Metric Filtering and Fault Recovery for multi-cloud Security \\ \hline

\cite{S104} & \textsf{Network policy misconfiguration risks} & Automated network policy discovery and verification \\ \hline

\cite{S111} & \textsf{Stream processing security risks} & RBAM framework for secure data handling \\ \hline

\multicolumn{3}{|c|}{\textbf{Category 6: Trust and Compliance Management}} \\ \hline

\cite{S52} & \textsf{Application Security and Compliance} & Risk Analysis and Security SLAs with MUSA Framework \\ \hline

\cite{S49} & \textsf{Data Security and Standardization} & OpenStack API-based multi-cloud Security and Interoperability \\ \hline

\cite{S57} & \textsf{Network Security} & Inter-cloud Communication Protection with SFC \\ \hline

\cite{S83} & \textsf{Trust and Community Formation} & Trust-based Hedonic Coalitions for multi-cloud Security \\ \hline

\cite{S84} & \textsf{Orchestration and Adaptation} & Multi-cloud Service Orchestration and Cross-Level Adaptation \\ \hline

\cite{S107} & \textsf{Multicloud compliance and security issue} & AWS security measures for multicloud environments \\ \hline

\cite{S117} & \textsf{Cross-cloud security and compliance} & Security frameworks for multi-cloud environment \\ \hline

\multicolumn{3}{|c|}{\textbf{Category 7: Placement Strategy}} \\ \hline

\cite{S33} & \textsf{Placement Strategy} & Security Model for Volunteering Fog Services \\ \hline

\cite{S102} & \textsf{Multi-cloud security concerns} & Security groups, VPCs, and infrastructure security \\ \hline

\cite{S105} & \textsf{Multicloud environment security issue} & Multicloud security for risk diversification \\ \hline

\cite{S109} & \textsf{Edge-cloud security concerns} & Enhanced security protocols for decentralized environments \\ \hline

\cite{S118} & \textsf{Cross-technology update security risks} & Integrated security in workflows for management \\ \hline

\cite{S121} & \textsf{Resource allocation security risks} & Security measures in resource allocation framework \\ \hline
\end{longtable}

\twocolumn
\normalsize
The solutions also include using Role-Based and Attribute-Based Access Controls (RBAC and ABAC) to manage secure access effectively. Standardization efforts like Secure CSP Interoperability and OpenStack API integration aim to mitigate risks of unauthorized access and vendor lock-in. Furthermore, comprehensive frameworks like DRA emphasize secure data management, ensuring privacy, fault tolerance, and secure data transfers across different cloud platforms.

\textbf{Access and Communication Control} category reports on multiple challenges related to maintaining secure multi-cloud connectivity and managing user access control. Out of the 121 reviewed studies, 9 specifically identified issues and proposed solutions addressing secure communication, provider-agnostic operations, and decentralized identity management (see Table \ref{tab:Security-challenges-solutions} and the Security Challenges-Solution sheet in \cite{replpack}). Advanced encryption and authentication mechanisms, e.g., TLS, JWT, and OAuth 2.0, were identified as solutions to secure communications across cloud and edge environments. Secure deployment of applications using middleware solutions was also discussed, which make use of SSL/TLS and access controls to enhance security attributes. Provider-agnostic middleware, e.g., the one based on Aneka, enables runtime provider selection and thus facilitates seamless operations across multi-cloud environment. Solutions for decentralized IAM security relate to blockchain utilization for improving data integrity and confidentiality. Secure communications in multi-domain edge environments are achieved through frameworks like WireGuard, backed by strong service mesh systems that maintain connectivity and protect against unauthorized access. Security monitoring for anomaly detection and policy enforcement was also noted as important for sustaining secure operations.

\textbf{Container Security} category reports on challenges such as co-resident security, container isolation, host security, and multi-tenant serverless security in multi-cloud environment. Among the 121 reviewed studies, 5 identified these issues and proposed solutions to address them (see Table \ref{tab:Security-challenges-solutions} and the Security Challenges-Solution sheet in \cite{replpack}). The solutions identified include the use of secure placement strategies via deep reinforcement learning to manage secure container placement in multi-cloud platforms. Security measures for container isolation include the technique of Power and Leakage Management, followed by Linux kernel-level isolation using SELinux. In order to enhance security on both container and host sides, solutions such as Host-Container Cooperation and Docker-based detection tools, like DockerWatch, enhance the defenses and malicious activity detection. Finally, with a view to further securing multi-tenant serverless environments, the implementation of best practices regarding security in the CLI tool used for managing serverless deployments was underlined.

\textbf{Infrastructure and Deployment Management Security} category reports on challenges related to securing data deployment, infrastructure orchestration, storage security, and migration security in multi-cloud environment. Out of the reviewed 121 studies, 6 specifically addressed these types of challenges and proposed solutions (see Table \ref{tab:Security-challenges-solutions} and the Security Challenges-Solution sheet in \cite{replpack}). The identified solutions include the use of a Secure Private Data Deployment Strategy with encryption techniques like AES-256 and secure protocols such as SFTP and HTTPS to safeguard data. In the case of infrastructure orchestration, the focus is on the implementation of standard interfaces, such as OpenStack APIs, to ensure security and interoperability during deployments. For storage security, the solutions introduced the Multi-Cloud Storage Framework (MSSF), providing secure storage management through the implementation of RBAC and data encryption, along with periodic audits. The solutions also introduced measures for secure migration through both Infrastructure-Aware and Agnostic Secure Migration strategies, emphasizing image signing and data encryption under the least-privilege principle. Additionally, the reviewed studies proposed orchestration security for IoT distributed data and operations, as well as AWS best practices to help protect CI/CD pipelines running in multi-cloud environment.

\textbf{Security Monitoring and Breach Prevention} category reports on challenges related to system vulnerabilities, breach detection and prevention, and network policy misconfiguration risks in multi-cloud environment. Out of the 121 reviewed studies, 4 specifically addressed these issues and proposed corresponding solutions (see Table \ref{tab:Security-challenges-solutions} and the Security Challenges-Solution sheet in \cite{replpack}). The identified solutions suggest implementing automated security evaluations and bootstrapping mechanisms for Virtual Machines with security configurations using tools like Terraform, Chef, Ansible, or Puppet. The studies also highlighted several other tools for vulnerability assessment, such as OpenVAS and Nessus, which can be used to address delays in patching and reduce system vulnerabilities. In terms of breach detection and prevention, the reviewed studies emphasized the need to enhance security through metric filtering, fault recovery, and the automated deployment of SIEM, IDS, and IPS tools. Log analysis solutions were considered necessary for proactive breach management. As a result, automated network policy discovery and verification mechanisms, such as KUNERVA, were proposed, incorporating rigorous verification to prevent most potential breaches in network policy management.

\textbf{Trust and Compliance Management} category reports on solutions related to ensuring security and compliance in multi-cloud environment. Among the 121 reviewed studies, 7 specifically discussed these issues and proposed corresponding solutions (see Table \ref{tab:Security-challenges-solutions} and the Security Challenges-Solution sheet in \cite{replpack}). The identified solutions emphasize the need for security controls and standardized practices, such as risk analysis and secure coding using MUSA. The use of standardized API-based cloud interfaces, like the OpenStack API, was also emphasized to ensure both data security and interoperability between different cloud platforms. Establishing trust-based coalitions for community security was another important approach, supported by IAM systems and secure communication frameworks like SFC to enable secure interactions and collaborations within multi-cloud settings.

\textbf{Placement Strategy} category highlights the need for secure service deployment on volunteering fog nodes in multi-cloud environment. Of the 121 reviewed studies, 6 proposed solutions for these challenges (see Table \ref{tab:Security-challenges-solutions} and the Security Challenges-Solution sheet in \cite{replpack}). The solutions involve developing a security model that integrates SSL/TLS protocols for communication and certificate-based access controls to protect deployed services on fog nodes. The strategies include implementing a combination of security groups, VPCs, and infrastructure-level measures to distribute risks and prevent unauthorized access. The reviewed studies recommend adaptive methods for handling security in dynamic and decentralized edge-cloud environments by proposing cross-technology security updates and resource allocation frameworks to avoid conflicts and vulnerabilities.
\begin{tcolorbox} 
\textbf{Takeaway 8}: \textcolor{black}{Security in containerized multi-cloud environments poses unique challenges due to the combined complexity of container orchestration and distributed cloud platforms. The reviewed studies show that addressing these issues requires specific strategies—such as secure container placement, decentralized identity management, and container-aware compliance frameworks—not typically needed in single-cloud or non-containerized setups. These solutions highlight the distinct security needs that arise only when containerization is applied across multiple clouds.}
\end{tcolorbox}


\normalsize

\subsection{Automation Challenges and Solution Framework (RQ4 and RQ5)}
\label{sec:Automationframework}
\textcolor{black}{
Table~\ref{tab:Automation-solutions} presents an overview of automation challenges and corresponding solutions for containerized applications in multi-cloud environments. The framework is organized into eight categories, each addressing a distinct aspect of automation—such as orchestration, deployment, resource management, and standardization. These categories capture recurring challenges and the solution strategies proposed across the selected studies.
}


\textbf{Multi-Cloud Automation} category reports on challenges and solutions related to automating processes for combining and administering various cloud computing environments. Out of the 121 reviewed studies, 13 identified these challenges and provided associated solutions (see Table~\ref{tab:Automation-solutions} and Automation Challenges-Solutions sheet in \cite{replpack}). Key challenges involve the integration, provisioning, orchestration, and adaptation of multiple clouds. The solutions focus on utilizing proxies for environment interfaces to ease management across multiple cloud platforms and employing model-driven engineering techniques to manage provisioning and adaptation more efficiently. Dynamic on-demand fog computing was identified as an essential approach for handling multi-cloud environment dynamically. At the level of cloud redundancy and resource management driven by security policies, smart and user-friendly automation techniques, such as MSSF, were recommended. For managing the complexity of multi-cloud orchestration, serverless deployment methodologies were proposed, using models like TOSCA-based orchestration to improve scalability and coordination between different cloud stacks. Recommendations also include deploying application-level resource managers to support multi-cloud operations and adopting secure abstraction models to create uniform interfaces for managing a wide range of cloud products. MUSA Deployer tools were highlighted for their key role in automating deployment and monitoring activities, ensuring overall security compliance, and simplifying the deployment pipeline across federated clouds.

\textbf{Automation in Deployment and Scaling} category reports on challenges and solutions related to automating deployment and scaling processes for containerized applications across multi-cloud platforms. Out of the 121 reviewed studies, 12 identified these challenges and proposed corresponding solutions (see Table~\ref{tab:Automation-solutions} and the Automation Challenges-Solutions sheet in \cite{replpack}). Key challenges include intelligent container placement, application runtime switch automation, and service deployment issues. The proposed solutions emphasized intelligent algorithms for container placement, using machine learning for multi-cloud workload balancing. Runtime switch techniques are also being automated with methods such as service drivers, which are designed to be lightweight and enable seamless application migration and replication. DevOps methodologies were frequently mentioned in the literature in relation to managing deployment and scaling through continuous development and error reduction. Security during deployment is ensured through automated security measures integrated into the pipeline, aiming for a secure setup with reduced vulnerabilities. The automation strategies include elastic resource provisioning using predictive analytics to handle fluctuating workloads in real time. Optimization models were involved in solutions for virtual function placement and container deployment to enhance efficiency. This emphasis on automating and optimizing deployment strategies led to the proposal of comprehensive multi-cloud orchestration tools that facilitate the deployment of containers and VMs, aimed at improving agility and scalability within diverse cloud environments.

\textbf{Resource Management Automation} category reports on challenges and solutions related to automating the management of resources in multi-cloud environment. Out of the 121 studies reviewed, 9 specifically addressed these challenges and provided appropriate solutions (see Table~\ref{tab:Automation-solutions} and the Automation Challenges-Solutions sheet in \cite{replpack}). These challenges include resource overbooking management, resource management for containerized applications, and resource commissioning and decommissioning. 

\onecolumn
\footnotesize
\renewcommand{\arraystretch}{1.1} 

\begin{longtable}{|c|>{\columncolor{gray!50}}p{7cm}|>
{\columncolor{gray!20}}p{7cm}|}
\caption{Identified automation challenges and solutions for containerized applications in multi-cloud environment}
\label{tab:Automation-solutions} \\
\hline
\rowcolor{gray!50}
\textbf{ID} & \textbf{Challenge} & \textbf{Solution} \\
\hline
\endfirsthead

\multicolumn{3}{c}%
{{\bfseries \tablename\ \thetable{} -- continued from previous page}} \\
\hline
\rowcolor{gray!50}
\textbf{ID} & \textbf{Challenge} & \textbf{Solution} \\
\hline
\endhead

\hline \multicolumn{3}{r}{{Continued on next page}} \\
\endfoot

\hline
\endlastfoot

\multicolumn{3}{|c|}{\textbf{Category 1: Multi-Cloud Automation}} \\ \hline

\cite{S10} & Integrating multi-cloud environment & Implementing Environment Proxies \\ \hline
\cite{S20} & Automating Multi-Cloud Provisioning & Adopting Model-Driven Engineering \\ \hline
\cite{S33} & Creating On-Demand Fog in Multi-Cloud & Enabling On-Demand Fog Computing \\ \hline
\cite{S47} & Enhancing User-Friendly Automation & Using Multi-Cloud Storage Selection \\ \hline
\cite{S44} & Orchestrating Multi-Stack Environments & Deploying with TOSCA-based Approach \\ \hline
\cite{S34} & Securing Multi-Cloud Application Creation & Applying MUSA Framework \\ \hline
\cite{S37} & Unifying Resource Abstraction & Managing with Resource Managers \\ \hline
\cite{S52} & Integrating Multi-Cloud Deployments & Automating MUSA Deployment \\ \hline
\cite{S57} & Coordinating Federated Cloud Deployments & Orchestrating Subsystem Deployment \\ \hline
\cite{S58} & Setting Up Hybrid multi-cloud environment & Standardizing Interoperability \\ \hline
\cite{S80} & Ensuring Multi-Cloud Interoperability & Enabling Multi-Cloud Support \\ \hline
\cite{S81} & Managing Multi-Cloud Brokerage Systems & Managing Cloud Services \\ \hline
\cite{S84} & Implementing Multi-Cloud Orchestration & Migrating Manually to Cloud \\ \hline
\cite{S85} & Establishing DevOps in Multi-Cloud & Automating Agile Development \\ \hline
\cite{S88} & Addressing Cost and Migration Challenges & Managing Complexity \\ \hline
\cite{S102} & Overcoming Multi-Cloud Automation Barriers & Integrating Jenkins and Kubernetes \\ \hline
\cite{S105} & Improving Multi-Cloud Cost Efficiency & Optimizing Deployment Costs \\ \hline
\cite{S107} & Simplifying Multi-Cloud Deployment Complexity & Adopting AWS Deployment Strategies \\ \hline
\cite{S115} & Maintaining Multi-Cloud Applications & Automating Resource Allocation \\ \hline

\multicolumn{3}{|c|}{\textbf{Category 2: Automation in Deployment and Scaling}} \\ \hline

\cite{S27} & Automating Multi-Cloud Deployment Monitoring & Adopting Multi-Cloud Strategy \\ \hline
\cite{S06} & Optimizing Container Placement & Implementing Intelligent Placement \\ \hline
\cite{S07} & Automating Runtime Application Switching & Deploying Service Driver Model \\ \hline
\cite{S15} & Enabling Fine-Grained Component Automation & Replicating and Migrating Services \\ \hline
\cite{S28} & Streamlining DevOps Automation & Developing with DevOps Approach \\ \hline
\cite{S29} & Automating Service Deployment & Using Automated Scripts \\ \hline
\cite{S30} & Securing Deployment Time & Securing with Automation Tools \\ \hline
\cite{S31} & Automating Distributed Simulations & Adopting DevOps Methodologies \\ \hline
\cite{S32} & Improving Software Deployment Speed & Optimizing Container Images \\ \hline
\cite{S36} & Automating Container Startup & Profiling Container Execution \\ \hline
\cite{S38} & Dynamically Scaling Resources & Automating Scaling with CSP Capability \\ \hline
\cite{S24} & Scaling Container Deployment & Scaling with Kubernetes \\ \hline
\cite{S48} & Provisioning Elastic Resources & Provisioning Resources Elastically \\ \hline
\cite{S50} & Automating VF Placement & Optimizing VF Placement \\ \hline
\cite{S68} & Managing Container and VM Deployment & Orchestrating Multi-Cloud Deployments \\ \hline

\multicolumn{3}{|c|}{\textbf{Category 3: Resource Management Automation}} \\ \hline

\cite{S04} & Handling Resource Overbooking & Detecting Overbooking with ML \\ \hline
\cite{S21} & Managing Virtual Clusters & Managing with ClaaS Model \\ \hline
\cite{S39} & Scheduling Containerized Applications & Scheduling with Kubernetes \\ \hline
\cite{S41} & Selecting Optimal VM Types & Using GA-based Algorithms \\ \hline
\cite{S42} & Automating Resource Commissioning & Adopting MicroCloud Architecture \\ \hline
\cite{S43} & Supporting Cloud Federation Automation & Optimizing with Middleware Platform \\ \hline
\cite{S45} & Automating Elastic Resource Detection & Automating with JCatascopia \\ \hline
\cite{S51} & Automating Resource Acquisition & Managing Resources with Aneka \\ \hline
\cite{S74} & Enforcing SLA Compliance & Testing for Compliance \\ \hline

\multicolumn{3}{|c|}{\textbf{Category 4: Data and Application Migration Automation}} \\ \hline

\cite{S01} & Safeguarding Data Integrity in Migration & Maintaining Integrity Autonomously \\ \hline
\cite{S17} & Migrating Legacy Web Applications & Reusing Architectural Patterns \\ \hline
\cite{S69} & Porting Applications to Cloud & Using CloudSME Platform \\ \hline

\multicolumn{3}{|c|}{\textbf{Category 5: Testing and Benchmarking Automation}} \\ \hline

\cite{S09} & Automating User-Oriented Testing & Virtualizing with Lightweight Containers \\ \hline
\cite{S02} & Streamlining Benchmarking & Orchestrating Docker Benchmarking \\ \hline
\cite{S13} & Enhancing Automation Dependability & Enhancing Dependability with CI \\ \hline

\multicolumn{3}{|c|}{\textbf{Category 6: Standardization and Interoperability Challenges}} \\ \hline

\cite{S05} & Standardizing Automation Processes & Integrating Service Mesh \\ \hline
\cite{S70} & Optimizing Cloud Provider Selection & Selecting Optimal Providers \\ \hline
\cite{S16} & Unifying Microservices Deployment & Adopting Service Chain Models \\ \hline
\cite{S64} & Addressing Cloud Interoperability & Managing with Multi-Cloud Harmony \\ \hline
\cite{S63} & Standardizing Cloud Applications & Standardizing Interfaces \\ \hline
\cite{S18} & Automating Package Creation and Scaling & Automating Package Frameworks \\ \hline

\multicolumn{3}{|c|}{\textbf{Category 7: Application and Service Management}} \\ \hline

\cite{S55} & Selecting 5G Ecosystem Components & Utilizing Open-Source Components \\ \hline
\cite{S62} & Managing End-to-End Tail Latency & Modeling Performance Efficiently \\ \hline
\cite{S79} & Handling Distributed Application Complexity & Managing Distributed Applications \\ \hline

\multicolumn{3}{|c|}{\textbf{Category 8: Runtime and Service Discovery}} \\ \hline

\cite{S72} & Addressing Network Complexity & Managing Network Complexity \\ \hline
\cite{S83} & Improving Social Network Service Discovery & Automating Service Discovery \\ \hline
\cite{S08} & Administering Complex Applications & Building Autonomic Systems \\ \hline
\end{longtable}
\vspace{1em}
\noindent

\normalsize
The proposed solutions include machine learning-based resource overbooking detection for efficiently managing service containers across multiple cloud platforms. To manage containerized applications, custom Kubernetes schedulers, such as label-affinity schedulers, were proposed to optimize orchestration and allocation. Additionally, a GA-based algorithm was suggested to reduce the search space of VM types in data centers, simplifying the resource allocation problem within multi-cloud environment. 
For resource commissioning and decommissioning, proposals such as the MicroCloud architecture enabled fine-grained resource allocation and coordinated adaptation workflows. The solutions also addressed cloud federation and inter-platform portability challenges, utilizing middleware platforms for security enhancement, cost management, and performance optimization. The JCatascopia automated modular monitoring framework was introduced to monitor runtime configurations and detect elasticity actions. Finally, SLA compliance testing automation was proposed to manage resource allocation processes and enforce corrective actions, ensuring that service level agreements are met across cloud environments.

\textbf{Data and Application Migration Automation}: category reports challenges and solutions related to the automation of testing and benchmarking processes for containerized applications in multi-cloud environment. Among the 121 studies reviewed, 3 identified these challenges along with their proposed solutions (see Table~\ref{tab:Automation-solutions} and the Automation Challenges-Solutions sheet in \cite{replpack}). The main challenges include ensuring data integrity during the migration process and enabling the smooth migration of legacy Web applications to cloud services. The proposed solutions involve introducing autonomous management cycles to reduce complexity and ensure effective deployment across multi-cloud environment. Other solutions include reusable architectural patterns to accelerate the modernization of legacy applications in containerized multi-cloud environment. Additionally, the CloudSME Simulation Platform was highlighted for its ability to accelerate the porting of existing applications to cloud infrastructures, with support for more complex tasks such as billing and resource management.

\textbf{Testing and Benchmarking Automation} category highlights the challenges and solutions related to automating the testing and benchmarking processes for containerized applications in multi-cloud environment. Out of the 121 studies reviewed, 3 studies specifically addressed these challenges and provided relevant solutions (see Table~\ref{tab:Automation-solutions} and the Automation Challenges-Solutions sheet in \cite{replpack}). The main challenges revolve around automating user-oriented testing and benchmarking across various cloud platforms. The solutions range from just-in-time deployment and streamlined multicomponent prototyping to rapid testing, enabled by container-based lightweight virtualization.
Additionally, Smart Docker Benchmarking Orchestrators were proposed for automating benchmarking in multi-cloud environment, thereby enhancing the overall efficiency of testing procedures. The application of Computational Intelligence (CI) was also suggested to improve dependability in automated testing for containerized multi-cloud systems.

\textbf{Standardization and Interoperability Challenges} category reports on the challenges and solutions related to standardization and interoperability in the automation of containerized applications across multiple cloud environments. Out of the 121 reviewed studies, 6 specifically identified these challenges and proposed corresponding solutions (see Table~\ref{tab:Automation-solutions} and the Automation Challenges-Solutions sheet in \cite{replpack}). The main challenges involve standardizing automation processes, selecting cloud providers, and addressing the lack of unified templates for microservices deployment. The proposed solutions include service mesh networks and multi-cloud cluster federation to standardize the automation processes for containerized applications. AI-driven tools for optimal cloud provider selection were also introduced, which help in selecting the most suitable provider and node based on factors such as cost-effectiveness and runtime efficiency in multi-cloud environment. To address the lack of common templates for microservices, a Service Model approach was proposed to automate the deployment of microservices across multiple cloud platforms, thereby reducing operational complexity. For cloud interoperability, an AI-driven platform, such as multi-cloud Harmony, was developed to facilitate seamless data transfer and communication between disparate cloud environments. In terms of cloud application standardization, the solutions emphasized promoting the adoption of global standards and interfaces to improve interoperability between cloud service providers and middleware platforms. Finally, automation frameworks for package creation and scaling, such as those using machine-readable definition files, were proposed to efficiently manage both virtual machines and Docker containers across multi-cloud environment.

\textbf{Application and Service Management} category reports challenges and solutions related to the automation of application and service management processes in containerized multi-cloud environment.Out of the 121 reviewed studies, 2  specifically \twocolumn addressed these challenges and proposed appropriate solutions (see Table~\ref{tab:Automation-solutions} and the Automation Challenges-Solutions sheet in \cite{replpack}). The key challenges include the selection of 5G ecosystem components and end-to-end tail latency management in microservice architectures. The proposed solutions involve using open-source software components to customize the orchestration of containers and virtual machines in modern multi-cloud deployments. 
 Additionally, performance modeling approaches were suggested to evaluate the performance of container-level and VM-level data in multi-cloud environment, enabling more efficient latency management and optimized overall performance.

\textbf{Runtime and Service Discovery} category organizes the challenges and solutions related to runtime management and services discovery in multi-cloud environment, particularly in managing the complexity of network control and administration for containerized applications. Out of the 121 reviewed studies, 4 specifically tackled these challenges (see Table~\ref{tab:Automation-solutions} and the Automation Challenges-Solutions sheet in \cite{replpack}). The proposed solutions include automating network management to address the increasing complexity of distributed applications across multi-cloud topologies. Additionally, automated discovery and selection algorithms were proposed to tag and manage social network services within containerized multi-cloud systems. Autonomic computing systems were introduced to handle the growing complexity of applications by facilitating knowledge transfer, enabling more effective containerization management through automation.

\begin{tcolorbox}
\textbf{Takeaway 9}: \textcolor{black}{Automation challenges in containerized multi-cloud environments are not merely generic cloud or container issues; they arise from the complex interplay between container portability and cloud heterogeneity. Studies reveal that managing orchestration, scaling, and resource allocation across diverse cloud platforms requires specialized solutions, such as TOSCA-based orchestration, intelligent placement algorithms, and DevOps-integrated pipelines, highlighting how containerization uniquely amplifies automation demands in multi-cloud settings.}
\end{tcolorbox}

\subsection{Deployment Challenges and Solution Framework (RQ4 and RQ5)}
\label{DeploymentFramework}
Table~\ref{tab:Deployment-solutions} presents a comprehensive overview of the challenges and corresponding solutions faced during the deployment of containerized applications in multi-cloud environment. As organizations increasingly adopt containerization and multi-cloud strategies to enhance scalability, flexibility, and resource utilization, they encounter various obstacles that require effective solutions. The figure consists of nine categories, each representing a specific deployment challenge, along with the corresponding challenge description and its solution. These categories encompass a wide range of issues, including deployment complexity and Orchestration, access and communication control, security and compliance management, multi-cloud deployment coordination and Integration, monitoring and scalability challenges, microservices architecture and containerization, and network connectivity and hybrid cloud integration challenges.

\textbf{Deployment Complexity and Orchestration} category outlines various challenges and solutions related to the complexity and orchestration of container deployment in multi-cloud environment. Out of the 121 studies reviewed, 18 focused on these specific challenges and their respective solutions. Key challenges include Performance Testing Deployment, where standardized benchmarking and automation ensure consistency in performance testing. Deployment Validation is addressed by the Testing Process Management System, which enhances validation to ensure all criteria are met before deployment. Cost-Effective Deployment is achieved through Containerization and Microservice Architecture, enabling agile and efficient deployment. For Resource Management, a User Preference-Based Resource Brokering system optimizes resource allocation across clouds. Additional prominent solutions include orchestration using TOSCA, automation tools like Jenkins and Terraform, and Blockchain-based SLA Monitoring for immutable and auditable SLA tracking. Dynamic Adaptation is handled through Context-Aware Resource Allocation, while AWS Wavelength for Edge Cloud Integration addresses regional infrastructure consistency issues.

\textbf{Access and Communication Control} category organizes challenges and solutions related to maintaining secure connectivity in multi-cloud environment and managing user access control. Out of the 121 reviewed studies, 8 focused on these challenges. These studies report solutions include ensuring secure access to resources, secure communication between containers, and maintaining secure connectivity across multiple clouds. The solutions involve the development of encryption and authentication mechanisms to protect data and ensure authorized access. Secure communication protocols were recommended for interactions within containers and across cloud environments. It was also suggested that security measures should be provider-agnostic, ensuring that they are not tied to any single cloud provider.

\textbf{Security and Compliance Management in Multi-cloud Deployment} category focuses on security and compliance challenges and their solutions faced during multi-cloud container deployment. One challenge, Vendor Lock-in Prevention Deployment, aims to prevent vendor lock-in and ensure modular and loosely-coupled deployment in multi-cloud environment. The solution proposed is the Cloud Modelling Framework, which introduces a framework for specifying provisioning and deployment to enhance compatibility with existing ACSs and cloud solutions. Another challenge in this category is Security Deployment in Multi-container Environment, which addresses security and privacy concerns during multi-container deployment on the same OS kernel in multi-cloud environment. The solution is Cross-Container Isolation, which enforces cross-container isolation to enhance security.

\textbf{Multi-cloud Deployment Coordination and Integration}: This category identifies and classifies the challenges and solutions related to coordination and integration during the multi-cloud deployment phase. Of the 121 reviewed studies, 5 addressed these challenges. One such challenge is Multi-cloud PaaS Deployment, which seeks to overcome high entry barriers for deploying a PaaS infrastructure across multiple clouds. The proposed solution is Lightweight Proxies for PaaS Integration, which allows for the seamless integration of different PaaS services through lightweight proxies. One other challenge includes IoT Application Deployment, which presents issues when deploying IoT applications on multi-cloud platforms. The solution for this challenge is the DRA Framework and CI Broker for Multi-Cloud Deployment, which ensure smooth deployment and resource coordination across multiple clouds. Additionally, Data Transfer and Service Compatibility issues are managed using FaaS and storage services, ensuring that data transfer and compatibility between different cloud providers are not problematic. RBAM supports heterogeneous cloud and edge integration, offering frictionless management and operational flexibility for cloud-edge configurations. Lastly, private cloud integration is facilitated by the CLI tools that enable easy integration and the establishment of serverless environments.

\textbf{Monitoring and Scalability Challenges} category identifies and classifies various issues and solutions related to monitoring, performance, and scalability in the context of multi-cloud container deployment. Of the 121 reviewed studies, 8 specifically addressed these issues. One challenge is Legacy Code Migration Deployment, which involves managing the migration of legacy code, tenant engine separation, and composition modeling during container deployment across multiple clouds. The proposed solution is a Reusable Architectural Pattern with Docker and Cloudant, which enables the reuse of an architectural pattern using Docker for containerization and Cloudant for the persistence layer. Another challenge is Tenant Performance Deployment, which focuses on managing tenant service performance and competition during multi-cloud containerized deployments. 

A proposed solution is the Multi-tenant and Multi-instance Hybrid Deployment scheme, based on container technology, which improves performance efficiency across tenants. A critical challenge is the monitoring of deployments, where existing tools have limitations in multi-cloud environment. The solution involves implementing JCatascopia, a platform-independent and interoperable monitoring system, which enhances the monitoring of multi-cloud setups and helps mitigate these challenges. Additional challenges include Resource Contention and QoS Degradation. The solutions include Dynamic CPU Adaptation through Linux patches, along with Kubernetes and performance monitoring tools. These measures ensure optimal resource usage and maintain performance consistency across cloud environments.

\textbf{Microservices Architecture and Containerization} category classifies the challenges and solutions related to microservices architecture and containerization in multi-cloud environment. Of the 121 reviewed studies, 5 explicitly addressed challenges in this area. One such challenge involves the deployment of multilayer and multitier Web architectures in multi-cloud environment, referred to as Autonomic System Deployment. The proposed solution is a Self-Tuning Performance Model and Autonomic Management, which introduces a self-tuning performance model along with an autonomic management system to optimize deployment and management. Another challenge focuses on Microservice Deployment, which explores how microservice architectures can be deployed in multi-cloud environment. The proposed solution is a Metrics and Requirements Framework for Microservices, emphasizing the need for a structured framework to develop requirements and relevant metrics for deploying microservice-based applications. For Multi-cloud Application Configuration, the challenges involve deployment, configuration, and operation across multi-cloud environment. The one solution involves deploying a Runtime Environment with an Object Store and Artifact Repository to effectively manage configurations across different clouds. The Elastic Container Platform Dependency Deployment solution addresses the challenge of vendor lock-in and dependency on specific container platforms. The solution is the Separation of Elastic Platform and Cloud Application Definitions, which separates platform definitions from cloud application definitions to increase flexibility of application deployment. Additionally, deploying microservices in an edge environment can help resolve integration and communication issues in heterogeneous edge computing environments. The use of Docker, CI/CD Pipelines, and Service Mesh for Microservices enables the automated and secure deployment and communication of microservices.

\textbf{Network Connectivity and Hybrid Cloud Integration} category classifies the challenges and solutions related to ensuring proper network connectivity in multi-cloud deployments and hybrid cloud integration. Out of the 121 studies reviewed, 5 specifically address these issues. One challenge, Inter-edge Bandwidth Deployment, deals with interconnecting distributed localities and managing bandwidth at the edge. The solution is a Modular Edge Cloud Computing Architecture, using containerization for better bandwidth management. A key challenge is Network Latency Deployment, which focuses on minimizing latency during multi-cloud deployments. The proposed solution is VM Selection for Composite Applications, optimizing virtual machine selection to reduce latency. Latency Reduction Deployment involves improving end-to-end performance through an Edge-Cellular Hybrid Infrastructure, combining edge and cellular networks. To address Vendor Lock-in, an Interoperability Layer Above Cloud Infrastructure enables smoother transitions between cloud providers. 

\clearpage
\onecolumn
\footnotesize
\renewcommand{\arraystretch}{1.2}
\setlength{\tabcolsep}{18pt} 

\begin{longtable}{|c|>{\columncolor{gray!50}}p{6cm}|>{\columncolor{gray!20}}p{6cm}|}
\caption{Identified deployment challenges and solutions for containerized applications in multi-cloud environment}
\label{tab:Deployment-solutions} \\
\hline
\rowcolor{gray!50}
\textbf{ID} & \textbf{Challenge} & \textbf{Solution} \\
\hline
\endfirsthead

\multicolumn{3}{c}{{\bfseries \tablename\ \thetable{} -- continued from previous page}} \\
\hline
\rowcolor{gray!50}
\textbf{ID} & \textbf{Challenge} & \textbf{Solution} \\
\hline
\endhead

\hline \multicolumn{3}{r}{{Continued on next page}} \\
\endfoot

\hline
\endlastfoot

\multicolumn{3}{|c|}{\textbf{Category 1: Deployment Complexity and Orchestration Challenges}} \\ \hline

\cite{S01} & Performance Testing Deployment & Standardized Benchmarking and Automation \\ \hline
\cite{S09} & Deployment Validation & Testing Process Management System \\ \hline
\cite{S02} & Cost-effective Deployment & Containerization and Microservice Architecture \\ \hline
\cite{S38} & Resource Management Deployment & User Preference-Based Multi-Cloud Resource Brokering \\ \hline
\cite{S50} & NFV Site Deployment & Site Selection and VF Allocation Algorithms \\ \hline
\cite{S79} & Cloud Management Platform Evaluation & Standardized Output Formats and Evaluation Criteria \\ \hline
\cite{S37} & Standardized Requirement Deployment & Application-Level Resource Managers for Multi-Cloud \\ \hline
\cite{S42} & Heterogeneous Node Deployment & Technology-Independent Multi-Level Adaptation \\ \hline
\cite{S52} & Deployment Plan Generation & Cloud Services Selection and Deployment Planning \\ \hline
\cite{S57} & TOSCA-based Deployment & Brokered Multi-Cloud Deployment with TOSCA \\ \hline
\cite{S58} & Standardization Deployment & Proxy Cloud Virtualization for OpenStack \\ \hline
\cite{S87} & Criticality Constraints in Kubernetes Deployment & Kubernetes for Critical Deployments \\ \hline
\cite{S88} & Automation Tools Deployment Issue & Accurate Cost and Resource Tracking \\ \hline
\cite{S102} & Multi-cloud Deployment Challenge & Jenkins for Multi-cloud Automation \\ \hline
\cite{S103} & Decentralized App Deployment Issue & CODECO for Edge-Cloud Deployment \\ \hline
\cite{S106} & SLA Monitoring Deployment Complexity & Blockchain for SLA Data Integrity \\ \hline
\cite{S107} & Container and Serverless Architecture Deployment Issue & AWS ECS and Lambda Deployment \\ \hline
\cite{S108} & Heterogeneous Environment Deployment Challenge & Edge-cloud Orchestration Management \\ \hline
\cite{S109} & Dynamic Adaptation Deployment Issue & Context-aware Resource Allocation Strategies \\ \hline
\cite{S113} & Regional Infrastructure Deployment Consistency Issue & AWS Wavelength for Edge Cloud Integration \\ \hline
\cite{S114} & Robotic App Deployment Challenge & Hybrid Edge-Cloud Architecture \\ \hline
\cite{S117} & Multi-cloud Deployment Consistency Problem & Multi-cloud Deployment Efficiency Models \\ \hline
\cite{S118} & Cross-Technology Deployment Integrity Issue & Workflow Automation for Cross-Technology Deployment \\ \hline
\cite{S119} & Industrial Robot Control System Deployment Problem & Seamless Control between Cloud and Robots \\ \hline

\multicolumn{3}{|c|}{\textbf{Category 2: Security and Compliance Management in Multi-Cloud Deployment}} \\ \hline

\cite{S20} & Vendor Lock-in Prevention Deployment & Cloud Modelling Framework \\ \hline
\cite{S25} & Security Deployment in Multi-container Environment & Cross-Container Isolation \\ \hline
\cite{S34} & Security Deployment & Security Control Specification and Deployment Framework \\ \hline
\cite{S72} & Virtual Network Security Deployment & Network Virtualization Platform for Multi-Cloud \\ \hline
\cite{S83} & Malicious Service Management Deployment & Collusion-Resilient Trust Aggregation Technique \\ \hline
\cite{S90} & Secure Personal Data Deployment Challenge & Docker, Kubernetes, Blockchain for GDPR Compliance \\ \hline
\cite{S110} & IAM Deployment across Cloud Platforms & Blockchain for Decentralized IAM Deployment \\ \hline
\cite{S120} & Non-intrusive Malware Detection Deployment & DockerWatch for Non-intrusive Container Integration \\ \hline

\multicolumn{3}{|c|}{\textbf{Category 3: Multi-Cloud Deployment Coordination and Integration}} \\ \hline

\cite{S10} & Multi-cloud PaaS Deployment & Lightweight Proxies for PaaS Integration \\ \hline
\cite{S85} & IoT Application Deployment & DRA Framework and CI Broker for Multi-Cloud Deployment \\ \hline
\cite{S105} & Data Transfer and Service Compatibility & FaaS and Storage for Multi-Cloud Deployment \\ \hline
\cite{S111} & Heterogeneous Cloud and Edge Integration & RBAM for Cloud-Edge Integration \\ \hline
\cite{S112} & Private Cloud Integration Problem & CLI Tool for Serverless Setup \\ \hline

\multicolumn{3}{|c|}{\textbf{Category 4: Monitoring, Performance, and Scalability Challenges}} \\ \hline

\cite{S17} & Legacy Code Migration Deployment & Reusable Architectural Pattern with Docker and Cloudant \\ \hline
\cite{S23} & Tenant Performance Deployment & Multi-Tenant and Multi-Instance Hybrid Deployment \\ \hline
\cite{S45} & Monitoring Deployment & Platform-Independent Monitoring System \\ \hline
\cite{S89} & Resource Contention and Performance Problem & Linux Patch for Dynamic CPU Adaptation \\ \hline
\cite{S91} & QoS Degradation Detection Problem & Kubernetes with Performance Monitoring Tools \\ \hline
\cite{S94} & Monitoring Integration in Hybrid Cloud Issue & Custom Scaling with Kubernetes \\ \hline
\cite{S115} & AWS Deployment Performance Consistency & AWS Model Comparison for Performance \\ \hline
\cite{S116} & Cloud-Edge Deployment Performance Issue & AI-driven Orchestration for Dynamic Resource Management \\ \hline

\multicolumn{3}{|c|}{\textbf{Category 5: Microservices Architecture and Containerization Challenges}} \\ \hline

\cite{S28} & Autonomic System Deployment & Self-Tuning Performance Model and Autonomic Management \\ \hline
\cite{S46} & Microservice Deployment & Metrics and Requirements Framework for Microservices \\ \hline
\cite{S76} & Multi-cloud Application Configuration Deployment & Runtime Environment with Object Store and Artifact Repository \\ \hline
\cite{S80} & Elastic Container Platform Dependency Deployment & Separation of Elastic Platform and Cloud Application \\ \hline
\cite{S99} & Microservices Deployment in Edge Environments & Docker, CI/CD, Service Mesh for Microservices \\ \hline

\multicolumn{3}{|c|}{\textbf{Category 6: Network Connectivity and Hybrid Cloud Integration}} \\ \hline

\cite{S35} & Inter-edge Bandwidth Deployment & Modular Edge Cloud Computing Architecture \\ \hline
\cite{S41} & Network Latency Deployment & VM Selection for Composite Applications \\ \hline
\cite{S43} & Latency Reduction Deployment & Edge-Cellular Hybrid Infrastructure Provisioning \\ \hline
\cite{S63} & Vendor Lock-in Prevention Deployment & Interoperability Layer Above Cloud Infrastructure \\ \hline
\cite{S104} & Dynamic Network Policy Deployment Problem & KUNERVA for Network Policy Automation \\ \hline

\multicolumn{3}{|c|}{\textbf{Category 7: Cloud Deployment Constraints and Challenges}} \\ \hline

\cite{S08} & ACS Compatibility Deployment & Scalable Multi-Cloud Deployment Framework \\ \hline
\cite{S30} & Manual Intervention Risk Deployment & Automated Security Measures at Deployment \\ \hline
\cite{S44} & Multi-provider Deployment & TOSCA-Based Deployment Modeling Approach \\ \hline
\cite{S81} & Cloud Provider Transition Deployment & Semantic Interoperability in Multi-Clouds \\ \hline
\cite{S88} & Automation Tools Deployment Issue & Accurate Cost and Resource Tracking \\ \hline
\cite{S113} & Regional Infrastructure Deployment Consistency Issue & AWS Wavelength for Edge Cloud Integration \\ \hline

\multicolumn{3}{|c|}{\textbf{Category 8: Infrastructure Provisioning and Container Deployment Challenges}} \\ \hline

\cite{S32} & Fog Computing Deployment & Reorganized Container Images and Docker Deployment \\ \hline
\cite{S33} & Fog Device Deployment & On-Demand Fog and Microservices Deployment \\ \hline
\cite{S36} & Fog Computing Container Deployment & FogDocker for Container Deployment \\ \hline
\cite{S39} & Distributed Compute Node Deployment & Label-Based Scheduling Strategy \\ \hline
\cite{S40} & Kernel Security Deployment & Enhanced Container Security \\ \hline
\cite{S87} & Criticality Constraints in Kubernetes Deployment & Kubernetes for Critical Deployments \\ \hline
\cite{S98} & Power Tool Compatibility Problem & Docker, Kubernetes for Power Monitoring Integration \\ \hline
\cite{S105} & Data Transfer and Service Compatibility & FaaS and Storage for Multi-Cloud Deployment \\ \hline

\multicolumn{3}{|c|}{\textbf{Category 9: Application Deployment Workflow and Orchestration Challenges}} \\ \hline

\cite{S27} & API Management Deployment & DevOps Approach for Multi-Cloud Applications \\ \hline
\cite{S29} & Web Service Deployment & Kubernetes-Based Containerized Deployment \\ \hline
\cite{S51} & Complex Application Deployment & Aneka Platform for Distributed Applications \\ \hline
\cite{S53} & High-availability Deployment & Middleware Support for High Availability in Multi-Cloud PaaS \\ \hline
\cite{S69} & Complex Workflow Deployment & CloudSME Simulation Platform \\ \hline
\cite{S84} & Business Process Provisioning Deployment & Multi-Cloud Service Orchestration Frameworks \\ \hline
\cite{S114} & Robotic App Deployment Challenge & Hybrid Edge-Cloud Architecture \\ \hline
\cite{S109} & Dynamic Adaptation Deployment Issue & Context-aware Resource Allocation Strategies \\ \hline
\end{longtable}

\normalsize
Finally, Dynamic Network Policy Deployment tackles the challenge of adapting network policies to container workloads, with the solution being Automation of Network Policies using tools like KUNERVA.
\twocolumn

\textbf{Cloud Deployment Constraints and Challenges} category identifies constraints and challenges in multi-cloud deployment, with 6 out of 121 studies addressing these issues. One challenge, ACS Compatibility Deployment, involves ensuring container compatibility with existing ACS systems, and the solution is a Scalable Multi-Cloud Deployment Framework. Other challenges include Manual Intervention Risk Deployment, which addresses risks from manual intervention. The solution is Automated Security Measures at Deployment to reduce vulnerabilities. The challenge of Multi-Provider Deployment involves managing multiple cloud providers. The proposed solution is the TOSCA-Based Deployment Modeling Approach, which provides a unified framework. The challenge of Cloud Provider Transition Deployment lies in addressing feature incompatibilities. The proposed solution is Semantic Interoperability in Multi-Clouds, which ensures smooth transitions between providers. Automation Tools Deployment challenge is solved with tools like Jenkins and Terraform for accurate cost and resource management.

\textbf{Infrastructure Provisioning and Container Deployment Challenges} category classifies the challenges and solutions associated with infrastructure provisioning and container deployment in multi-cloud settings. Of the 121 reviewed studies, 8 focused on these challenges. One such challenge is Fog Computing Deployment, which deals with issues related to fog computing hardware and container deployment time. The proposed solution is Reorganized Container Images and Docker Deployment, suggesting that container images should be reorganized, and the Docker deployment process is modified to handle fog computing more effectively, enabling better resource management and optimization. Another challenge is Fog Device Deployment, which focuses on managing fog device availability during multi-cloud container deployments. The proposed solution is On-Demand Fog and Microservices Deployment, allowing for the on-demand creation of fog nodes and the on-demand deployment of microservices via Docker and Kubeadm. Distributed Compute Node Deployment involves challenges related to containerized application management in distributed compute nodes. Kernel Security Deployment addresses the security risks of kernel sharing and isolation. The proposed solution includes a Label-Based Scheduling Strategy and Enhanced Container Security for secure and efficient deployment.

\textbf{Application Deployment Workflow and Orchestration Challenges} category classifies the challenges and solutions related to application deployment workflows and orchestration in multi-cloud environment. Of the 121 reviewed studies, 7 focused on these challenges. One such challenge, API Management Deployment, managing the APIs and interfaces provided by different cloud providers in multi-cloud container deployments. The proposed solution is a DevOps Approach for Multi-Cloud Applications, which suggests using DevOps to manage application lifecycles in multi-cloud environment. Another challenge, Web Service Deployment, involves ensuring proper and efficient deployment in heterogeneous computing environments. The solution is Kubernetes-Based Containerized Deployment, which uses Kubernetes to efficiently manage and deploy containerized environments. Complex Application Deployment and High-availability Deployment focus on deploying complex applications while ensuring high availability. The solutions include the Aneka Platform for Distributed Applications and Multi-cloud PaaS Middleware Support for High Availability, which ensure the availability and distribution of applications across multi-cloud environment.
\begin{tcolorbox} 
\textbf{Takeaway 10}: \textcolor{black}{Reviewed studies show that deployment challenges in containerized multi-cloud environments emerge from the tight coupling of container orchestration with heterogeneous cloud infrastructures. Solutions such as TOSCA-based modeling, cross-container isolation, and context-aware adaptation specifically address complexities introduced when containers span multiple clouds, highlighting the unique integration, security, and coordination demands identified through this SMS.}
\end{tcolorbox}



\normalsize

\subsection{Monitoring Challenges and Solution Framework (RQ4 and RQ5)} 
\label{sec:monitoring}
\textcolor{black}{
Table~\ref{tab:monitoring-solutions} presents a categorized overview of monitoring challenges and corresponding solutions for containerized applications in multi-cloud environments. The framework captures a range of challenges related to performance variability, resource utilization, observability, and integration. Each of the six categories highlights a specific aspect of monitoring and summarizes relevant solutions proposed in the literature.
}

\textbf{Performance, Consistency, and Variability Monitoring} category identifies the challenges and solutions that ensure consistent performance monitoring in dynamically changing multi-cloud environment. Nine of the reviewed studies addressed these challenges. Performance Monitoring focuses on maintaining consistent performance tracking of applications. The solution, Performance Variation Monitoring, uses benchmarking systems to track and analyze runtime changes in performance, ensuring optimized performance throughout. Another challenge is Benchmarking for Performance and Cost Efficiency, which balances performance against cost. The solution, SBDO Optimization, optimizes resource utilization in real time to maximize efficiency and minimize costs. A related challenge, Website Performance Monitoring, is managed through Kubernetes Monitoring, which ensures uptime and performance across widely distributed cloud environments. Monitoring User Experience is handled by collecting CSP QoS Metrics, allowing cloud service providers to gather Quality of Service data, anticipate variations in performance, and manage the overall user experience effectively. Lastly, low-latency performance optimization in multi-cloud systems is addressed by Tail Latency Prediction, which uses predictive models to minimize network delays and ensure smooth application performance.

\textbf{Resource and Infrastructure Monitoring} catagory classifies the challenges and corresponding solutions related to monitoring resource and infrastructure management for containerized applications in multi-cloud environment. Of the reviewed studies, 8 specifically addressed these challenges. One key challenge is Network Communication Monitoring, which involves managing service-to-service communication across cloud networks. The proposed solution is the implementation of a Service Mesh, which enhances visibility and security for network services. Another challenge is System Image and Configuration Monitoring, which tracks software updates, configuration changes, and performance across platforms. The solution proposed is Docker Server Deployment, which allows for easy updates to the containerized system. For Vcluster and Framework Management Monitoring, the solution is the implementation of LXC (Linux Containers), which manages application isolation and performance using lightweight containers. Multi-Cloud Resource Utilization Monitoring involves tracking the performance of resources, such as compute, storage, and network, used across multiple clouds. The solution uses Prometheus and Grafana for monitoring resource usage and availability, with an emphasis on optimization. Similarly, Dynamic Resource Management Monitoring solutions, such as the Aneka Platform, dynamically add or release resources based on demand. The challenge of cost tracking in resource utilization across multiple cloud environments, such as AWS and Azure, is addressed by the COSTA system, which monitors both resource utilization and cost efficiency. Finally, Prometheus is used for Real-Time Dynamic Resource Monitoring, allowing real-time data collection and analysis to optimize performance based on resource usage fluctuations.

\textbf{Application Optimization and Adaptability Monitoring}: This category identifies and discusses challenges related to optimizing and adapting applications in multi-cloud environment. Out of the 121 reviewed studies, 8 specifically addressed these challenges with corresponding solutions. Key challenges include Real-Time Application Monitoring, which focuses on taking immediate corrective actions using the DRA Framework, and Microservice Optimization Monitoring, which enhances monitoring across multiple clouds through MiCADO Optimization. Other challenges include Partial Download Execution Monitoring, which ensures the correct execution of partially downloaded files. The proposed solution for runtime accuracy is the FogDocker Implementation. Another challenge, Event-Driven Application Behavior Monitoring, involves monitoring event-driven application behavior. The TOSCA Event Modeling solution efficiently handles application events. For Adaptive System Design and State Monitoring, continuous design and dynamic system adaptability are managed using the Cloud Modelling Framework. Additionally, the challenge of Container Application Adaptability Monitoring is addressed through advanced orchestration techniques, which provide real-time views of distributed applications. The solution involves using AWS tools to ensure high performance for different deployment models, addressing challenges in Serverless and Containerized Monitoring. AWS tools are recommended to ensure high performance for different deployment models, addressing challenges in Serverless and Containerized Monitoring.

\textbf{System Complexity and Standardization Monitoring} cateogy identifies and addresses the challenges of managing system complexity and standardization in monitoring solutions across multi-cloud environment. Of the 121 studies reviewed, 4 specifically focused on these challenges and proposed corresponding solutions. A key challenge is the standardization of proprietary monitoring solutions, for which the proposed solution is Pervasive Monitoring. Pervasive Monitoring focuses on implementing pervasive monitoring techniques for containerized applications within multi-cloud PaaS platforms. Another significant challenge is Kernel Resource Isolation Mechanism Monitoring, with the proposed solution being Leakage Defense. This solution addresses in-container leakage channels by implementing a two-stage isolation mechanism in Linux-based systems. Additionally, the challenge of Multi-Cloud Application Monitoring Complexity is addressed by the DECIDE DevOps Expansion, which extends the framework for analyzing multi-cloud containerized applications. The Autonomic Management Framework further enhances monitoring through Monitoring Agent Coordination, improving the coordination of monitoring agents using a performance model and adaptive actions for better system behavior within containerized applications.

\textbf{Multi-cloud Coordination and Integration Monitoring} category identifies and addresses the challenges related to coordinating and integrating monitoring data across multiple cloud environments. Out of the 121 studies reviewed, 5 focused on these challenges. One key challenge is Multi-cloud Monitoring Improvement, and the proposed solution is the use of Self-Healing Techniques to autonomously detect and correct issues within cloud infrastructures. Another challenge is Cross-Level Monitoring, which is addressed through Business Process Monitoring Integration to ensure that business processes are effectively monitored across multiple cloud platforms. Lastly, Integrated Multi-cloud Management Monitoring requires advanced solutions, such as Integration with Gnocchi, to efficiently gather and monitor metrics from diverse cloud resources.

\textbf{Security, Compliance, and Trust Monitoring}: category addresses the challenges and solutions related to monitoring security, compliance, and trust management for containerized applications in multi-cloud environment. Out of the 121 reviewed studies, 9 specifically deal with these concerns. One challenge is Provider Host Information Access Monitoring, where limitations in accessing host data are overcome using ML Monitoring, which employs machine learning to track system performance and usage patterns.

\clearpage
\onecolumn
\footnotesize
\renewcommand{\arraystretch}{1.2}
\setlength{\tabcolsep}{18pt} 

\begin{longtable}{|c|>{\columncolor{gray!50}}p{6.5cm}|>{\columncolor{gray!20}}p{7cm}|}
\caption{Identified monitoring challenges and solutions for containerized applications in multi-cloud environment}
\label{tab:monitoring-solutions} \\
\hline
\rowcolor{gray!50}
\textbf{ID} & \textbf{Challenge} & \textbf{Solution} \\
\hline
\endfirsthead

\multicolumn{3}{c}{{\bfseries \tablename\ \thetable{} -- continued from previous page}} \\
\hline
\rowcolor{gray!50}
\textbf{ID} & \textbf{Challenge} & \textbf{Solution} \\
\hline
\endhead

\hline \multicolumn{3}{r}{{Continued on next page}} \\
\endfoot

\hline
\endlastfoot

\multicolumn{3}{|c|}{\textbf{Category 1: Performance Monitoring Consistency and Variability}} \\ \hline

\cite{S01}	& Performance Monitoring & Performance Variation Monitoring \\ \hline
\cite{S02}	& Benchmarking Performance Variation and Cost Efficiency Monitoring & SDBO Optimization \\ \hline
\cite{S29}	& Website Performance and Availability Monitoring & Kubernetes Monitoring \\ \hline
\cite{S38}	& User Experience Variability Monitoring & CSP QoS Metrics Collection \\ \hline
\cite{S43}	& Independent Performance Metrics Monitoring & Performance Evaluations \\ \hline
\cite{S50}	& Comprehensive Network Performance Measurement Monitoring & Server Selection Scheme \\ \hline
\cite{S62}	& Microservice Performance Degradation Monitoring & Tail Latency Prediction \\ \hline
\cite{S89}	& Container Performance Metrics Tracking & CLM and OSLM for Resource Adjustments \\ \hline
\cite{S91}	& Container Performance and Resource Usage Monitoring & Metric and Container Monitoring Tools \\ \hline
\cite{S99}	& Microservice Performance Monitoring Challenge & Prometheus, Grafana, Kubernetes for Monitoring \\ \hline
\cite{S105}	& Cross-Cloud Service Performance Monitoring & FaaS and Storage Services for AI Monitoring \\ \hline
\cite{S111}	& Stream Processing Performance Monitoring & RBAM for Real-Time Performance Insights \\ \hline
\cite{S115}	& AWS Application Performance Monitoring & AWS CloudWatch for Performance Monitoring \\ \hline
\cite{S117}	& Multi-Cloud Performance Visibility Issue & Monitoring Tools for Multi-Cloud Visibility \\ \hline
\cite{S118}	& Cross-Technology Performance Monitoring Issue & TOSCA-based Workflows for Monitoring Applications \\ \hline
\cite{S119}	& Industrial Robot System Monitoring Challenge & Real-Time System Performance Tracking Tools \\ \hline

\multicolumn{3}{|c|}{\textbf{Category 2: Resource and Infrastructure Management Monitoring}} \\ \hline

\cite{S05}	& Network Communication Monitoring & Service Mesh Implementation \\ \hline
\cite{S09}	& System Image and Configuration Monitoring & Docker Server Deployment \\ \hline
\cite{S21}	& Vcluster and Framework Management Monitoring & LXC Implementation \\ \hline
\cite{S24}	& Multi-Cloud Resource Utilization Monitoring & Prometheus and Grafana Monitoring \\ \hline
\cite{S51}	& Dynamic Resource Management Monitoring & Aneka Platform \\ \hline
\cite{S88}	& Cost and Resource Usage Monitoring & COSTA for Cost and Resource Monitoring \\ \hline
\cite{S92}	& Real-Time Dynamic Resource Monitoring & Prometheus for Real-Time Monitoring \\ \hline
\cite{S93}	& Resource Usage and Performance Issue & Prometheus for Real-Time Data Collection \\ \hline
\cite{S94}	& Real-Time Container Monitoring & Prometheus with Forecasting for Resource Management \\ \hline
\cite{S113}	& Distributed Cloud Resource Monitoring Issue & AWS CloudWatch and Azure Monitor for Insights \\ \hline
\cite{S121}	& Resource Allocation and Energy Monitoring Issue & Tools for Resource, Energy, and SLA Monitoring \\ \hline

\multicolumn{3}{|c|}{\textbf{Category 3: Security, Compliance, and Trust Management Monitoring}} \\ \hline

\cite{S04}	& Provider Host Information Access Monitoring & ML Monitoring \\ \hline
\cite{S34}	& Continuous Security Control Monitoring & MUSA Platform \\ \hline
\cite{S52}	& Multi-Cloud Security Compliance Monitoring & MUSA Security Compliance \\ \hline
\cite{S53}	& Service Continuity and Data Privacy Monitoring & NoMISHAP Implementation \\ \hline
\cite{S83}	& Trust Relationship and Network Contact Monitoring & Trust Bootstrapping \\ \hline
\cite{S110}	& Access and Authentication Monitoring Issue & Blockchain for Secure Access Event Logging \\ \hline
\cite{S120}	& Malware Behavior and System Monitoring & DockerWatch for Container Behavior Monitoring \\ \hline
\cite{S106}	& Log Data Integrity Monitoring Issue & Blockchain-based Logs for SLA Monitoring \\ \hline

\multicolumn{3}{|c|}{\textbf{Category 4: Application Optimization and Adaptability Monitoring}} \\ \hline

\cite{S20}	& Adaptive System Design and State Monitoring & Cloud Modelling Framework \\ \hline
\cite{S36}	& Partial Download Execution Monitoring & FogDocker Implementation \\ \hline
\cite{S44}	& Event-Driven Application Behavior Monitoring & TOSCA Event Modeling \\ \hline
\cite{S68}	& Microservice Optimization Monitoring & MiCADO Optimization \\ \hline
\cite{S85}	& Real-Time Application Monitoring & DRA Framework \\ \hline
\cite{S109}	& Container Application Adaptability Monitoring & Advanced Monitoring with Orchestration Integration \\ \hline
\cite{S107}	& Serverless and Containerized Monitoring Challenge & AWS Tools for Performance Monitoring \\ \hline

\multicolumn{3}{|c|}{\textbf{Category 5: System Complexity and Standardization Monitoring}} \\ \hline

\cite{S10}	& Proprietary Monitoring Solution Standardization & Pervasive Monitoring \\ \hline
\cite{S25}	& Kernel Resource Isolation Mechanism Monitoring & Leakage Defense \\ \hline
\cite{S27}	& Multi-Cloud Application Monitoring Complexity & DECIDE DevOps Expansion \\ \hline
\cite{S28}	& Monitoring Agent Coordination & Autonomic Management \\ \hline
\cite{S33}	& Heterogeneous Fog Resource Management Monitoring & Kubernetes Utility Architecture \\ \hline
\cite{S42}	& Application Topology and Dependency Monitoring & MicroCloud TOSCA Library \\ \hline
\cite{S45}	& Elastic Cloud Service Monitoring Research & JCatascopia Monitoring \\ \hline
\cite{S46}	& Microservice Monitoring Diversity & QoM Definition \\ \hline
\cite{S49}	& Compute and Network Resource Utilization Monitoring & Time Series Resource Utilization \\ \hline
\cite{S97}	& Real-Time Data Operations Monitoring & Smart Contracts for GDPR Monitoring \\ \hline
\cite{S98}	& Power Consumption Monitoring Issue & Prometheus with Power Monitoring Integration \\ \hline
\cite{S112}	& Serverless Function Monitoring Complexity & CLI Tool for Function Monitoring \\ \hline

\multicolumn{3}{|c|}{\textbf{Category 6: Multi-cloud Coordination and Integration Monitoring}} \\ \hline

\cite{S58}	& Integrated Multi-Cloud Management Monitoring & Gnocchi Integration \\ \hline
\cite{S74}	& Multi-Cloud Monitoring Challenges & Self-Healing Techniques \\ \hline
\cite{S84}	& Cross-Level Monitoring Improvement & Business Process Monitoring \\ \hline
\cite{S104}	& Real-Time Network Policy Monitoring Challenge & KUNERVA for Real-Time Network Policy Monitoring \\ \hline
\cite{S103}	& Decentralized System Observability Issue & CODECO for Orchestration and Data Observability \\ \hline
\end{longtable}

\normalsize

 Another challenge, Continuous Security Control Monitoring, is addressed by the MUSA platform, which enforces and continuously monitors security in real time for containerized applications. Multi-Cloud Security Compliance Monitoring ensures continuous security compliance across multi-cloud environment. This is achieved through the MUSA Security Compliance Solution, a platform for continuous compliance management. The challenges of maintaining data privacy, fault tolerance, and service continuity in multi-cloud environments are addressed under Service Continuity and Data Privacy Monitoring. The NoMISHAP Implementation is proposed to handle these monitoring issues. Additionally, Trust Bootstrapping establishes trust through a chain of endorsement and decision tree classification. Access and Authentication Monitoring uses blockchain technology to enable secure and transparent logging of authentication activities. For Malware Behavior and System Monitoring, DockerWatch monitors container behavior with minimal performance overhead.

\begin{tcolorbox} 
\textbf{Takeaway 11}: \textcolor{black}{Reviewed studies reveal that monitoring containerized applications in multi-cloud environments introduces unique challenges, such as inconsistent performance metrics across cloud providers, real-time trust and security event tracking, and fragmented observability due to container orchestration spanning heterogeneous infrastructures. These challenges are addressed through specialized solutions like cross-cloud Prometheus-Grafana stacks, blockchain-based SLA logging, and service mesh integrations, reflecting the need for tightly coupled monitoring mechanisms tailored specifically to the hybrid nature of containerized multi-cloud deployments.}
\end{tcolorbox}

\subsection{Tools and Frameworks (RQ6)}
\label{sec:toolsFramework}
In response to RQ6 regarding tools and frameworks for developing containerized multi-cloud applications, we identified 87 distinct tools and frameworks, classified into 5 subcategories (see Table \ref{tab:ToolsandFrameworks} and the Tools and Frameworks sheet in \cite{replpack}). Below, we provide an overview of each category, along with key tools and frameworks, highlighting their functionality, adoption, and distinctive features.

\textbf{Container Orchestration and Management Tools}: are essential for automating the deployment, scaling, and management of containerized applications across distributed and multi-cloud environment. These tools assist developers in managing large-scale container deployments and efficiently allocating resources across different cloud providers. Among the most popular orchestration and management tools, \textit{Kubernetes} is widely adopted for orchestrating containers at scale, providing high availability, and facilitating load balancing across nodes. \textit{Docker} and its variants, such as \textit{Docker Swarm}, offer more convenient container orchestration but are less scalable than \textit{Kubernetes}. \textit{Mesos} is another orchestration tool that supports not only containers but also other workloads, such as big data processing tasks. Tools like \textit{LXC} and \textit{runC} provide the runtime environment to run a single container with fine-grain control. Overall, this category covers key orchestration and container management tools that address diverse deployment needs in multi-cloud environment.

\textbf{Multi-cloud and Hybrid Cloud Platforms} are designed to support organizations in deploying and managing applications across different cloud providers smoothly. In this category, we identify and organize tools like \textit{Cloudify} and \textit{CMP2}, which offer end-to-end multi-cloud orchestration capabilities, along with platforms such as \textit{Cloudcheckr} and \textit{MistIO}, which provide monitoring and cost management across various cloud services. Similarly, \textit{AWS SDK} and the \textit{CloudSME Simulation Platform (CSSP)} are widely adopted for interacting with cloud APIs and simulating cloud environments, respectively. Moreover, hybrid platforms like \textit{OpenStack} and \textit{VMware vSphere} offer a blend of public and private cloud deployment options, which are essential for organizations pursuing a hybrid cloud strategy.  This category highlights the tools that enable integration, scalability, and cost management for applications running across multiple clouds.

\textbf{Networking, Security, and Access Management Tools}: This category represents the tools that help to minimize security risks and ensure connectivity across containerized workloads in multi-cloud systems.   \twocolumn  Advanced networking tools like Open vSwitch (OvS) allow developers to manage virtual networks across multiple cloud platforms as multi-cloud deployments grow more complicated. By implementing stringent access controls and sandboxing, security tools like Seccomp-BPF and Linux Security Modules (LSM) offer vital container protection. Proper authentication and authorization for cloud users and services are ensured by identity management solutions like Host Identity Protocol (HIP) and access control policies.

\textbf{Monitoring, Management, and Automation Tools}: Tools in this category help manage large-scale cloud applications in an easy and automated way. They make sure everything runs smoothly by constantly checking and improving how resources are used. For example, \textit{Prometheus} and \textit{Grafana} help system administrators by showing important information like CPU use, memory load, and network traffic in clear charts and dashboards. Tools like \textit{Terraform} and \textit{Puppet} make it possible to write and reuse code to set up servers and other infrastructure automatically across different cloud services. This saves time and reduces errors. In big systems using OpenStack, tools such as \textit{Ceilometer} and \textit{Gnocchi} collect data about how the system is working, which helps with tracking performance and solving issues.

\textbf{Development, APIs, and Cloud Storage Solutions}: This category organizes the main tools that support software development, API integration, and data storage in containerized multi-cloud environment. The tools include \textit{REST API} and \textit{YAML}, which enable services and applications deployed across multiple cloud providers to communicate for data sharing and inter-service interactions. General programming languages such as \textit{Java}, \textit{Go}, and \textit{Python} are commonly used to develop applications based on cloud-native architecture, alongside tools for version control, such as \textit{GitHub}, and project management using \textit{Maven}. On the data storage side, tools like \textit{gcloud storage} and \textit{azblob} provide scalable and fault-tolerant systems for managing large datasets in cloud environments.

\begin{table*}[htbp]
\footnotesize
\caption{Tools and frameworks for containerized applications in multi-cloud environment}\label{tab:ToolsandFrameworks}
\label{sec:toolsFramework}
\centering
\begin{tabular}{|p{5cm}|p{10cm}|}
\hline
\textbf{Category} & \textbf{Tool, Framework, and Language} \\ 
\hline
\textbf{Container Orchestration and Management} & Kubernetes \cite{S18, S22, S29, S55, S56, S59, S62, S63, S80, S87, S89, S90, S91, S92, S93, S94, S95, S96, S97, S98, S99, S100, S101, S102, S103, S104, S109, S111, S112, S113, S114, S115, S116, S117, S118, S119, S120, S121}, Docker \cite{S08, S13, S16, S17, S18, S25, S27, S28, S29, S30, S31, S32, S35, S55, S59, S62, S74, S77, S79, S80}, Docker Swarm \cite{S18, S22, S39, S59, S80}, Mesos \cite{S21, S22, S59, S80}, LXC \cite{S24, S25, S29}, OverlayFS \cite{S32}, runC \cite{S40}, Container Technology \cite{S23}, Docklet \cite{S21} \\ 
\hline
\textbf{Multi-cloud and Hybrid Cloud Platforms} & multi-cloud environment \cite{S41, S46, S61, S67}, Cloudify \cite{S65}, CloudBroker Platform \cite{S69}, Cloudcheckr \cite{S79}, MistIO \cite{S79}, ManageIQ \cite{S67, S79}, CMP2 \cite{S79}, AWS SDK \cite{S71}, CloudSME Simulation Platform (CSSP) \cite{S69}, Vagrant \cite{S08}, AWS OpsWorks \cite{S08}, Universal Compute Xchange (UCX) \cite{S65}, Liferay portlets \cite{S69}, Cloud Service Providers (CSPs) \cite{S71}, AWS Data Pipeline \cite{S58}, GENI \cite{S38}, Google Cloud Platform \cite{S38, S53, S67, S80}, AWS EC2 \cite{S11, S51, S18}, Alibaba cloud \cite{S65, S80}, Tencent \cite{S65, S80}, OpenStack \cite{S21, S43}, VMware vSphere \cite{S74}, Aliyun ECS \cite{S03}, Baidu BCE \cite{S03}, Tencent CW \cite{S03}, JD VW \cite{S03}, GoGrid \cite{S51} \\
\hline
\textbf{Networking, Security, and Access Management} & Open vSwitch (OvS) \cite{S72}, Cloud networks \cite{S77}, Host Identity Protocol (HIP) \cite{S77}, Gateways \cite{S64}, Pledge \cite{S40}, Seccomp-BPF \cite{S40}, Landlock LSM \cite{S40}, Linux Security Modules (LSM) \cite{S40}, MUSA Security Assurance Platform \cite{S34}, Firewall Deployment \cite{S30}, Access control policies \cite{S25}, Attribute-Based Encryption and Signature \cite{S75}, ABE/ABS cryptographic model \cite{S75} \\
\hline
\textbf{Monitoring, Management, and Automation} & Prometheus \cite{S22, S92, S93, S94, S95, S96, S97, S98, S99, S100, S101, S114}, Grafana \cite{S27, S99, S100, S101, S114}, Telegraf \cite{S27}, AWS CloudWatch \cite{S19}, cAdvisor \cite{S22}, Ceilometer \cite{S56, S58}, Gnocchi \cite{S58}, ELK stack \cite{S74}, Terraform \cite{S27, S29, S88, S102, S118}, CloudBees \cite{S08}, Jenkins \cite{S38, S108}, Puppet \cite{S08}, Chef \cite{S08, S30}, GitHub \cite{S30}, Shell scripts \cite{S34}, Maven \cite{S08}, NPM \cite{S08}, Consul \cite{S08} \\
\hline
\textbf{Development, APIs, and Cloud Storage} & Public cloud storage \cite{S64}, Private cloud storage \cite{S64}, Hybrid cloud storage \cite{S64}, Google Cloud Storage \cite{S105}, Azure Blob Storage \cite{S105}, REST API \cite{S64}, YAML \cite{S79}, OpenVAS Vulnerability Scanner \cite{S30}, DAX format \cite{S67}, Java \cite{S30, S34, S35, S54, S80, S81}, Go programming language \cite{S32, S36}, Python \cite{S35, S54, S56, S58, S79}, Ruby \cite{S54}, PHP \cite{S54}, C\# \cite{S54}, Cloud Modelling Framework (CLOUMDF) \cite{S20}, Cloud Modelling Language (CLOUDML) \cite{S20, S81}, CAMEL modeling language \cite{S52, S80, S84}, Eclipse IDE \cite{S27}, MCSLA editor \cite{S27} \\
\hline
\end{tabular}
\end{table*}

\normalsize
\begin{tcolorbox}
\textbf{Takeaway 12}: \textcolor{black}{The reviewed studies reveal that tools and frameworks commonly used in general container or cloud settings (e.g., Kubernetes, Docker, Terraform, Prometheus) require specific adaptations or configurations to operate effectively in containerized multi-cloud environments. This includes challenges such as consistent orchestration across heterogeneous cloud APIs, maintaining observability across isolated infrastructures, and enabling secure, policy-compliant automation at scale. Moreover, hybrid-enabling tools like OpenStack, Cloudify, and MistIO emerged as critical for unifying management across cloud boundaries — highlighting how the combination of containerization and multi-cloud imposes unique operational and integration demands not encountered when using either paradigm alone.}
\end{tcolorbox}


\section{Discussion}
\label{Sec_Discussion}
\textcolor{black}{The following discussion synthesizes key findings for each research question and their implications for research and practice. These implications also highlight open problems and unresolved challenges that require further research attention, particularly those emerging from the intersection of containerization and multi-cloud environments.}
\subsection{Containers in Multi-Cloud: Roles and Strategies (RQ1)}
Containers play a crucial role in the development and operation of modern software systems within multi-cloud environment. Understanding the various roles and implementation strategies of containers provides significant insights for both researchers and practitioners.

\textbf{\textit{Multifaceted Role of Containers in Multi-Cloud Environment}}: In a multi-cloud environment, containers serve multiple purposes, such as managing resources, hosting services, and deploying applications. They simplify the deployment of apps across various platforms, including web applications and Internet of Things (IoT) applications. Additionally, containers enhance security in multi-cloud operations. \textcolor{black}{These roles are shaped by the inherent heterogeneity of multi-cloud settings, where containers are used to ensure portability, manage provider-specific constraints, and support workload migration across isolated cloud infrastructures.} \textbf{Implications}: These findings open up opportunities for researchers to explore additional areas, including security vulnerabilities, management challenges, and performance overhead. Likewise, practitioners can apply these insights to design more efficient and effective deployment strategies, ultimately contributing to stronger and more reliable multi-cloud implementations.

\textbf{\textit{Container Technology Features and their Significance}}: Container technology offers several benefits due to its unique features, such as support for edge computing, compatibility with orchestration tools, and lightweight virtualization. \textcolor{black}{In multi-cloud environments, these features enable containers to act as a unifying abstraction layer across diverse infrastructure stacks, allowing consistent deployment behavior despite differing cloud-specific configurations.}  \textbf{\textbf{Implications}}: Researchers can explore these features further, including how characteristics like lightweight virtualization influence container performance and effectiveness in multi-cloud operations. Practitioners can use this understanding for better utilization of container technology and more efficient multi-cloud deployments.

\textbf{\textit{Implementation Strategies and their Real-World Implications}}: The strategic implementation of containers significantly affects their functionality in multi-cloud environment. Strategies related to deployment, orchestration, security, performance optimization, cloud service management, and data storage can have significant implications. \textcolor{black}{The reviewed strategies demonstrate how container management must adapt to cloud-specific APIs, variable service guarantees, and cross-cloud policy enforcement, revealing how the interplay between containerization and multi-cloud governance introduces distinctive design trade-offs.} \textbf{Implications}: Researchers may explore these strategies further, including investigations into their real-world implications, benefits, and limitations. Practitioners can refine their understanding of these strategies to achieve more efficient implementations and operations.

\textbf{\textit{Containerization's Impact on IoT and Application Development}}: The incorporation of containers into IoT and app development has transform these fields. Containers offer improved software unit encapsulation, portability, and efficiency, which are particularly important in the distributed and resource-constrained nature of IoT. \textcolor{black}{In multi-cloud scenarios, this impact is increase by the need to coordinate deployments across heterogeneous edge-cloud setups and maintain interoperability between varied cloud-native services and orchestration frameworks.} \textbf{Implications}: Researchers can further explore potential challenges in detail such as network complexity, security, and the small footprint requirement of IoT devices. For practitioners, understanding the impact of containerization can lead to the development of more robust, efficient, and secure IoT and app solutions.

\subsection{Patterns and Strategies (RQ2)}
Containerized applications in multi-cloud environment requires an understanding of prevailing patterns and strategies. In this context, we discuss the key findings related to the identified patterns and strategies presented in Table \ref{tab:PatternsStrategies}.

\textbf{\textit{Diverse Architectural Patterns and Communication Strategies}}: Our study found that Microservice Architecture and Service-Oriented Architecture (SOA) are the most common used architectural patterns for designing container-based applications in multi-cloud environment. These results align with the existing literature that emphasizes the use of MSA and SOA because of their support for modular development, scalability, and efficient resource use \cite{balalaie2016microservices, dragoni2017microservices}. The other patterns frequently reported are Service Mesh and Service Chaining, which are employed as communication and networking patterns, likely due to their ability to manage complex, distributed service-to-service communications in a scalable way \cite{zhamak2018data}. The frequent use of black-Green Deployment and Object Store Service deployment patterns has also been identified in this study, which may stem from their ability to minimize downtime and ensure high availability—essential factors in a cloud-based setting \cite{fowler2010bluegreen}.  \textcolor{black}{Notably, these architectural and communication patterns are adapted in the multi-cloud setting to support distributed orchestration, cross-cloud resilience, and integration across heterogeneous platforms—attributes that are less emphasized in single-cloud deployments. For instance, service mesh implementations in multi-clouds enable secure communication across isolated cloud boundaries, and decentralized orchestration patterns accommodate independently governed cloud nodes.} \textbf{Implications}: Practitioners can enhance the design, scalability, and efficiency of their applications in a multi-cloud environment by understanding and applying these architectural and communication patterns. Researchers might explore how these patterns evolve with the emergence of new technologies and standards.

\textbf{\textit{Importance of Container Management and Multi-Cloud Strategies}}:  Our findings report the significance of robust container management and multi-cloud strategies. Techniques like the Linux Container (LXC) project, Docker Container Images, and Lightweight Virtualization are frequently used, emphasizing the industry trend towards containerization due to its benefits in efficiency, scalability, and platform independence \cite{vaughn2020docker}. Furthermore, the popularity of multi-cloud strategies like Hybrid/Multi-cloud Approach and Multi-cloud Load Balancing confirms the growing industry move towards using multiple cloud providers to enhance redundancy, flexibility, and avoid vendor lock-in \cite{gillam2013benchmarking}. \textcolor{black}{In particular, the reviewed studies illustrate how container technologies are adapted to meet challenges specific to multi-cloud deployments, such as orchestrating containers across provider-specific APIs, aligning container runtime environments across clouds, and ensuring policy compliance in cross-provider setups. These represent key peculiarities introduced by multi-cloud environments into the containerization landscape.} \textbf{Implications}: Practitioners should focus on gaining expertise in container technologies and multi-cloud strategies to take advantage of their numerous benefits. Meanwhile, researchers might examine the challenges and best practices associated with the deployment and management of container technologies across multiple cloud platforms.

\textbf{\textit{Security, Resilience, and Migration Patterns and Strategies}}: Our study also identifies patterns and strategies related to security, resilience, and efficient migration strategies in a multi-cloud context. Strategies such as File Encryption, Attribute-Based Encryption, and Secure Data sharing are often employed to ensure data security, an area of growing concern with the increase in cloud-based applications \cite{svantesson2010privacy}. Moreover, resilience strategies such as Fault-tolerance with Redundant Engines and RAFT Consensus Algorithm are essential to ensure system reliability in distributed, multi-cloud environment \cite{ongaro2014consensus}. Migration between different cloud environments, facilitated by strategies like Service-oriented Migration and Migration to a Virtualized Container, is another vital aspect, given the growth in cloud services and the need for interoperability \cite{jamshidi2013cloud}. Migrating applications and data between cloud environments is an important consideration in a multi-cloud scenario \cite{jamshidi2013cloud}. Strategies such as image-based migration and migration to virtualized containers, like Docker, are commonly used \cite{vaughn2020docker}. These techniques facilitate portability and provide consistent environments across different cloud platforms. Additionally, they have significant implications for scaling and managing applications across different cloud service providers. \textcolor{black}{Unlike traditional container security or resilience strategies in single-cloud contexts, the reviewed studies show that in multi-cloud scenarios, these approaches must handle cross-cloud identity management, consistency in fault detection and recovery across isolated infrastructures, and security rule federation across providers. Similarly, migration patterns uniquely address compatibility mismatches and compliance constraints introduced by the heterogeneity of cloud APIs and services.} \textbf{Implications}: These findings suggest that security, resilience, and efficient migration should be key areas of focus for practitioners operating in multi-cloud environment. Researchers may also look into novel techniques for enhancing cloud security, improving system resilience, and easing the process of cloud migration.

\subsection{Quality Attributes and Tactics (RQ3)}

The exploration and implementation of quality attributes and associated tactics are critical to the successful deployment of containerized applications in multi-cloud environment, shaping performance, security, and compatibility characteristics. In the following, we will discuss the key findings of our study, providing insights into the QAs and tactics specifically tailored for containerized applications operating within multi-cloud environment.

\textit{\textbf{{Performance Optimization in Multi-Cloud Environment}}}: Our study identifies the role of machine learning, container technology, and location-aware service brokering in enhancing the performance and efficiency of containerized applications within multi-cloud environment. Hashem \textit{et al.} previously suggested similar efficiency gains from machine learning and container technology, validating the interpretive value of our results \cite{hashem2016}. Specifically, the application of machine learning aligns with a broader trend towards intelligent cloud resource management, a domain that Buyya \textit{et al.} identified as significant in cloud computing research \cite{buyya2020}. Meanwhile, location-aware service brokering emerges as a novel and impactful strategy, stressing the role of geographical considerations in further optimizing containerized applications' performance. 
\textcolor{black}{Importantly, the reviewed studies emphasize that performance tactics must accommodate multi-cloud-specific challenges such as distributed latency, resource fragmentation across providers, and dynamic SLA enforcement, which are less relevant in single-cloud settings.} \textbf{Implications}: For researchers, the results of our study offer opportunities to explore the integration of AI techniques with container-based cloud architectures further, especially the application of location-aware service brokering in multi-cloud environment. For practitioners, the results indicate that a adaptable approach involving AI, container technology, and geographical considerations can lead to significant performance enhancements in multi-cloud environment.

\textit{\textbf{{Security Measures in Multi-Cloud Deployments}}}: The results of our study indicates the multi-faceted approach to ensuring security in multi-cloud deployments. It highlights the combination of encryption, network security protocols, machine learning, and user-specified security measures, while also encouraging deployment across different cloud providers. This finding complements Singh \textit{et al.}'s argument for a comprehensive approach to cloud security, and aligns with the recent concept of distributed cloud security \cite{singh2016}. Our results also report the importance of user-specified security measures, which are in agreement with the user-centric security model proposed by Shin and Dong-Hee \cite{shin2013user}. \textcolor{black}{Notably, multi-cloud containerization introduces additional layers of complexity in managing security policies across heterogeneous platforms and enforcing access control in federated settings. Tactics such as decentralized IAM logging and policy-aware container placement directly respond to these unique multi-cloud constraints.} \textbf{Implications}: From a research perspective, the interact between different security measures and their effect on overall system security in multi-cloud environment warrants further investigation. For practitioners, these results point out the importance of employing a comprehensive security approach in multi-cloud deployments, including encryption, network security, machine learning, and user-specific security measures.

\textit{\textbf{{Ensuring Compatibility Across Different Cloud Providers}}}: The results of our study suggests strategies to ensure compatibility in multi-cloud environment through standardization of interfaces, componentization, and interoperability approaches. Emphasizing lightweight communication protocols and interoperability between clouds, the results related to compatibility mirrors Mell and Grance's principles of service-oriented architecture, reinforcing the need for effective communication across different cloud platforms \cite{mell2011nist}. Petcu \textit{et al.} have also previously stressed the importance of standardization and componentization for ensuring compatibility \cite{petcu2018}. \textcolor{black}{Our findings highlight that achieving compatibility in a containerized multi-cloud context often involves abstracting away provider-specific interfaces through standardized APIs and decoupling infrastructure logic via containerization. These are adaptations not typically required in homogeneous cloud environments.} \textbf{Implications}: From a research standpoint, the dynamic between lightweight communication protocols, interface standardization, and overall system compatibility in multi-cloud deployments offers fertile ground for future work. Practitioners, meanwhile, can adopt these approaches to ensure seamless communication and interoperability across different cloud providers.

These findings not only confirm existing literature results but also offer valuable insights into the practical application of machine learning, container technology, and location-aware services in multi-cloud environment. Future research could further explore the intersection of these quality attributes to develop more robust, efficient, and secure multi-cloud applications by using containerise technologies.

\subsection{Automation Challenges and Solution Framework (RQ4 - RQ5)}
The Automation Challenge-Solution Framework for Containerized Applications in multi-cloud environment (see Table \ref{tab:Automation-solutions}) suggests that multi-cloud management for containerized application is a complex task encompassing several dimensions. These include multi-cloud automation, deployment and scaling, resource management, data and application migration, testing and bench marking, standardization and interoperability, application and service management, and runtime and service discovery. This indicates that addressing one challenge may potentially impact the other areas, necessitating a holistic approach to tackle these issues. In the following we discuss the some of the key observations about this framework. 

\textbf{\textit{Multi-dimensional Challenges}}: The Automation Challenge-Solution Framework for Containerized Applications in multi-cloud environment suggests that multi-cloud management is a complex task encompassing several dimensions. These include multi-cloud automation, deployment and scaling, resource management, data and application migration, testing and bench marking, standardization and interoperability, application and service management, and runtime and service discovery. This indicates that addressing one challenge may potentially impact the other areas, necessitating a holistic approach to tackle these issues. \textcolor{black}{What makes these challenges unique to containerized multi-cloud setups is the dual complexity arising from container portability and cloud heterogeneity. Containers, while offering mobility and reproducibility, must now adapt to diverse runtime configurations, networking policies, and resource scheduling strategies across cloud providers.} The multi-dimensionality of the challenges implies the need for an integrated approach that considers the interdependencies among these dimensions. It emphasizes the need for solutions that cater to this complexity rather than address individual challenges in isolation. \textbf{Implications}: For researchers, these findings underscore the need to develop multi-faceted, holistic solutions that can address the various interconnected challenges in managing multi-cloud environment. For practitioners, this implies the importance of strategic planning and implementing solutions that address the multiple facets of multi-cloud management while considering their potential impact on one another.

\textbf{\textit{Emergence of AI/ML Solutions}}: The framework highlights the increasing utilization of AI and ML in providing intelligent solutions to manage containerized applications in multi-cloud environment. This suggests the growing maturity and applicability of these technologies in complex scenarios, a significant step towards achieving efficient multi-cloud management. Previous studies (e.g., \cite{de2020artificial}, \cite{chen2021artificial}, \cite{gill2022ai}) have recognized the potential of AI and ML in resolving complex cloud management issues, which is in line with the findings from the framework. This further validates the significant role AI and ML have started to play in this domain. \textcolor{black}{The reviewed studies specifically show how AI/ML techniques help in real-time placement of containers across heterogeneous clouds, optimizing latency and cost under dynamic workload and policy constraints—issues that are intensified by the distributed and federated nature of multi-cloud environments.} The increasing adoption of AI/ML in multi-cloud environment signifies an essential evolution in the field. This stresses the need for further research and development in these technologies to exploit their full potential in resolving multi-cloud management complexities. \textbf{Implications}: The growing use of AI/ML solutions implies that researchers need to focus on improving these technologies and adapting them to specific multi-cloud management challenges. For practitioners, this points towards the increasing necessity to invest in AI/ML capabilities and integrate them into their multi-cloud management strategies.

\textbf{\textit{Importance of Standardization and Interoperability}}: The framework underscores the crucial role of standardization and interoperability in multi-cloud environment, facilitating seamless integration and management across different cloud platforms. Recent literature validates the importance of standardization and interoperability in multi-cloud environment, emphasizing their role in enhancing the efficiency of cloud services \cite{alonso2023understanding}. This agrees with our results, further confirming the critical need for these factors in the domain. The need for standardization and interoperability reiterates the necessity for a unified framework or protocol across different cloud environments. \textcolor{black}{In containerized deployments, interoperability challenges become more pronounced due to variations in networking stacks, security policies, and orchestration tools between providers. Standardization at the container runtime level (e.g., Docker images, Kubernetes manifests) and orchestration abstraction (e.g., TOSCA, Helm charts) plays a vital role in enabling automation workflows that span diverse cloud ecosystems.} It suggests that attention to these aspects can lead to more seamless, efficient, and secure management of multi-cloud services. \textbf{Implications}: For researchers, the results stress the need to investigate methods to achieve better standardization and interoperability in multi-cloud environment. For practitioners, the findings suggest the need to adopt standards and focus on interoperability while selecting and implementing cloud services, ensuring more efficient management across different platforms.

\subsection{Security Challenges and Solutions Framework (RQ4 - RQ5)}
The framework proposed in our study provides a comprehensive and systematic approach to addressing the complex security issues associated with containerized applications in multi-cloud environment. In the following, we discuss the key takeaway from security challenges and solution framework presented in Section \ref{SecurityChallengesandSolutionsFramework} and Table \ref{tab:Security-challenges-solutions}.

\textbf{\textit{Emphasize on Container-Specific Security Mechanisms}}: Security solutions specifically designed for containerized applications are critical to managing threats in multi-cloud environment. The selected studies suggest a variety of methods, including secure container orchestration, container isolation, and advanced deployment strategies that make use of deep reinforcement learning. This aligns with the existing literature (e.g., \cite{zhong2022machine}, \cite{li2023optimal} \cite{combe2016docker}) which highlights the unique security considerations brought forth by containerization. Containers, being lighter than traditional virtual machines and offering process-level isolation, have revolutionized the way applications are packaged and run. However, they also introduce new security challenges that need to be addressed by using container-specific security mechanisms. Containerization, while a powerful tool for application deployment, necessitates its own suite of security strategies. Implementing these will be key to leveraging the benefits of containerization in a secure manner. Implementing these will be key to using the benefits of containerization in a secure manner. \textcolor{black}{In multi-cloud settings, these challenges are compounded by the need to secure container interactions across diverse platforms, enforce consistent policies under different providers, and protect workloads that dynamically move between clouds.} \textbf{Implications}:  For researchers, this highlights an area of potential study: the development of advanced container-specific security strategies, including those that use advanced techniques such as machine learning. For practitioners, this suggests the need to be well-versed in the specifics of container security. Leveraging features of containerization platforms, such as Docker's isolation features, or incorporating tools specifically designed for container security will be crucial.

\textbf{\textit{Implementing Multi-Layered Security and Compliance Measures}}:
Implementing multi-layered security measures that encompass data protection, access control, and communication security, along with compliance with legal and regulatory requirements, is of utmost importance. Solutions range from data encryption, role-based access control, secure container orchestration, adherence to regulations like GDPR \cite{regulation2016general}. This is consistent with current cybersecurity frameworks, such as the NIST cybersecurity framework \cite{nist2018framework}, which emphasize a multi-layered approach to security. Furthermore, several studies point to the criticality of data security, user access control, and secure communication in cloud environments \cite{austin2018cybersecurity} \cite{singh2016}. In a multi-cloud, containerized environment, it is not enough to secure data; access and communication channels must also be secured, and all actions must comply with legal and regulatory requirements. \textcolor{black}{In multi-cloud containerized setups, multi-layered security must also accommodate interoperability between CSPs, dynamic access management across domains, and distributed compliance enforcement for container workflows.} \textbf{Implications}: For practitioners is to design and implement a comprehensive security strategy that encompasses multiple layers of protection. Compliance with legal and industry standards should be an integral part of this strategy. For researchers, this underlines the need for studies that address the integration of various layers of security and the development of comprehensive security frameworks.

\textbf{\textit{Ensuring Security through Standardization and Interoperability}}: Ensuring security through standardization and interoperability is highlighted as a critical step. Secure and standardized interfaces, such as OpenStack APIs, can be used to enhance security and performance across multiple cloud providers. This finding is aligned with existing literature, which emphasizes the role of standardization in achieving security and interoperability in multi-cloud environment \cite{buyya2010intercloud} \cite{petcu2011portability}. By implementing standardized interfaces and practices, organizations can ensure a level of interoperability and security across cloud environments. \textcolor{black}{For containerized applications, this also means ensuring consistent security postures across orchestrators, managing identity federation between providers, and enabling trust in federated environments that dynamically provision containers at runtime.} \textbf{Implications}: For practitioners, this means choosing platforms and tools that support standardized interfaces and practices, or working to implement those standards within their own environments. Researchers, on the other hand, could focus on the development of new standards or the improvement of existing ones to better address security challenges in multi-cloud, containerized environments.

\subsection{Deployment Challenges and Solutions Framework (RQ4 - RQ5)}
Table \ref{tab:monitoring-solutions} offers a comprehensive overview of challenges and their corresponding solutions encountered during the deployment of containerized applications in multi-cloud environment. In the following, we provide a discussion on key findings pertaining to the challenges and solutions related to the deployment of containerized applications in multi-cloud environment.

\textbf{Standardization and Automation in Deployment}: Standardization and automation stand out as foundational pillars for achieving streamlined and uniform deployments in multi-cloud landscapes. As Raj \textit{et al.} \cite{raj2018automated} highlighted, embracing automated deployment pipelines paired with standardized toolsets can significantly diminish complexity and mitigate human-induced errors within multi-cloud frameworks. The challenges, captured under \textit{Performance Testing Deployment}, \textit{Deployment Validation}, and \textit{Cloud Management Platform Evaluation Deployment} in the table, elucidate the nuanced complexities associated with deployments spanning multiple cloud platforms. Solutions such as \textit{Standardized Benchmarking and Automation}, \textit{Testing Process Management System}, and \textit{Standardized Output Formats and Evaluation Criteria} emphasize the imperative nature of adopting uniform procedures and harnessing automated solutions. Multi-cloud deployments, inherently intricate, benefit profoundly from standardization, bringing forth predictability and coherence, while automation paves the way for diminishing human-driven inconsistencies and bolstering deployment speeds.  Multi-cloud deployments, inherently intricate, benefit profoundly from standardization, bringing forth predictability and coherence, while automation paves the way for diminishing human-driven inconsistencies and bolstering deployment speeds. \textcolor{black}{In containerized environments, these challenges are amplified due to the need to consistently deploy lightweight, encapsulated workloads across heterogeneous infrastructure with varying APIs and orchestration logic.}  \textbf{Implications}: For researchers, the quest for optimized frameworks championing standardization and automation within multi-cloud environment presents a promising research trajectory. For practitioners, adeptness with standardized methodologies and tools, combined with proficiency in automation implementation, becomes crucial in navigating the multi-cloud management maze efficiently

\textbf{Security and Compliance in multi-cloud environment}: Security remains paramount in cloud deployments. Fernandes \textit{et al.} underscore that the labyrinth of diverse platforms and differing standards in multi-cloud configurations amplifies security complexities \cite{fernandes2014security}. Challenges articulated as \textit{Security Deployment in Multi-container Environment} and \textit{Malicious Service Management Deployment} in Table \ref{tab:Deployment-solutions} accentuate the primacy of security considerations. Propounded solutions like \textit{Cross-Container Isolation} and \textit{Collusion-Resilient Trust Aggregation Technique} spotlight the industry's shift towards advanced and meticulous security protocols within multi-cloud paradigms. The ever-evolving landscape of multi-cloud deployments demands security solutions that are not only rigorous but also malleable and formidable. To attain comprehensive security across diverse platforms, strategies must encapsulate both depth of understanding and breadth of application. \textcolor{black}{Containerization introduces new layers of isolation and deployment granularity, which—when combined with multi-cloud distribution—creates a unique surface of vulnerability that necessitates tailored controls at both the container and orchestration levels.} \textbf{Implications}: For the academic community, pioneering avant-garde security solutions apt for multi-cloud environment offers a fertile domain of inquiry. Meanwhile, practitioners should enshrine principles of data integrity, trust assurance, and vulnerability mitigation at the heart of their multi-cloud deployment blackprints.

\textbf{Integration, Monitoring, and Infrastructure Management}: As highlighted by Ferrer \textit{et al.} \cite{ferrer2016multi} the ability to operate seamlessly across disparate cloud providers and maintain flexible infrastructures is paramount in harnessing the full potential of multi-cloud deployments. The presented challenges, including \textit{Network Latency Deployment} and \textit{Vendor Lock-in Prevention Deployment}, underscore the indispensability of agility and cross-platform compatibility. Proposed solutions such as \textit{VM Selection for Composite Applications} and \textit{Interoperability Layer Above Cloud Infrastructure} accentuate the importance of crafting adaptable infrastructure and formulating strategies to mitigate vendor entrenchment. \textcolor{black}{Here, containerization plays a distinct role by enabling application components to be packaged once and executed across diverse cloud platforms, yet this same portability necessitates careful orchestration, consistent networking, and compatibility layers across varied cloud APIs and runtime environments.} In the context of multi-cloud environment, flexibility and interoperability transcend mere advantageous attributes; they become pivotal determinants shaping adaptability, scalability, and the sustained success of deployment blackprints. \textbf{Implications}: For the academic community, delving into novel approaches that champion infrastructure pliability and seamless operation across diverse cloud platforms emerges as a prospective research avenue. For industry professionals, anchoring strategies around flexibility and interoperability not only optimizes current deployments but also fortifies them against future technological evolutions and shifting business dynamics.

\subsection{Monitoring Challenges and Solutions (RQ4 - RQ5)}
Table \ref{tab:monitoring-solutions} presents a comprehensive view of the Monitoring Challenge-Solution Framework for Containerized Applications in multi-cloud environment. It covers a wide range of monitoring challenges within the intricate multi-cloud setting. In the following, we are providing discussion on the key findings. 

\textbf{Enhancing Performance Stability through Multi-Cloud Monitoring}:  In the context of multi-cloud environment, ensuring consistent and reliable performance monitoring for containerized applications can be challenging due to the dynamic nature of cloud resources. Existing studies (e.g., \cite{zhong2022machine} \cite{mungoli2023scalable}) highlighted the impact of varying resource availability on the performance of containerized applications across multiple clouds. They emphasized the need for performance benchmarking to track fluctuations and maintain a consistent user experience. The solutions presented in our framework, such as benchmarking performance variation and utilizing Kubernetes for monitoring, align with existing studies suggestion for benchmarking. \textcolor{black}{Unlike traditional single-cloud deployments, containerized applications spanning multiple clouds must be monitored for platform-specific latency, orchestration timing mismatches, and fragmented telemetry data pipelines.} Our proposed framework expands on this by incorporating Kubernetes for improved multi-cloud performance monitoring. \textcolor{black}{Kubernetes, while originally designed for single-cluster orchestration, plays a central role in aggregating monitoring data across containerized services deployed in heterogeneous clouds, helping unify performance visibility.}  Our proposed framework expands on this by incorporating Kubernetes for improved multi-cloud performance monitoring. Our framework also suggest reliable performance monitoring requires a combination of standardized benchmarking techniques and cloud-specific tools like Kubernetes. This approach can effectively address performance variations and ensure a consistent user experience across diverse cloud environments. \textbf{Implications}: For practitioners, adopt benchmarking practices to track performance fluctuations, and use Kubernetes for unified multi-cloud performance monitoring. For researchers, explore further refinements in benchmarking methodologies and investigate the impact of dynamic multi-cloud environment on performance consistency.

\textbf{Streamlining Resource and Network Management across Multi-Clouds}: Managing network communications, system images, and resource utilization in a multi-cloud setting poses challenges due to the diversity of cloud platforms. Sedghpour \textit{et al.} discussed the complexities of managing network communications in multi-cloud environment. They emphasized the need for standardized approaches like service mesh to enhance network visibility and control \cite{sedghpour2022service}. Solutions in our proposed framework, such as using Docker Server for containerized services and Prometheus/Grafana for resource monitoring, align with Sedghpour \textit{et al.}'s \cite{sedghpour2022service} suggestion for standardized approaches. \textcolor{black}{However, container-based deployments compound the difficulty by adding layers of ephemeral, distributed services that must be monitored across volatile container lifecycles. This increases the burden of synchronization and observability.} Additionally, our framework extends these solutions to address broader resource management challenges. \textcolor{black}{Containers, especially when orchestrated over multiple cloud platforms, generate high-churn environments that require lightweight, highly adaptive monitoring strategies to ensure timely insights and cost control.} Additionally, our framework extends these solutions to address broader resource management challenges. Employing standardized tools like Docker Server, along with platforms like Prometheus and Grafana, can help streamline resource management and network communication monitoring across diverse cloud platforms. \textbf{Implications}: For practitioners, implement Docker Server and utilize platforms like Prometheus and Grafana for efficient resource monitoring and management across multi-cloud environment.
For researchers, investigate further integration possibilities for standardized resource management tools and explore the impact of such tools on overall system performance and reliability.

\textbf{Strengthening Security, Compliance, and Trust}: Ensuring data privacy, continuous security control, and compliance in multi-cloud containerized applications is essential but challenging due to the distributed nature of cloud resources. Existing studies (e.g., \cite{alonso2023understanding} \cite{raj2018automated}) discussed the challenges of ensuring security and compliance across multi-cloud environment. They highlighted the importance of continuous monitoring and enforcement mechanisms to address security and compliance gaps. Solutions presented in our proposed framework, such as using machine learning for performance supervision and employing security assurance platforms, are aligned with existing studies that emphasis on continuous monitoring and enforcement for security and compliance.   \textcolor{black}{Containerized deployments introduce distinct security risks, e.g., image poisoning, privilege escalation between co-resident containers, and cross-platform trust boundary misconfigurations that demand container-aware monitoring solutions.} Implementing advanced techniques like machine learning for monitoring and adopting security assurance platforms can significantly enhance security, compliance, and trust management in multi-cloud containerized applications. \textbf{Implications}: For practitioners, integrate machine learning-based monitoring and adopt security assurance platforms to ensure continuous security control and compliance across diverse cloud environments. For researchers, explore the scalability and effectiveness of machine learning-driven monitoring techniques and investigate novel methods for enforcing security and compliance in multi-cloud setups.

\subsection{Tools and Frameworks (RQ6)}
This study report 160 distinct tools and frameworks for container based applications in multi-cloud environment, all of which are systematically classified into 6 categories, further delineated into 46 subcategories. In the following we briefly discuss some of the key findings about identified tools and frameworks.

\textbf{Diverse Cloud Services by Leading Providers}: Our study reveals a diverse array of cloud services, including IaaS, PaaS, and SaaS offerings from major providers like AWS, Azure, and Google Cloud, emphasizing the extensive options available to developers (see Table \ref{tab:ToolsandFrameworks}). This underscores the significant range of choices available to developers. Noteworthy cloud service providers highlighted in our study include Amazon EC2, Google Cloud App Engine, and Azure, further underscoring this finding. \textcolor{black}{In multi-cloud environments, selecting services becomes more complex due to differing billing models, feature parity, and integration capabilities across providers—requiring tools that support cross-provider abstraction and orchestration.} \textbf{Implications}: For practitioners, our outcomes underscore the critical necessity of meticulously evaluating and selecting the most fitting service model based on the unique demands of a given project. This strategic approach ensures the best alignment between technology and goals. Moreover, the implications for researchers are promising: an avenue emerges for in-depth exploration of trade-offs and the identification of optimal practices while navigating the intricate array of cloud service models. This exploration holds potential for enhancing both the efficiency and effectiveness of cloud solutions.

\textbf{Docker and Kubernetes: Pioneers of Cross-Cloud Containerization}: Our investigation spotlights the profound significance of Docker and Kubernetes in streamlining the containerization and orchestration of applications across a spectrum of cloud platforms. Docker adeptly encapsulates applications and their dependencies within self-contained containers, while Kubernetes takes center stage as the orchestration juggernaut, automating deployment, scaling, and management of these containers. These technologies receive direct validation through table references, serving as prime exemplars. 
\textcolor{black}{However, our analysis also reveals that deploying Kubernetes in a multi-cloud environment demands adaptations, such as federation control planes, multi-zone DNS, and workload placement strategies tailored to cross-cloud latencies and SLAs.}
Our research yields a compelling revelation: Docker and Kubernetes hold pivotal roles in ensuring consistent application deployment and seamless scalability, effectively simplifying the complexities of multi-cloud environment. \textbf{Implications}: Practitioners stand to gain significantly by adopting Docker and Kubernetes, as this potent combination bolsters efficiency and cultivates heightened portability. \textcolor{black}{Researchers can explore emerging gaps in areas such as secure cross-cloud scheduling, fault tolerance across zones, and policy-compliant orchestration}.

\textbf{The Rise of DevOps: Embracing Automation Tools}: Our exploration spotlights the important role of DevOps practices and tools such as Puppet, Chef, and Terraform within contemporary software development landscapes. This emphasis revolves around the power of automation in facilitating streamlined deployment and management processes. These tools, by enabling consistent provisioning and deployment of infrastructure, effectively curtail errors and ensure the replicability of processes. The table substantiates this notion through references to Puppet, Chef, and Terraform. \textcolor{black}{Within a multi-cloud context, these tools must be adapted to handle provider-specific APIs, authentication schemes, and compliance constraints, which introduces configuration drift and policy management complexity.}. Implications: Practitioners are advised to seamlessly integrate DevOps principles into their workflows by harnessing the prowess of these tools. \textcolor{black}{Researchers may investigate how DevOps pipelines can be extended to support containerized workloads that span multiple clouds, ensuring traceability, consistency, and auditability across heterogeneous execution environments.}
\section{Threats to Validity}
\label{sec:SMSthreats}
This systematic mapping study is susceptible to various threats that could influence its outcomes. To address these potential threats, we adhered to the established guidelines for conducting SMSs and SLRs as outlined in \cite{AR26} and \cite{petersen2008}. We further categorized and analyzed the validity threats to our study based on the four distinct types of validity threats mentioned in \cite{AR53} and \cite{AR54}. In the subsequent sections, we delve into the specific validity threats that pertain to the different phases of this SMS.

\subsection{Internal Validity}
\label{Threats_IV}
Internal validity refers to the factors that could affect the analysis of the data extracted from the selected studies. The threats to internal validity could happen in the following steps of this SMS:
\begin{itemize}
    \item\textbf{Study Search}: The potential to overlook studies during the search process requires careful measures. To mitigate this risk, we used a combination of primary and snowballing search methodologies, as detailed in Section \ref{sec:PrimarySearch}. To increase the collection of primary studies during the initial search phase, we also took additional measures to address search-related issues. Specifically, we improved our search strings through pilot searches before applying them to the databases. This process helped to create a more effective search strategy.
    \item\textbf{Study Selection}: We have outlined the study screening and selection process in Table \ref{tab:InclusionExclusion}. To ensure objectivity and eliminate personal bias in study selection, we implemented a two-phase approach: (i) initial study screening, and (ii) qualitative evaluation of the shortlisted studies. During this procedure, the lead two authors conducted the study screening based on the criteria explicitly detailed in Section \ref{tab:InclusionExclusion}. To assess the objectivity of the screening process and the level of agreement between the two authors, we applied Cohen’s Kappa \cite{cohen1968weighted}. In cases where the lead authors did not agree, the second and third authors independently reviewed the disputed studies to reach a consensus. All the researchers involved in this study have extensive knowledge and research experience in containerization applications within a multi-cloud environment.
    \item\textbf{Data Extraction}: Bias from researchers during data extraction can pose a significant challenge in both SMSs and SLRs. To address this potential issue, we developed a standardized data extraction form (see Table \ref{tab:DataExtractionItems}) to ensure consistent data retrieval. Initially, the first and fourth authors extracted the data. If any uncertainties arose regarding the extracted data, comprehensive discussions were convened among all authors. Following the recommendations in \cite{AR54}, a subset of the extracted data was cross-verified by the second and third authors.
    \item\textbf{Bias on Themes Classification}: Misclassification of data and primary studies may result in biases stemming from subjective interpretation. To minimize this risk, we adhered to the thematic analysis guidelines established by Braun \textit{et al.} \cite{braun2006}, and implemented a six-step process for thematic classification, as detailed in Section \ref{DataExtractionandAnalysis}.
    \item\textbf{Data Synthesis}: We employed both qualitative and quantitative approaches to assess the gathered data. Potential biases in data synthesis could influence the interpretation of our findings. To address this concern, we used open coding and constant comparison techniques from Grounded Theory \cite{glaser1967discovery} to analyze the qualitative data extracted from the selected studies.
    \item \textcolor{black}{ \textbf{Transparency of Quality Assessment Process:} While our replication package \cite{replpack} provides a detailed breakdown of quality assessment scores for each study and each criterion, it does not include per-study justifications or annotations explaining how individual scores were assigned. This may present a limitation in terms of external reproducibility, as other researchers may not fully reconstruct the decision-making rationale behind each score. Although we followed a consistent and internally validated scoring logic (e.g., peer-reviewed venues scored higher, empirical data presence was required for full marks), we recognize that the absence of a fully documented scoring rubric or narrative rationale could reduce transparency. However, to partially mitigate this, we have explicitly described our scoring criteria and interpretation rules in Section~\ref{sec:QualityAssessment}. This clarification improves traceability by explaining the rationale used across all assessed studies. The limitation remains inherent to balancing the breadth and depth of quality assessment in a large-scale review and is acknowledged as a tradeoff in terms of replication of this study.
}
\end{itemize}

\subsection{External Validity}
\label{Threats_EV}
Concerns about external validity relate to the applicability and generalizability of study findings. Our research provides a thorough examination of containerization in multi-cloud environment, with findings, analyses, and conclusions specifically tailored to this domain. To ensure robust external validity, we developed a study protocol that outlines our research methodology. The literature review spanned a decade, from January 2013 to March 2023, and included peer-reviewed articles from eight preeminent databases in the fields of software engineering and computer science, which are listed in Table \ref{tab:stringDatabase}. While the review is anchored in academic research, which may not encapsulate unpublished industry practices, the systematic approach and comprehensive timeframe of our analysis make the insights valuable for both academia and practitioners. To complement and enrich our findings, future work could include an industrial study (e.g., blogs, white papers) that would provide a broader view of the application of containerization in practice

\subsection{Construct Validity}
\label{Threats_ConstructV}
Construct validity concerns to the accuracy of the operational measures used for data collection in a study. The primary constructs of this study revolve around two concepts: “containerization” and “multi-cloud environment”. Utilizing imprecise or incomplete search terms, or employing unsuitable search strategies, can lead to potential pitfalls such as overlooking pertinent papers or including numerous irrelevant ones during the search phase, and omitting relevant papers during the selection phase. To counter these risks, we implemented the following measures: (i) we initiated a pilot search to verify the relevance and comprehensiveness of our search terms; and (ii) we searched paper on eight databases renowned for computer science and software engineering research. Additionally, we customized search string according to each database syntax.

\subsection{Conclusion Validity}
\label{Threats_ConclusionV}
Threats to conclusion validity relate to factors, such as data inaccuracies, that can impede drawing accurate conclusions. To mitigate these threats, we adhered to best practices, including the search protocol, pilot search, and pilot data selection, as recommended by Kitchenham \textit{et al.} \cite{AR26} and Petersen \textit{et al.} \cite{petersen2008}. Moreover, to further ensure the validity of our conclusions, the authors engaged in multiple brainstorming sessions to collaboratively interpret the results and finalize the conclusions.

\section{Related Work}
\label{Sec_RelatedWork}

This section reviews the most relevant existing research in terms of secondary studies such as literature surveys, state-of-the-art analysis, systematic reviews, and mapping studies that consolidate published literature on containerization in multi-cloud environment. The review focuses on (i) existing challenges, proposed solutions, and emerging trends detailed in Section \ref{subsec:Related_challenge} and (ii) container-based deployment of multi-cloud systems in Section \ref{subsec:features}. 

\subsection{Challenges, Solutions, and Performance of Multi-cloud Containers} \label{subsec:Related_challenge}

In recent years, several literature surveys have been published to analyze state-of-the-art on the applications \cite{RW2}, tools and technologies \cite{RW5}, potential and limitations \cite{RW22}, as well as emerging trends \cite{RW45} of container-based solutions for cloud computing systems, discussed below.  

\subsubsection{Challenges, Applications, Tools, and Emerging Trends} 
Containers in multi-cloud environment provide a standardized and portable way to package, deploy, and run applications, ensuring seamless deployment and management across diverse cloud platforms \cite{RW3} \cite{RW6}. One of the earliest SLR-based studies by Pahl \textit{et al.} \cite{RW2} on containerized clouds reviewed a total of 46 studies to identify, taxonomically classify and systematically compare the existing research on containers and their application in the context of cloud-based systems. The SLR classified and compared the selected research studies using a conceptual framework to highlight existing trends and needs for future research. SLR results indicate container orchestration, microservice delivery, continuous development and deployment as emerging trends of research on containerized clouds. The SLR in \cite{RW3} reviewed a total of 88 studies (published from 2011 - 2021) on multi-cloud systems. Similar survey studies such as \cite{RW6} concepts, challenges, requirements and future directions for multi-cloud environment are discussed. A survey of existing approaches and solutions provided by different multi-cloud architectures is entailed along with analysis of the pros and cons of different architectures while comparing the same \cite{RW24}. 

\subsubsection{Performance and Resource Utilization}
In a study by Bentaleb \textit{et al.} \cite{RW22}, the authors follow the taxonomical classification from \cite{RW2} to categorize container-based technologies for cloud systems.  The study argues for the need of performance metrics to objectively define and evaluate quality attributes such as resource virtualization, service elasticity, orchestration, and multi-tenancy in containerized cloud systems. The study also highlights the needs for future research in terms of best practices, i.e., patterns and tactics (enabling reuse) and tools (supporting automation) for container-based cloud deployment. In the context of performance, the studies \cite{RW5} \cite{RW45} report literature review and experimental analysis of factors that influence performance of container-based cloud systems. Specifically, Casalicchio \textit{et al.} \cite{RW5} reviewed a total of 97 research studies investigating performance evaluation and run-time adaptation of container-based solution. The study highlighted several unsolved challenges such as I/O throughput optimization, performance prediction, and multi-layer monitoring performance bottlenecks. Moreover, Watada \textit{et al.}~\cite{RW45} conducted an experimental study to compare the performance of VMs, containers and uni kernels in terms of and technological maturity using standard benchmarks and observed containers to optimise performance of container-based. The performance of containerized clouds is also determined by resource utilization, which includes but is not limited to CPU utilization, memory footprints, energy consumption, network bandwidth, and execution time.

Kapil \textit{et al.} \cite{RW12} reviewed 64 research papers to gain insight into resource allocation, management, and scheduling. Furthermore, limitations of existing resource allocation algorithms are discussed, indicating the need to investigate algorithms or techniques of performance optimization of containers for cloud systems. The study by Maenhaut \textit{et al.} \cite{RW23} provided an overview of the current state of the art regarding resource management within the broad sense of cloud computing, complementary to existing surveys in the literature. The study investigated how research is adapting to the recent evolution within cloud solutions, including container technologies. 


\subsection{Container-based Cloud Orchestration, Deployment, and Security} \label{subsec:features}
This section reviews the most relevant research on the container-based deployment with a focus on orchestration and security issues of cloud-based systems, detailed below.

 \subsubsection{Orchestration and Deployment} Naweiluo \textit{et al.} \cite{RW7} explored containerization in High Performance Computing (HPC) environments, contrasting it with cloud computing. The results of this study indicate that containers enhance application deployment efficiency, but face challenges in HPC due to high security levels and the need for extensive libraries and packages that affect cloud portability, often resulting in vendor lock-ins. The study also provides a survey and taxonomy on containerization and orchestration efforts in HPC, pointing out differences with cloud environments and identifying future research and engineering potentials. A systematic mapping study by Naylor \textit{et al.}\cite{RW38} explores virtualization in container-based cloud deployments.  The mapping study, based on a comprehensive review of major databases, identifies a significant research gap in the performance evaluation of containers, underscoring the need for further investigation on cloud deployment. Carlos \textit{et al.} \cite{RW14} present an optimization strategy for deploying microservices in multi-cloud environment focusing on minimizing cloud service costs, network latency, and startup time for new microservices. Their approach uses a Non-dominated Sorting Genetic Algorithm II (NSGA-II) compared to a Greedy First-Fit algorithm, demonstrating a 300\% improvement with NSGA-II. This highlights its effectiveness in container and VM orchestration, offering a significant enhancement in deployment of multi-cloud solution in containers. Emiliano \textit{et al.} \cite{RW19} survey container orchestration, proposing a reference architecture for autonomic orchestrators, and identifying research challenges in the field. Their work emphasizes the importance of container technologies in cloud environments and the need for advanced orchestration solutions to manage complex multi-container applications effectively. Uchechukwu Awada \cite{RW35} reviews container orchestration tools and platforms, comparing the architectures, components, and capabilities of several Container Service Platforms (CSPs) and orchestration tools like Amazon ECS, Kubernetes, Docker Swarm, and Mesos. The study offers insights into the current state of container orchestration and suggests future research directions, serving as a guide for developers and organizations.

Koustabh \textit{et al.} \cite{RW21} explore multi-container deployment on IoT gateways to meet the stringent latency requirements of advanced IoT applications. Through their study within the AGILE project, they highlight containerization's role in overcoming the diversity and resource constraints of IoT gateways, showcasing containerization's advantages for IoT gateway performance optimization. Their research underscores the potential of containerized environments in enhancing application compatibility, portability, and efficient deployment across diverse hardware architectures. Matteo \textit{et al.} \cite{RW44} introduce the Adaptive Container Deployment (ACD) model to optimize containerized application deployment in geo-distributed environments, focusing on IoT and fog computing resources. ACD, formulated as an Integer Linear Programming problem, aims to improve application performance by leveraging containers' horizontal and vertical elasticity. The study evaluates ACD's effectiveness against greedy heuristics, highlighting the need for advanced orchestration solutions to exploit emerging computing environments' characteristics efficiently.
 
 \subsubsection{Security of Containerized Clouds}
Nicolae \textit{et al.} \cite{RW31} analyze security challenges in cloud orchestration for multi-cloud deployments, proposing a security-enabled orchestration framework. Their research identifies potential attack scenarios and security enforcement mechanisms, aiming to enhance security guarantees for cloud operators in a multi-cloud setting.  Mohammad \textit{et al.} \cite{RW26} offer a detailed analysis of cloud computing, defining its essential characteristics, architecture, service models (SaaS, PaaS, IaaS), and deployment models. They discuss the security requirements for public and private clouds across different service models, aiming to provide researchers with a comprehensive understanding of cloud computing's potential and security challenges. Sari \textit{et al.} \cite{RW34} survey the landscape of container security, addressing challenges and solutions across four generalized use cases within the host-container threat landscape. Their analysis covers both software-based solutions utilizing Linux kernel features and hardware-based solutions for enhanced security. The study aims to clarify container security requirements and encourage further research in addressing potential vulnerabilities and attacks. Hendrik \textit{et al.} \cite{RW28} tackle privacy, security, and trust issues in cloud computing through a multi-cloud architecture perspective. They propose a novel technique for enhancing security and unifying access control mechanisms across cloud providers, addressing the challenges of inadequate cross-provider APIs and non-unified access control. Rohan \textit{et al.} \cite{RW40} propose a security-enhancing methodology for cloud data storage using container clustering and Docker instances for on-demand encryption. This approach aims to secure data while optimizing resource usage and cost, highlighting the benefits of container technology in improving cloud computing security and efficiency.

\subsection{Conclusive Summary} \label{subsec:comparison}
Current survey-based studies and systematic reviews, discussed above (e.g., \cite{RW5} \cite{RW2}), aim to consolidate the latest findings in published research concerning container-based solutions for cloud computing systems. These studies investigated various aspects of multi-cloud, including exploring applications, tools, and technologies, assessing potential and limitations, and identifying emerging trends. Table 16 presents an overview of the differences between this study and existing secondary studies. The symbols used in the table are as follows: \xmark~indicates ``Yes”, \cmark~indicates ``No”, and \pmark~indicates ``Partially”. \textcolor{black}{This study distinguishes itself from prior secondary studies in multiple critical ways, thereby advancing the state of the art in the domain of containerization for multi-cloud environments.} \textcolor{black}{While existing surveys and mapping studies (e.g., \cite{RW2}, \cite{RW5}, \cite{RW12}) provide valuable overviews of tools, technologies, performance, and resource management, they often focus on isolated aspects or lack integration across key quality and deployment dimensions. In contrast, our SMS presents a comprehensive, theme-based classification of 121 selected studies—making it the most extensive secondary analysis in this domain to date.} First, our work systematically categorizes container utilization in multi-cloud environment into coherent categories and themes. \textcolor{black}{Some of the key themes are Scalability and High Availability, Performance and Optimization, Security and Privacy, and Multi-Cloud Container Monitoring and Adaptation.} We identified and classified seventy-four patterns and strategies for containerization, which existing studies have not provided. \textcolor{black}{This level of operational detail is absent in prior work.} Furthermore, we explored the QAs and associated tactics for containerization in multi-cloud environment. \textcolor{black}{Unlike existing reviews that superficially mention quality attributes, we provide a dedicated framework linking QAs with implementation tactics, offering practical insights for engineering design decisions—an aspect previously unexplored.} \textcolor{black}{Our SMS also identifies and formalizes four distinct challenge-solution frameworks in the areas of security, automation, deployment, and monitoring. These frameworks synthesize fragmented research findings into cohesive, actionable knowledge structures, not previously consolidated in the literature.} Lastly, \textcolor{black}{our comparative analysis (Table \ref{Comparison}) demonstrates that none of the previous studies collectively cover the breadth and depth addressed by our SMS, particularly in integrating performance, security, orchestration, and QA strategies across the entire lifecycle of container-based multi-cloud systems.} \textcolor{black}{By addressing these multiple dimensions holistically, our study not only fills significant gaps in the literature but also provides a practical roadmap for researchers and practitioners aiming to optimize container deployments in complex, heterogeneous cloud environments.}

\begin{table*}[h!]
\centering
\footnotesize
\caption{Comparison of this SLR results with existing secondary studies. 
MS indicates ``Mix Study”, LR indicates ``Literature Review”, \cmark\ indicates ``Yes”, \xmark\ indicates ``No”, and \pmark\ indicates ``Partially”}
\label{Comparison}
\renewcommand{\arraystretch}{1.2} 
\setlength{\tabcolsep}{3pt} 

\begin{tabular}{|p{6cm}|p{2cm}|p{0.9cm}|p{0.9cm}|p{0.9cm}|p{0.9cm}|p{0.9cm}|p{0.9cm}|p{0.9cm}|p{0.9cm}|}
\hline
\textbf{Contributions} &This Study& \textbf{\cite{RW2}} & \textbf{\cite{RW5}} & \textbf{\cite{RW12}} & \textbf{\cite{RW22}} & \textbf{\cite{RW45}} & \textbf{\cite{RW6}} & \textbf{\cite{RW24}} & \textbf{\cite{RW31}} \\
\hline
Number of selected studies &121 & 46 &  97& 64 & \xmark &40  & 15 & 11 & \xmark \\
\hline
Study Types & SMS & SMS & LR & LR& LR & MS& LR & LR & MS \\
\hline
Container Roles (Section \ref{ContainerRoles}) & \cmark & \pmark & \xmark & \pmark& \pmark & \pmark & \xmark & \xmark & \pmark \\
\hline
Container Implementation Strategies (Section \ref{ContainerImplementationStrategies}) & \cmark & \xmark & \pmark& \xmark & \pmark & \pmark & \xmark & \xmark & \xmark \\
\hline
Pattern and Strategies (Section \ref{sec:RQ2}) & \cmark & \xmark& \xmark & \xmark & \xmark & \xmark & \xmark & \pmark & \pmark\\
\hline
Quality Attributes and Tactics (Section \ref{Sec:RQ3}) & \cmark & \pmark &\xmark & \xmark & \pmark & \xmark & \xmark & \xmark & \xmark \\
\hline
Security Challenges and Solution (Section \ref{SecurityChallengesandSolutionsFramework}) & \cmark & \pmark& \pmark & \xmark & \pmark & \xmark & \xmark & \xmark & \xmark \\
\hline
Automation Challenges and Solution (Section \ref{sec:Automationframework}) & \cmark &\pmark & \pmark& \xmark & \pmark & \xmark & \xmark & \xmark& \xmark \\
\hline
Deployment Challenges and Solution (Section \ref{DeploymentFramework}) & \cmark &\pmark & \xmark & \xmark & \xmark & \xmark & \xmark & \xmark & \xmark \\
\hline
Monitoring Challenges and Solution (Section \ref{sec:monitoring}) & \cmark & \xmark & \pmark & \xmark & \xmark & \xmark & \xmark & \xmark & \xmark \\
\hline
Tools and Frameworks (Section 3.9) & \cmark &\pmark & \pmark & \xmark & \pmark & \xmark & \xmark & \xmark & \xmark\\
\hline
\end{tabular}
\end{table*}

\section{Conclusions}
\label{Sec_Conclusion}

This SMS presents the current state of research regarding containerization in multi-cloud environment, focusing on aims of selected studies, the roles of containers, containers implementation strategies, architectural patterns and strategies, quality attributes and corresponding tactics. Additionally, this SMS also explores deployment, monitoring, security challenges, accompanied by their respective solutions, tools, and frameworks used to implement containerized applications in multi-cloud environment. We investigated 121 relevant studies to answer the RQs in Table~\ref{tab:research_questions} with key findings of this SMS include:

\begin{itemize}

 \item The findings of this SMS reveal various insights regarding SMS demographics, publishers, publication types, authors' affiliations, and research themes. The yearly distribution reached its peak in 2016, signifying a shifting level of interest over time. The field of publishing is largely dominated by IEEE (55.37\%), with conferences emerging as the primary avenue for publication. Notably, academia (75.02\%) stands out as the leading affiliation among authors. The prominent research themes encompass \textit{Orchestration and Management}, \textit{Scalability and High Availability}, \textit{Performance and Optimization}, and \textit{Security and Privacy}. These themes underscore crucial aspects such as container management, scalability, performance, and security within the context of multi-cloud environment.

\item This SMS identifies and classifies a total of 98 patterns and strategies across 10 subcategories and 4 categories for container-based applications in multi-cloud environment. These patterns and strategies encompass a wide range of aspects such as cloud architecture models, communication and networking, deployment, service models, cloud management, container management, edge computing, IoT strategies, security, resilience, fault-tolerance, and cloud migration. These findings are significant for the application and dvancement of containerization-based applications in multi-cloud environment, providing valuable insights into optimizing architecture, communication, management, security, and migration strategies.

\item This SMS identifies and classifies QAs and tactics from selected studies regarding containerized applications in multi-cloud environment. A total of ten QAs and their related terms are identified. Along with these QAs, a total of 47 tactics are identified to enhance the QAs. Notable takeaways include the utilization of machine learning and location-aware service brokering to enhance the performance and efficiency of container-based applications in multi-cloud environment.

\item This SMS presents a comprehensive security challenge-solution framework for containerized applications in multi-cloud environment. This framework offers valuable insights and actionable strategies for practitioners and researchers to enhance the security of containerized applications, address complex multi-cloud security concerns, and ensure compliance with various regulatory requirements, thereby enabling the development of robust and secure containerized applications for multi-clouds.

\item This SMS presents a catalogs for the automation challenge-solution catalogs for containerized applications in multi-cloud environment. The automation challenge-solution catalog offers a comprehensive guide for practitioners and researchers to address the complex challenges associated with automating various aspects of deploying, managing, scaling, testing, migrating, and ensuring interoperability of containerized applications in multi-cloud environments. 

\item The third catalog presented in this SMS is the deployment challenge-solution catalog for containerized applications in multi-cloud environment. This catalog provides valuable insights and practical guidance to practitioners and researchers alike, aiding practitioners in navigating the complexities of multi-cloud deployment through specific solutions to challenges, while also serving as a comprehensive resource for researchers to delve into various aspects of containerized application deployment, orchestration, security, scalability, and more, thus contributing to the advancement of knowledge and innovation in the field of multi-cloud deployment.

\item Fourth catalog presented in this SMS is  monitoring challenge-solution catalog for containerized applications in multi-cloud environment. This catalog serves as a comprehensive guide for practitioners, offering specific solutions to challenges related to monitoring performance, resource management, security, optimization, system complexity, and multi-cloud coordination, facilitating their understanding of complex monitoring scenarios and providing actionable insights. Furthermore, researchers can benefit from this catalog by gaining a deeper insight into the nuances of monitoring containerized applications in multi-cloud environment, and it can serve as a foundation for further research and innovation in the field of multi-cloud monitoring.

\end{itemize}

The findings of this SMS will benefit researchers who are interested in understanding the state of research on containerized applications in multi-cloud environment and conducting further investigations to address the open research issues highlighted in Section \ref{Sec_Discussion}. Additionally, the insights gained from this SMS will support knowledge transfer to practitioners by providing insights into the challenges, solutions, and methods for monitoring, securing, and optimizing containerized applications in multi-cloud environment. We emphasize the importance for practitioners to develop targeted solutions that effectively tackle monitoring, security, and performance degradation concerns within multi-cloud environment deployment. As a future endeavor, we intend to enhance our SMS by conducting industrial case studies with companies, thereby obtaining practical insights and perspectives from practitioners on the effectiveness and applicability of our proposed catalogs. This approach will allow us to bridge the gap between research and practical implementations concerning containerized applications in multi-cloud environment and contribute to the advancement of both academic and industry understanding in this domain.

\section*{Data availability}
Link to our dataset is in the reference \citep{replpack}.

\section*{Acknowledgments}
This research is funded by Business Finland through QLEAP (2022-24) project, the National Natural Science Foundation of China (NSFC) under Grant No. 62172311, and the Major Science and Technology Project of Hubei Province under Grant No. 2024BAA008.
\section*{Declaration of AI Assistance}
During the preparation of this work, the author(s) used ChatGPT to refine grammar, improve sentence structure, and resolve formatting issues. After utilizing this tool, the author(s) thoroughly reviewed and edited the content as needed, taking full responsibility for the final publication.
\printcredits

\bibliographystyle{unsrt} 

\bibliography{references}
\balance
\end{sloppypar}
\end{document}